\documentclass[10pt,preprint]{aastex}
\usepackage[]{natbib}
\usepackage{epsfig}
\usepackage{longtable} 

\newcommand{\Msolar}{M$_{\odot}$}
\newcommand{\Rsolar}{R$_{\odot}$}
\newcommand{\kms}{km s$^{-1}$}

\newcommand{\PRV}{P$_{RV}$}
\newcommand{\PPM}{P$_{PM}$}

\newcommand{\SBtwo}{15}  
\newcommand{\SBone}{70}  
\newcommand{\orb}{98}     
\newcommand{\orbm}{85}    
\newcommand{\orbnm}{13}
\newcommand{\magn}{10.8$\leq$V$\leq$16.5}         
\newcommand{\mrv}{-42.36 $\pm$~0.04 \kms} 

\begin{document}

\title{WIYN Open Cluster Study. XXXVI. Spectroscopic Binary Orbits in NGC 188}
\shorttitle{WOCS: Spectroscopic Binary Orbits in NGC 188}

\author{Aaron M. Geller\footnote{Visiting Astronomer, Kitt Peak National Observatory, National Optical Astronomy Observatory, which is operated by the Association of Universities for Research in Astronomy (AURA) under cooperative agreement with the National Science Foundation.}~~and Robert D. Mathieu$^1$,}
\affil{Department of Astronomy, University of Wisconsin - Madison, WI 53706 USA}
\and
\author{Hugh C. Harris}
\affil{United States Naval Observatory, Flagstaff, AZ, 86001, USA}
\and
\author{Robert D. McClure}
\affil{Dominion Astrophysical Observatory - Herzberg Institute of Astrophysics - National Research Council \\ 5071 W. Saanich Road, Victoria, BC V9E 2E7 Canada}
 \shortauthors{Geller et al.}

\begin{abstract}

We present \orb~spectroscopic binary orbits resulting from our ongoing radial-velocity survey of the 
old (7 Gyr) open cluster NGC 188.  All but \orbnm~are high-probability cluster members 
based on both radial-velocity and proper-motion membership analyses.  \SBtwo~of these 
member binaries are double lined.  Our stellar sample spans a magnitude range of \magn~(1.14-0.92 \Msolar)
and extends spatially to 17 pc ($\sim$13 core radii).  All of our binary orbits have
periods ranging from a few days to on the order of 10$^3$ days, and thus are hard binaries 
that dynamically power the cluster.   For each binary, we present the orbital solutions and 
place constraints on the component masses.  Additionally, we discuss a few binaries of note
from our sample, identifying a likely blue straggler - blue straggler binary system (7782),
a double-lined binary with a secondary star which is under-luminous for its mass (5080), 
two potential eclipsing binaries (4705 and 5762), and two binaries which are likely members 
of a quadruple system (5015a and 5015b).
    
\end{abstract}


\section{Introduction}

Within an open cluster, dynamical interactions with hard binaries\footnote{\footnotesize A hard binary
is defined as having an internal energy that is much greater than the energy of the relative motion 
of a single star moving within the cluster \citep{heg74}. 
For solar mass stars in a cluster with a one-dimensional velocity dispersion equal to 
1 \kms, all hard binaries have periods less than $\sim$10$^5$ days.} provide energy to the cluster, 
and can foster a complex interplay of stellar evolution, stellar dynamical exchanges, mass transfer, 
and even stellar collisions.  Such interactions have the potential to result in the formation of ``anomalous'' 
stars that defy standard stellar evolutionary theory, such as blue stragglers (BSs). 
Recent $N$-body simulations \citep[e.g.,][]{hur05} are beginning to illuminate the likely formation 
mechanisms of such anomalous stars within open clusters, and it has become clear that the binary population 
plays a significant role.  Detailed studies of open cluster binary populations are critical to constrain such models 
so that we can study cluster evolution as well as the formation mechanisms of anomalous stars.
Furthermore, accurate and comprehensive surveys of binary populations are essential 
for our understanding of the onset of mass transfer, tidal interactions, initial and present-day mass 
functions, stellar dynamics, and even star formation processes.  

Radial-velocity (RV) surveys offer an efficient way to identify single\footnote{In the following, as in \citet{gel08} (Paper 1)
we use the term ``single'' to identify stars with no significant RV variation; 
a star is termed single if the standard deviation of its RV measurements is less than four times our precision.
Certainly, many of these stars are also binaries, although generally with longer periods and/or lower total mass than the binaries
identified in this study.} and binary open cluster members 
as well as to solve for binary orbital solutions.  Open clusters are 
ideally suited for such surveys as they offer a coeval sample of stars that are generally easily accessible 
through ground-based observations using even modest-sized telescopes.  Spectroscopic binary 
surveys have been carried out for a few well known clusters (e.g., Hyades\, \citet{deb00};
Praesepe, \citet{mer99}, \citet{abt99}, \citet{deb00}; Pleiades, \citet{mer92}; and M67; \citet{mat90}).
Today, the advent of multi-object spectrographs permits surveys of larger stellar samples in more distant 
open clusters, allowing us to explore binary populations as a function of age, stellar density, metallicity and stellar mass.

We present \orb~binary orbits in the old (7 Gyr) open cluster NGC 188, derived from our 
ongoing RV survey of the cluster, covering a magnitude range of \magn~(1.14-0.92 \Msolar), a 1\degr~diameter 
region on the sky (roughly 13 core radii\footnote{We adopt a core radius of 1.3 pc \citep{bon05} at a
distance of 1.9 kpc, which corresponds to 2.35 arcminutes on the sky.}) and, for some binaries, a timespan of 
up to thirty years.  This survey of NGC 188 is part of the  WIYN Open Cluster Study \citep[WOCS;][]{mat00}.
Our detectable binaries all have periods ranging from a few days to on the order 
of 10$^3$ days.  Given an internal velocity dispersion of 0.64 $\pm$ 0.04 \kms~\citep[][hereafter Paper 1]{gel08}, 
these binaries constitute much of the hard-binary population that dynamically powers the cluster.  
In Paper 1, we describe our observations, data reduction and the precision of 
our measurements.  We also provide RV membership probabilities (\PRV) for stars observed $\geq$3 times 
and identify RV variable stars.  In this second paper in the series, we present our complete current RV 
database on the cluster (Section~\ref{data}). In Section~\ref{orbits}, we provide the 
\SBone~single-lined (SB1) and \SBtwo~double-lined (SB2) binary cluster-member orbital solutions derived from this survey.
For each binary, we provide the plotted orbital solution,
tabulated orbital parameters, and constraints on the component masses. In Section~\ref{anom} we discuss a 
few binaries of note, including a likely blue straggler - blue straggler binary system (7782),
a SB2 binary with a secondary star which is under-luminous for its mass (5080), 
two potential eclipsing binaries (4705 and 5762), and two binaries which are likely members 
of a quadruple system (5015a and 5015b).  Finally, in the Appendix we provide the orbital solutions and parameters for the \orbnm~field 
binaries that we have serendipitously discovered over the course of our survey.
The third paper in this series will study the binary frequency of the cluster and analyze the binary distributions in period, 
eccentricity and secondary mass.  With the data analyzed in this series of papers, we will gain a detailed understanding of the 
cluster dynamics, the properties of the hard-binary population and their influence on the formation of anomalous stars like BSs,
and thereby provide valuable constraints for future $N$-body models of NGC 188.

\section{Data} \label{data}

Our NGC 188 stellar sample spans a magnitude range of \magn~and a 1\degr~diameter region on the sky.
Our magnitude limits include solar-mass main-sequence stars, subgiants, giants, and BSs, and
our spatial coverage extends radially to $\sim$13 core radii.  The IDs and coordinates for our 
stellar sample are taken from the \citet{pla03} proper-motion (PM) study.
As explained in Paper 1, our full RV database is composed of two data sets, one from WIYN\footnote{\footnotesize
The WIYN Observatory is a joint facility of the University of Wisconsin - Madison, Indiana University, Yale 
University, and the National Optical Astronomy Observatories.} and one from 
the Dominion Astrophysical Observatory (DAO).  The DAO dataset is composed of RVs measured at the DAO 1.2m and the Palomar 5m
telescopes both converted to the DAO Radial-Velocity Spectrometer (RVS) system.
Here we present our complete current RV database for each of our observed stars in the field of NGC 188 in Table~\ref{RVtable}.
We include in this table both cluster members and nonmembers as well as stars without sufficient observations to derive membership 
information.  We refer the reader to Paper 1 for thorough descriptions of our stellar sample and its completeness, and where 
we provide our findings on cluster membership and velocity variability.

We show data for two stars, one SB2 binary and one single star, in Table~\ref{RVtable}, and provide the full table electronically. 
For individual RV measurements, we list the reduced Heliocentric Julian Date (HJD-2400000 d), 
the observatory at which the observations were taken, using "W" for WIYN, "D" for DAO and "P" for Palomar, 
the measured RV, and the cross-correlation peak height for WIYN measurements as a guide to the 
quality of measurement (with a maximum value of 1; see Paper 1 for a detailed description of the precision of our data as 
a function of the peak height).  For the binaries with orbital solutions, we also provide the residual (O-C), 
derived as the observed minus the expected RV from the orbital solution, and the phase. 
For SB2 binaries with orbital solutions, we provide RVs and cross-correlation peak heights (where available) for 
both stars and their respective residuals.

Observations taken at the WIYN 3.5m range in date from October 1995 through August 2008.
Observations made at the DAO 1.2m range in date from February 1980 through November 1996.  
All observations prior to 1980 were taken at the Palomar 5m, with the earliest 
observations taken in December 1973.
We have found no zero-point offset between the WIYN and DAO data sets (Paper 1), 
and have thus integrated both sets of measurements without modification into the single RV data set presented here.  
The precision of the WIYN data is 0.4 \kms~and of the DAO data is 1.0 \kms~(Paper 1).

\begin{deluxetable}{l c c c c c c c c c}
\tabletypesize{\scriptsize}
\tablewidth{0pt}
\tablecaption{Radial-Velocity Data Table\label{RVtable}}
\tablehead{\colhead{ID} & \colhead{HJD-2400000} & \colhead{T} & \colhead{RV$_1$} & \colhead{Correlation} & \colhead{(O-C)$_1$} & \colhead{RV$_2$} & \colhead{Correlation} & \colhead{(O-C)$_2$} & \colhead{Phase} \\
\colhead{} & \colhead{(days)} & \colhead{} & \colhead{(km s$^{-1}$)} & \colhead{Height$_1$} & \colhead{(km s$^{-1}$)} & \colhead{(km s$^{-1}$)} & \colhead{Height$_2$} & \colhead{(km s$^{-1}$)} & \colhead{}}
\startdata
 3344 &           &   &         &      &        &         &      &        &       \\
      & 50328.945 & W &   -63.3 & 0.75 &   ~0.4 &   -20.1 & 0.63 &   -0.4 & 0.571 \\
      & 50331.848 & W &   -63.2 & 0.43 &   ~0.5 &   -18.7 & 0.32 &   ~1.0 & 0.594 \\
      & 50331.941 & W &   -64.6 & 0.51 &   -0.9 &   -19.1 & 0.39 &   ~0.6 & 0.595 \\
      & 50613.922 & W &   -44.8 & 0.59 &   -0.3 &         &      &        & 0.825 \\
      & 50614.746 & W &   -43.1 & 0.93 &   ~0.3 &         &      &        & 0.831 \\
      & 50615.660 & W &   -42.6 & 0.56 &   -0.5 &         &      &        & 0.838 \\
      & 50616.672 & W &   -41.9 & 0.90 &   -1.3 &         &      &        & 0.846 \\
      & 50616.895 & W &   -42.4 & 0.94 &   -2.0 &         &      &        & 0.848 \\
      & 50653.902 & W &   -19.8 & 0.71 &   ~1.3 &   -66.3 & 0.60 &   -0.8 & 0.141 \\
      & 50703.820 & W &   -62.9 & 0.66 &   ~0.2 &   -21.2 & 0.53 &   -0.9 & 0.535 \\
      & 50714.910 & W &   -63.0 & 0.61 &   ~0.3 &   -19.0 & 0.49 &   ~1.1 & 0.623 \\
      & 50780.637 & W &   -20.5 & 0.58 &   ~0.9 &   -66.4 & 0.51 &   -1.2 & 0.143 \\
      & 50797.961 & W &   -42.2 & 0.85 &   -0.3 &         &      &        & 0.280 \\
      & 50815.594 & W &   -56.9 & 0.74 &   ~0.2 &   -30.5 & 0.67 &   -3.8 & 0.419 \\
      & 50918.957 & W &   -35.5 & 0.49 &   ~0.1 &   -47.4 & 0.48 &   ~2.4 & 0.237 \\
      & 50920.941 & W &   -37.8 & 0.55 &   ~0.2 &   -48.7 & 0.52 &   -1.3 & 0.252 \\
      & 50921.941 & W &   -40.2 & 0.41 &   -1.0 &   -47.4 & 0.40 &   -1.3 & 0.260 \\
      & 50976.676 & W &   -59.3 & 0.62 &   ~0.8 &   -23.9 & 0.53 &   -0.4 & 0.693 \\
      & 50996.750 & W &   -40.3 & 0.84 &   -0.7 &   -46.5 & 0.82 &   -0.9 & 0.852 \\
      & 51127.672 & W &   -32.2 & 0.78 &   ~0.7 &   -54.5 & 0.70 &   -1.6 & 0.887 \\
      & 51128.648 & W &   -28.6 & 0.83 &   ~2.8 &   -54.3 & 0.73 &   ~0.2 & 0.895 \\
      & 51174.980 & W &   -40.3 & 0.89 &   -1.0 &   -46.8 & 0.85 &   -0.8 & 0.261 \\
      & 51176.012 & W &   -42.1 & 0.93 &   -1.6 &         &      &        & 0.269 \\
      & 51179.039 & W &   -41.6 & 0.77 &   ~2.2 &         &      &        & 0.293 \\
      & 51186.973 & W &   -50.8 & 0.75 &   ~0.5 &   -33.4 & 0.70 &   -0.3 & 0.356 \\
      & 51243.863 & W &   -45.0 & 0.67 &   ~2.5 &   -39.3 & 0.66 &   -2.2 & 0.806 \\
      & 51355.922 & W &   -58.6 & 0.80 &   ~1.5 &   -25.4 & 0.68 &   -1.9 & 0.692 \\
      & 53389.984 & W &   -52.3 & 0.52 &   -0.6 &   -29.8 & 0.61 &   ~2.8 & 0.777 \\
 3359 &           &   &         &      &        &         &      &        &       \\
      & 50614.777 & W &   -42.7 & 0.91 &        &         &      &        &       \\
      & 50616.922 & W &   -42.9 & 0.91 &        &         &      &        &       \\
      & 50653.930 & W &   -42.6 & 0.89 &        &         &      &        &       \\
      & 50703.836 & W &   -42.2 & 0.86 &        &         &      &        &       \\
      & 50714.953 & W &   -43.0 & 0.67 &        &         &      &        &       \\
      & 50780.648 & W &   -42.5 & 0.85 &        &         &      &        &       \\
      & 50797.992 & W &   -42.2 & 0.88 &        &         &      &        &       \\
      & 50815.625 & W &   -41.8 & 0.92 &        &         &      &        &       \\
      & 50857.012 & W &   -42.0 & 0.75 &        &         &      &        &       \\
      & 50918.977 & W &   -42.2 & 0.63 &        &         &      &        &       \\
      & 50920.961 & W &   -41.5 & 0.83 &        &         &      &        &       \\
      & 50976.695 & W &   -42.3 & 0.77 &        &         &      &        &       \\
      & 50996.777 & W &   -42.5 & 0.77 &        &         &      &        &       \\
      & 51498.625 & W &   -42.2 & 0.95 &        &         &      &        &       \\
      & 51921.875 & W &   -42.3 & 0.85 &        &         &      &        &       \\
      & 52462.945 & W &   -42.0 & 0.92 &        &         &      &        &       \\
      & 53074.992 & W &   -43.0 & 0.75 &        &         &      &        &       \\
\enddata
\\
(This table is available in its entirety in a machine-readable form in the online journal. A portion is shown here for guidance regarding its form
and content)
\end{deluxetable}

\section{Spectroscopic Binary Orbits} \label{orbits}

In the following section, we present our \orbm~orbital solutions of the binary members of NGC 188.
We first discuss our \SBone~SB1 binaries and then our \SBtwo~SB2 binaries.  For both sets, we provide the tabulated 
orbital parameters, plotted orbit curves and component mass estimates.

\subsection{Single-Lined Orbital Solutions} \label{SB1}

For each SB1 binary, we solve for the orbital solution using the data given in Table~\ref{RVtable}.
We provide the plotted orbital solutions in Figure 1; for each binary we plot the orbit in the top panel
and the RV residuals in the bottom panel.  In Table~\ref{SB1tab} we provide the orbital elements for 
each binary in two rows, where the first row includes the binary ID, the orbital period ($P$), the number of orbital cycles observed, 
the center-of-mass RV ($\gamma$), the orbital amplitude ($K$), the eccentricity ($e$), the longitude of periastron ($\omega$), 
a Julian Date of periastron passage ($T_\circ$), the projected semi-major axis ($a \sin i$), the mass function ($f(m)$), the rms 
residual velocity from the orbital solution ($\sigma$), and the number of RV measurements ($N$).  
Where applicable, the second row contains the respective errors on each of these values.
In Table~\ref{SB1masstab}, we present physical properties for each SB1, including the WOCS ID, 
the $V$ magnitude and the $(\bv)$ color \citep[both from][]{ste04}, the radial distance from the cluster center (in arcminutes), 
the RV membership probability (\textit{\PRV}; from Paper 1), the PM membership probability 
\citep[\textit{\PPM}; from][]{pla03}, a photometric estimate for the mass of the primary ($M_1$), a lower limit for the mass of the 
secondary ($M_2$ min), and finally a photometric estimate for the mass of the secondary ($M_2$).  

The photometric estimates for the primary and secondary masses are derived simultaneously across the available $UBVRI$ photometry
for each binary using a photometric deconvolution technique.  We use the observed $(U\!-\!V)$, $(\bv)$, $(V\!-\!R)$, and $(V\!-\!I)$ 
colors, where available, and $V$ magnitudes (as compiled by \citet{ste04}) along with a 7 Gyr, solar-metallicity Padova 
isochrone\footnote{For the isochrone, we set $E(\bv) =$ 0.025 and $(m-M)_V =$ 11.23 \citep{for07}.} 
\citep{gir02} to produce a set of synthetic binaries.  
This set of binaries contains primary stars within a range of masses whose magnitudes extend from the observed $V$ magnitude to 
this magnitude plus 0.75 (as this would be the contribution from an equal mass companion) and, for each 
primary star, a set of secondary stars of equal or lesser mass.  
The component masses of the synthetic binary that has a composite $V$ magnitude and colors in the available photometric bands that most closely 
match the observed $V$ magnitude and colors, in both color-magnitude and color-color space, are taken as the photometric primary and 
secondary mass estimates.

We only attempt to quote masses for the main-sequence, sub-giant and giant binaries.  We caution the reader that,
for binaries with mass ratios $\lesssim$0.5, the photometric masses are less certain, as solar-type binaries with these low 
mass ratios fall very near to the isochrone \citep[e.g.,][]{hur98}.  Also the morphology of the isochrone near the turnoff, makes 
the masses for binaries in this region more sensitive to selection of the distance modulus.
In certain cases (e.g., when the observed binary lies directly on the isochrone to within the photometric errors, or the binary is found 
blueward of the main-sequence or redward of the giant branch), we cannot derive reliable mass estimates in the manner described above.
For such cases, we use the observed $V$ magnitude to estimate an upper limit on the  mass of the primary star.  We have found that the 
secondary must be at least 2.5 magnitudes fainter than the primary at a central wavelength of 5250 \AA~(the central wavelength of the 
WIYN spectra) for the binary to be observed as single lined.  Thus, in these cases, we use this resulting upper 
limit on the $V$ magnitude for the secondary to derive the upper limit on its mass (and note this in the table).  
Finally, for all SB1 binaries we use the primary mass estimate along with the orbital mass function to derive a lower limit 
on the secondary mass.  

For two binaries, 4965 and 4688, we notice a clear trend with time in the residuals of the orbital solutions fit to the observed RVs.
We assume that this trend is due to the presence of an additional long-period companion (or companions). 
Therefore, for each of these two binaries, we fit a polynomial function (of first and second order, respectively) to the residuals,
subtract this fit from the observed RVs, and refit the orbit to these corrected RVs.  There is no trend in the resulting residuals from 
the corrected orbital solutions for either of these binaries.  We note that all of the orbital parameters derived from the corrected 
orbital solutions agree with those of the uncorrected orbital solutions to within the errors, except for two parameters in 
4688; the orbital amplitude, $K$, increased from 6.7 $\pm$ 0.5 in the uncorrected orbit to 9.6 $\pm$ 1.7 in the corrected orbit, 
and the orbital eccentricity, $e$, increased from 0.57 $\pm$ 0.05 in the uncorrected orbit to 0.70 $\pm$ 0.04 in the corrected orbit. 
We show the corrected orbital solution plots in Figure 1 and parameters in Table~\ref{SB1tab}.
In our RV data table, Table~\ref{RVtable}, we include the observed RVs and the residuals to the corrected orbital solutions.  

Curiously, this SB1 photometric deconvolution technique has yielded three cases where we would expect to see the secondary.  Binaries 
4524 and 4843 lie well blueward of the giant branch, and binary 4390 lies well redward of the main sequence.  
We also note that some spectra of 4710 reveal an additionally component for which we have no current explanation.  This binary 
is located near the main-sequence turnoff.  
The rest of the mass estimates yield luminosity ratios in which we indeed would not expect to observe the secondary star, given 
our observing setup.

We use a Monte Carlo technique to estimate the mean uncertainty on our mass estimates, assuming this uncertainty to be derived from two 
main sources: the uncertainties on the photometry and on the isochrone fit.  For binaries in which we can estimate masses from the 
photometric deconvolution technique, we find a mean uncertainty for the primary mass of 0.09 \Msolar~and on the secondary of 0.14 \Msolar.
The standard deviations about these means are 0.15 \Msolar~and 0.20 \Msolar, respectively.
Uncertainties on the minimum secondary masses are found in a similar manner, using the derived primary 
mass uncertainty along with the error on the mass function resulting from the orbital solution, and result in a mean uncertainty 
of 0.04 \Msolar~with a standard deviation about the mean of 0.10 \Msolar. 
Finally, for binaries in which we can only give limits on the primary and secondary masses, we note that the mean uncertainty on the 
$V$ magnitudes for all binaries is 0.011 magnitudes.  For solar-type stars, a shift of this amount to the observed magnitude of a 
main-sequence star results in a shift in mass of 0.003 \Msolar.

\begin{center}
\begin{longtable}{ccc}
\epsfig{file=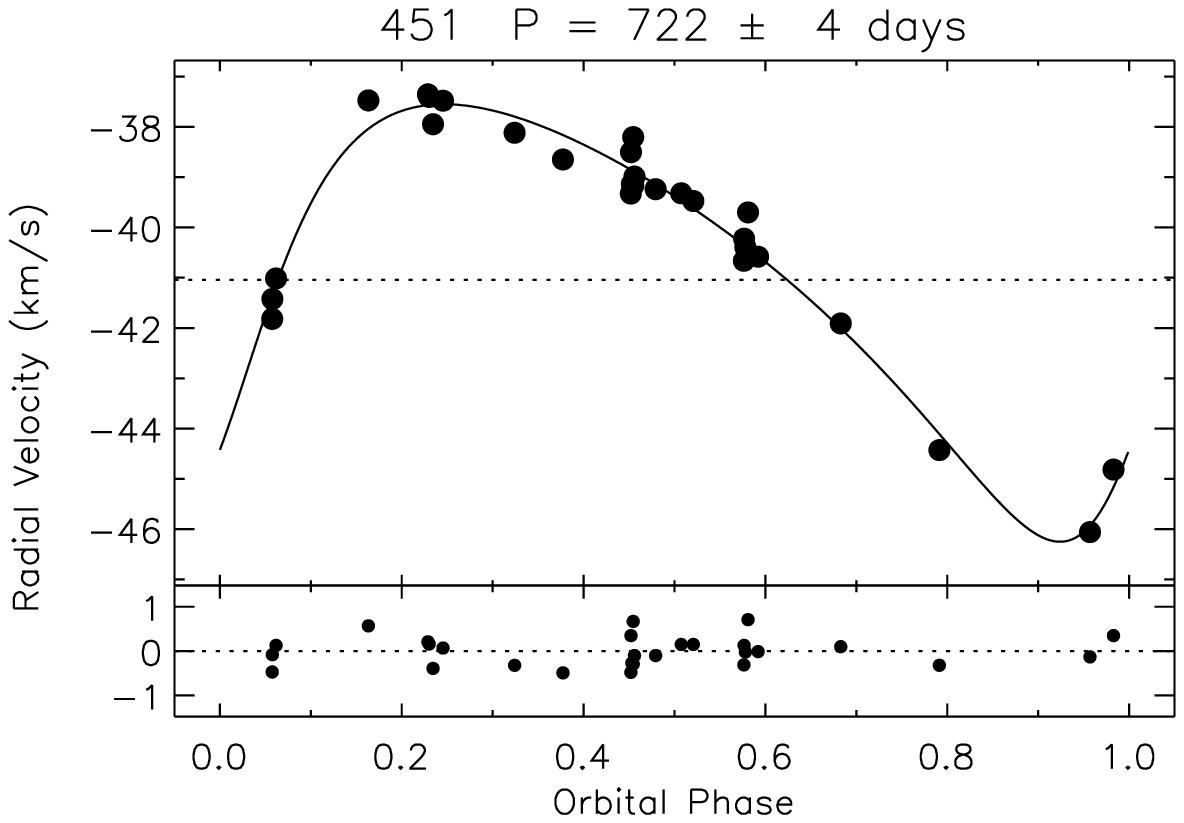,width=0.3\linewidth} & \epsfig{file=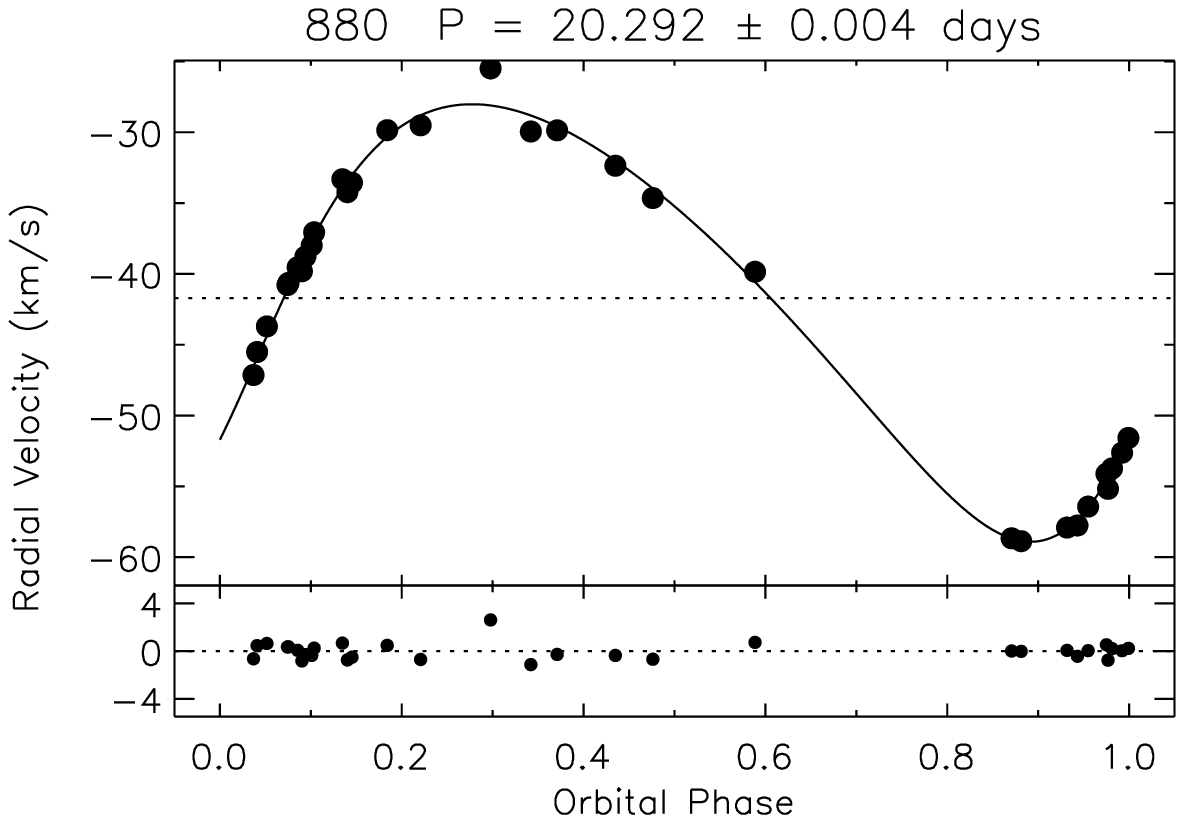,width=0.3\linewidth} & \epsfig{file=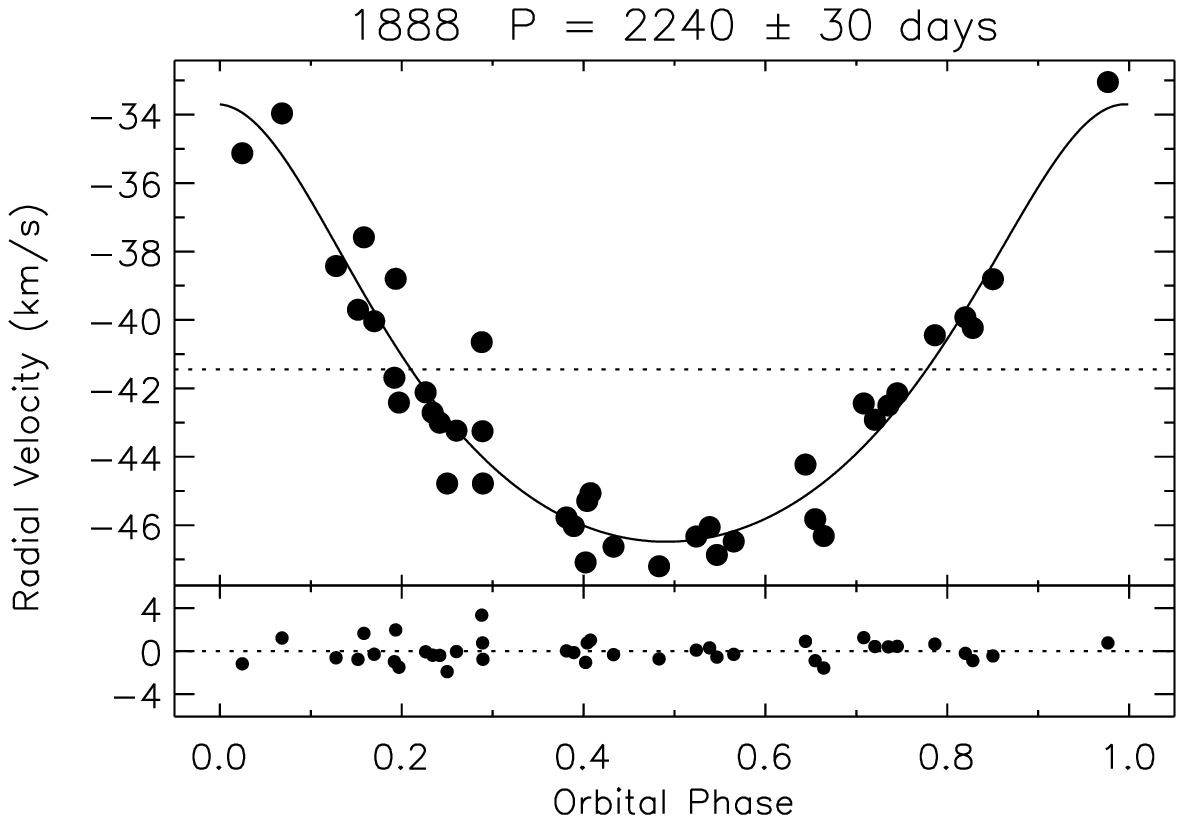,width=0.3\linewidth} \\
\epsfig{file=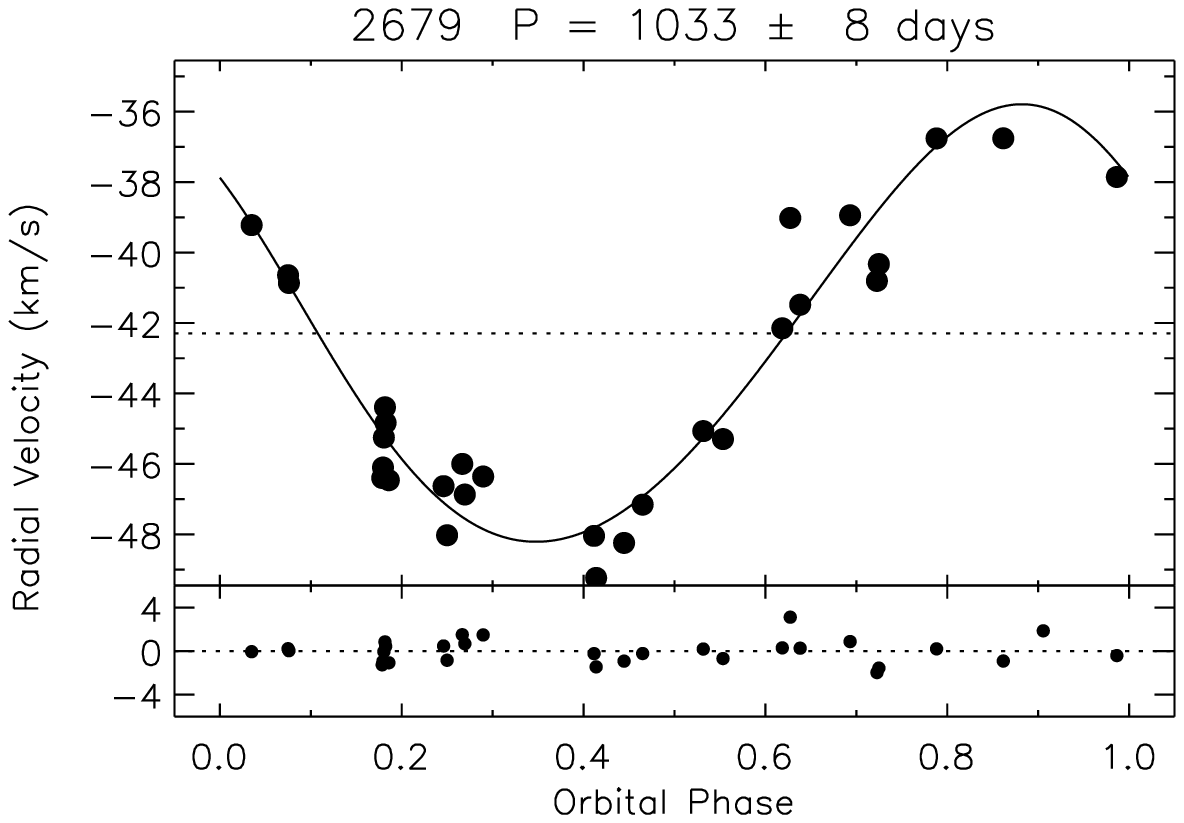,width=0.3\linewidth} & \epsfig{file=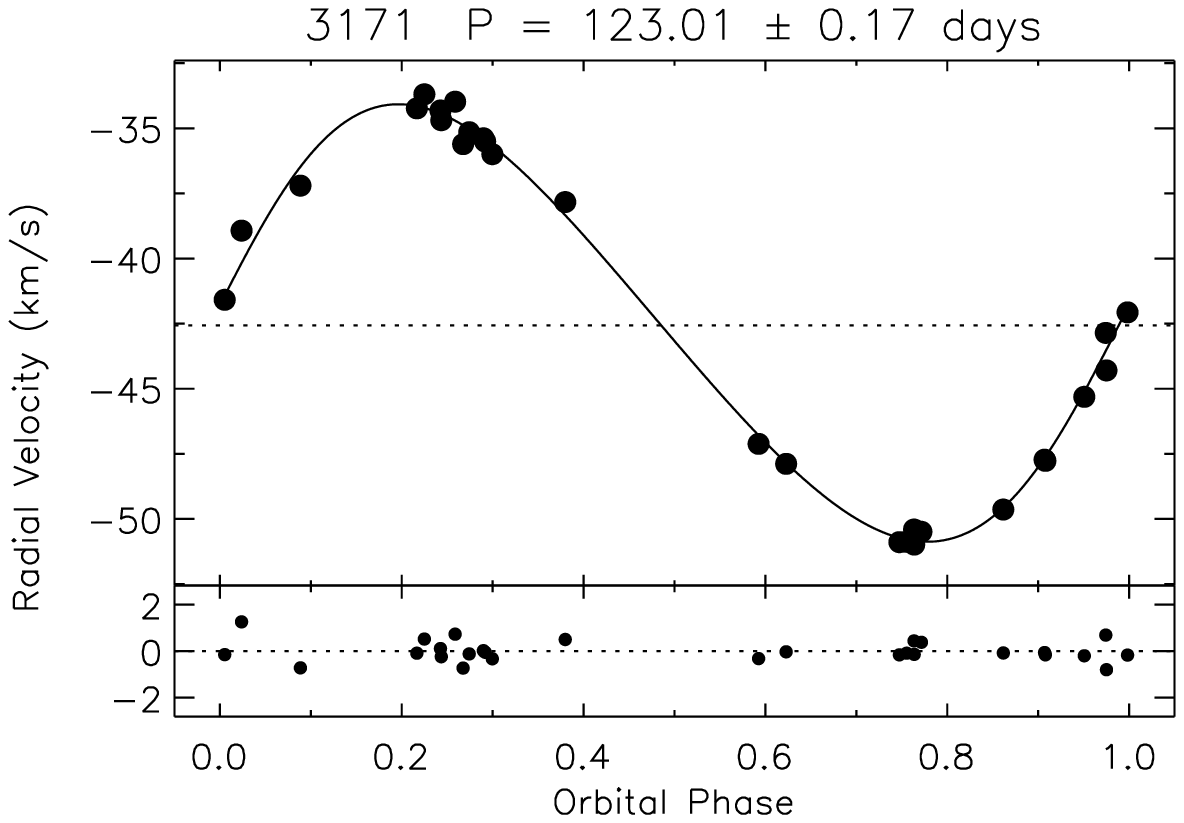,width=0.3\linewidth} & \epsfig{file=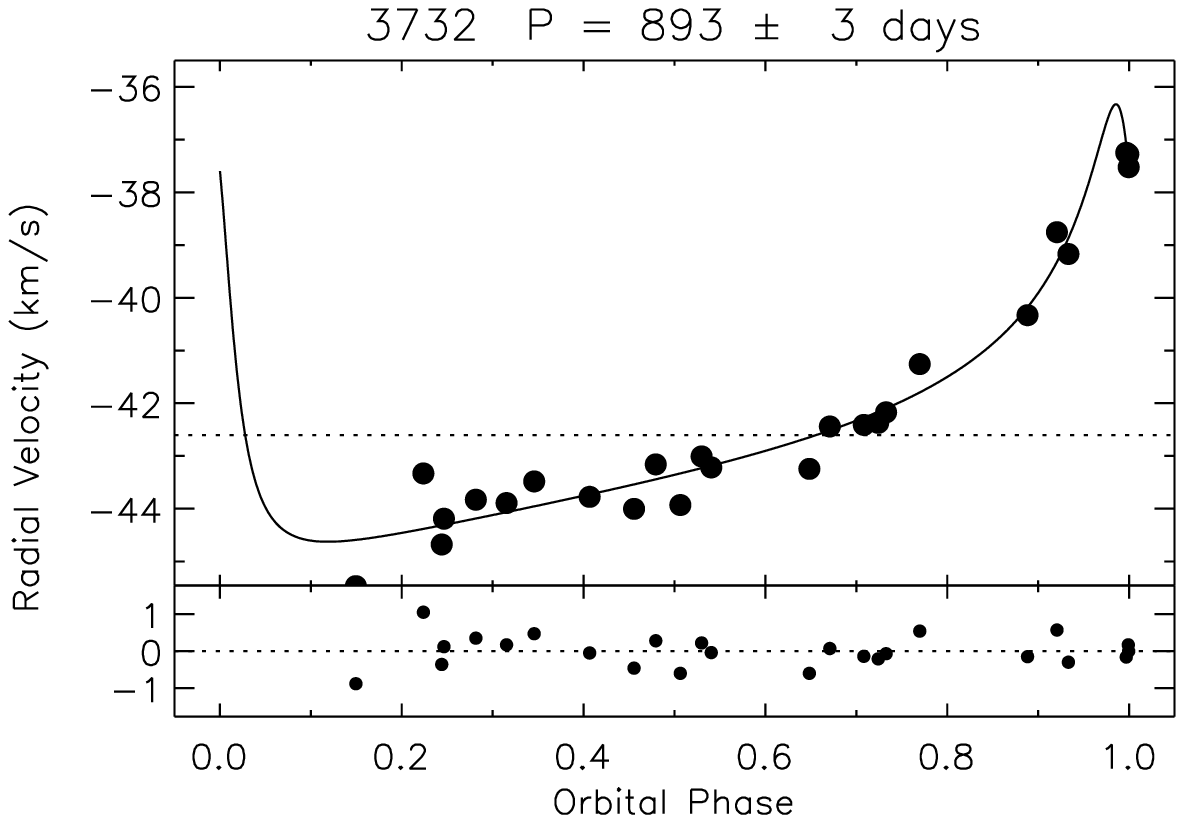,width=0.3\linewidth} \\
\epsfig{file=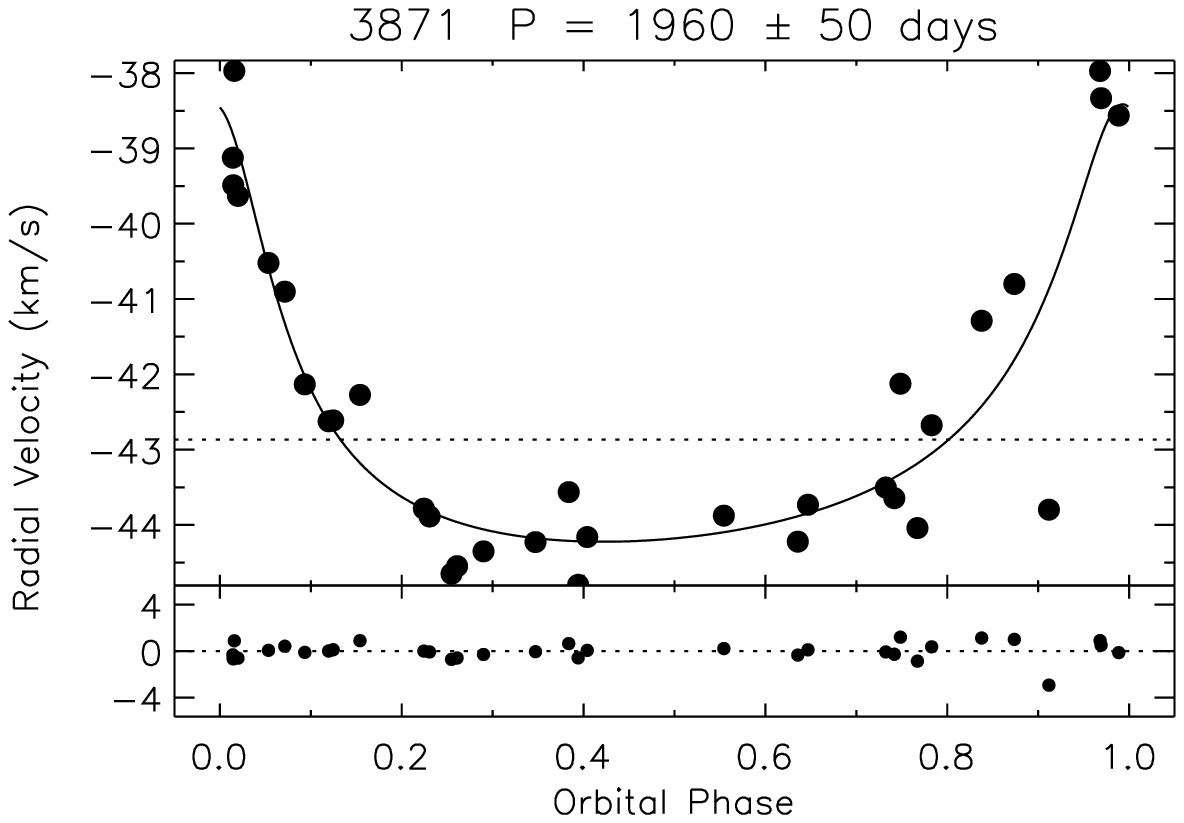,width=0.3\linewidth} & \epsfig{file=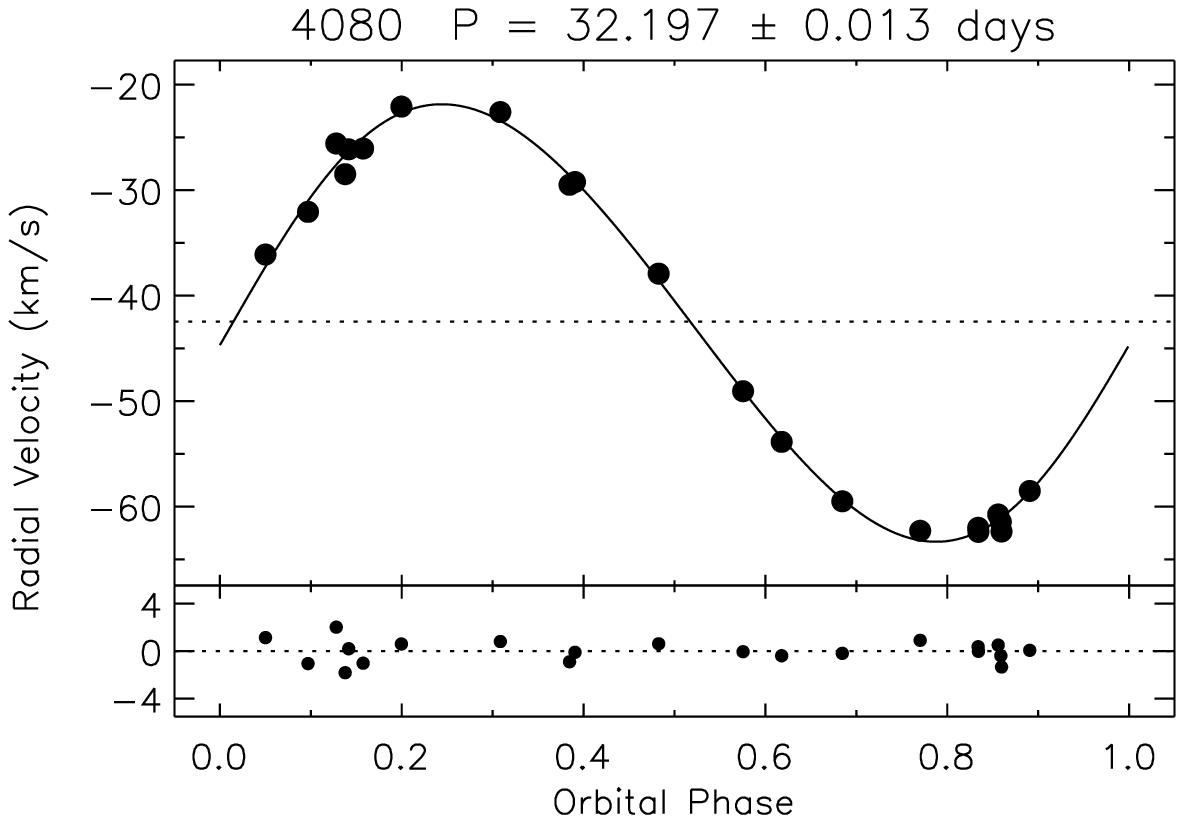,width=0.3\linewidth} & \epsfig{file=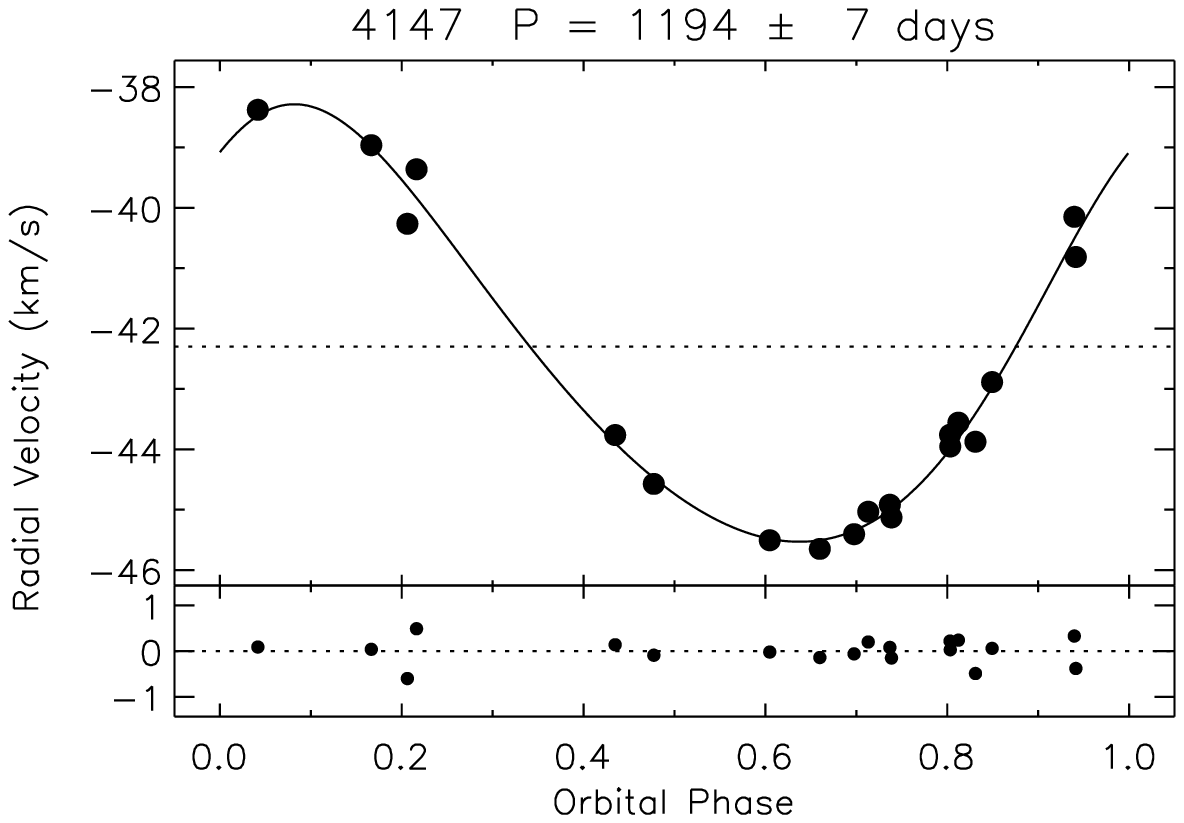,width=0.3\linewidth} \\
\epsfig{file=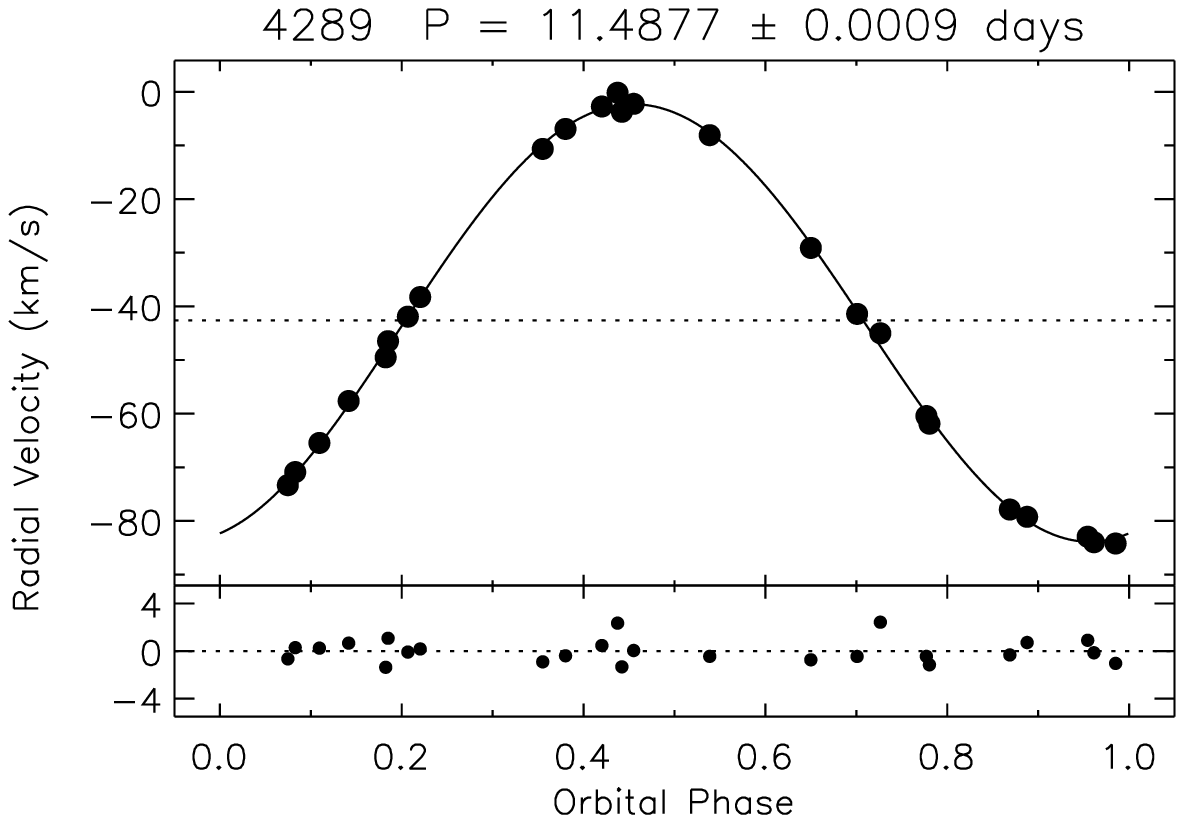,width=0.3\linewidth} & \epsfig{file=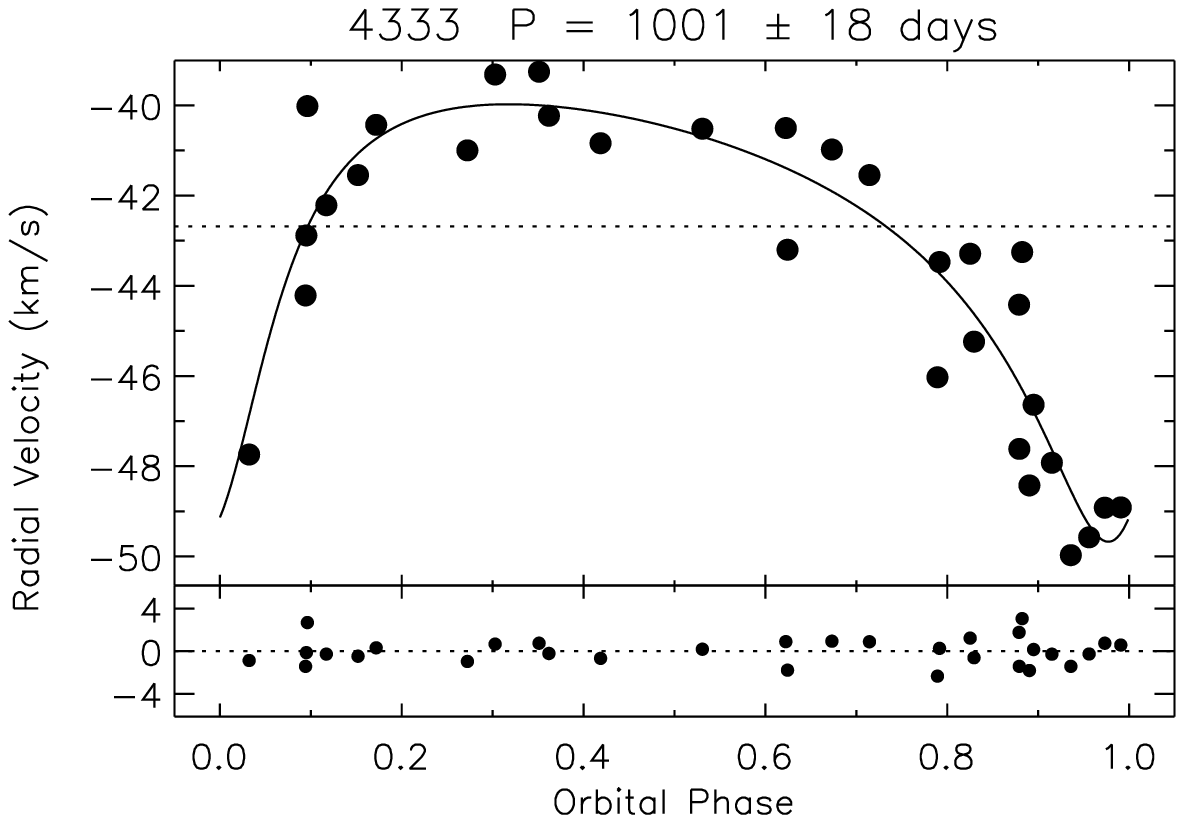,width=0.3\linewidth} & \epsfig{file=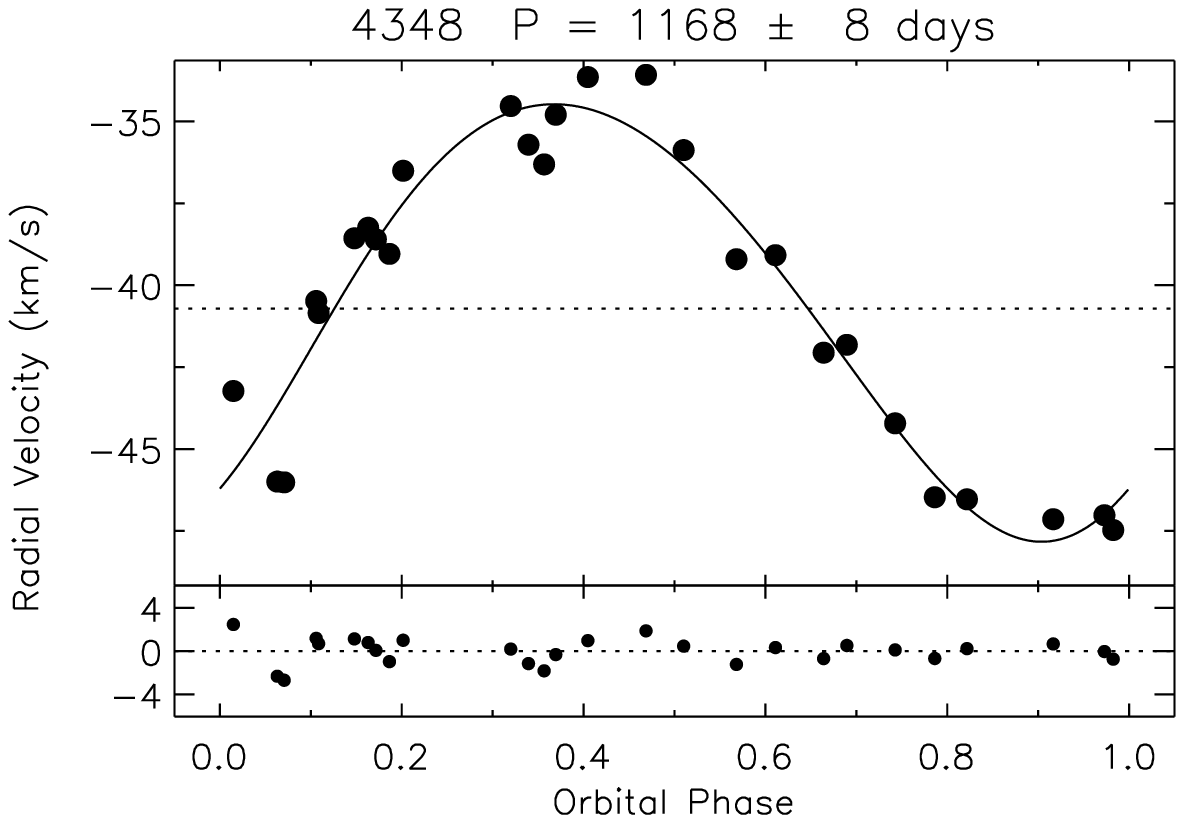,width=0.3\linewidth} \\
\epsfig{file=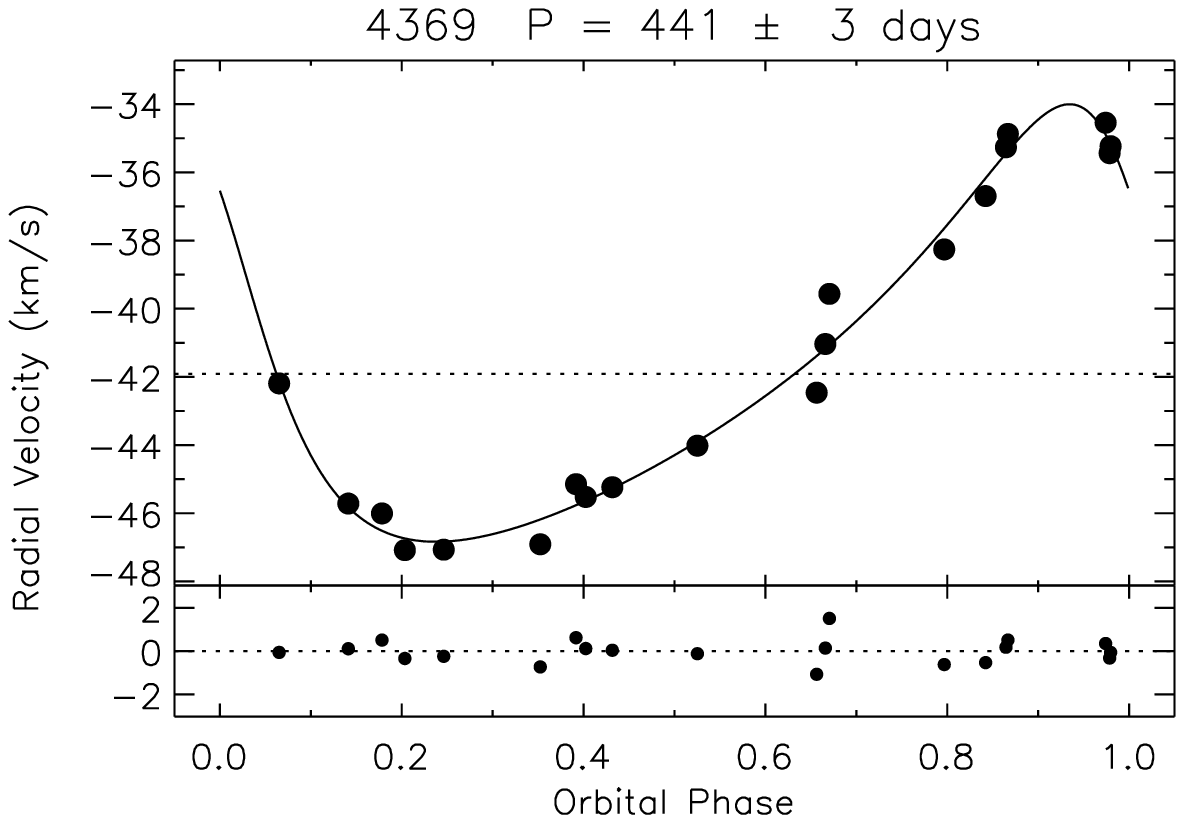,width=0.3\linewidth} & \epsfig{file=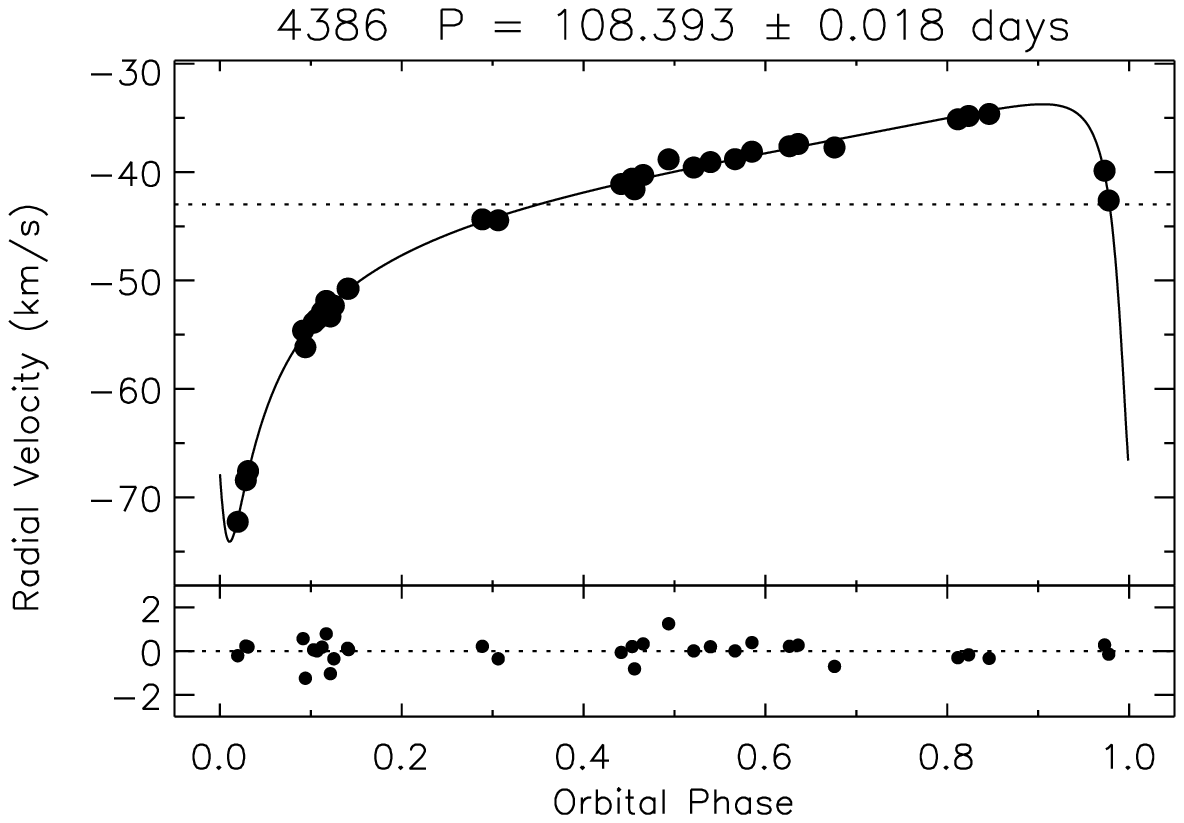,width=0.3\linewidth} & \epsfig{file=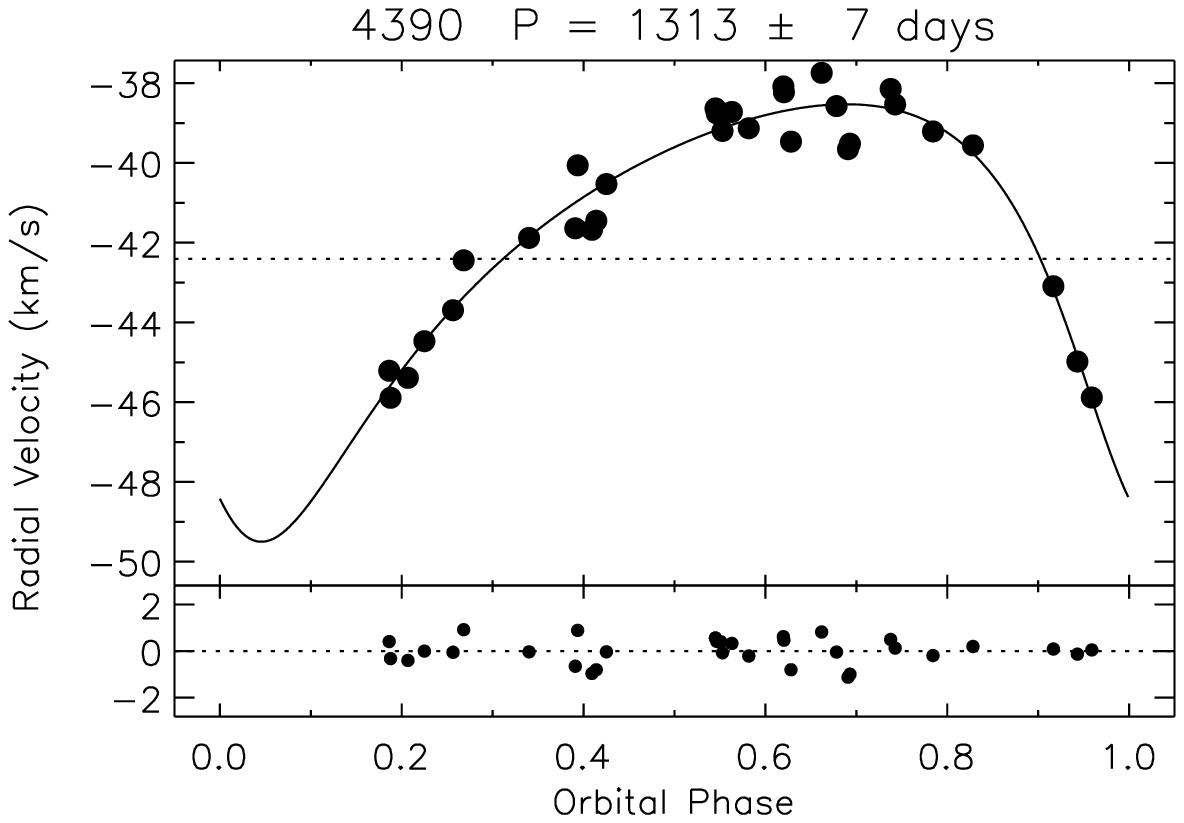,width=0.3\linewidth} \\
\epsfig{file=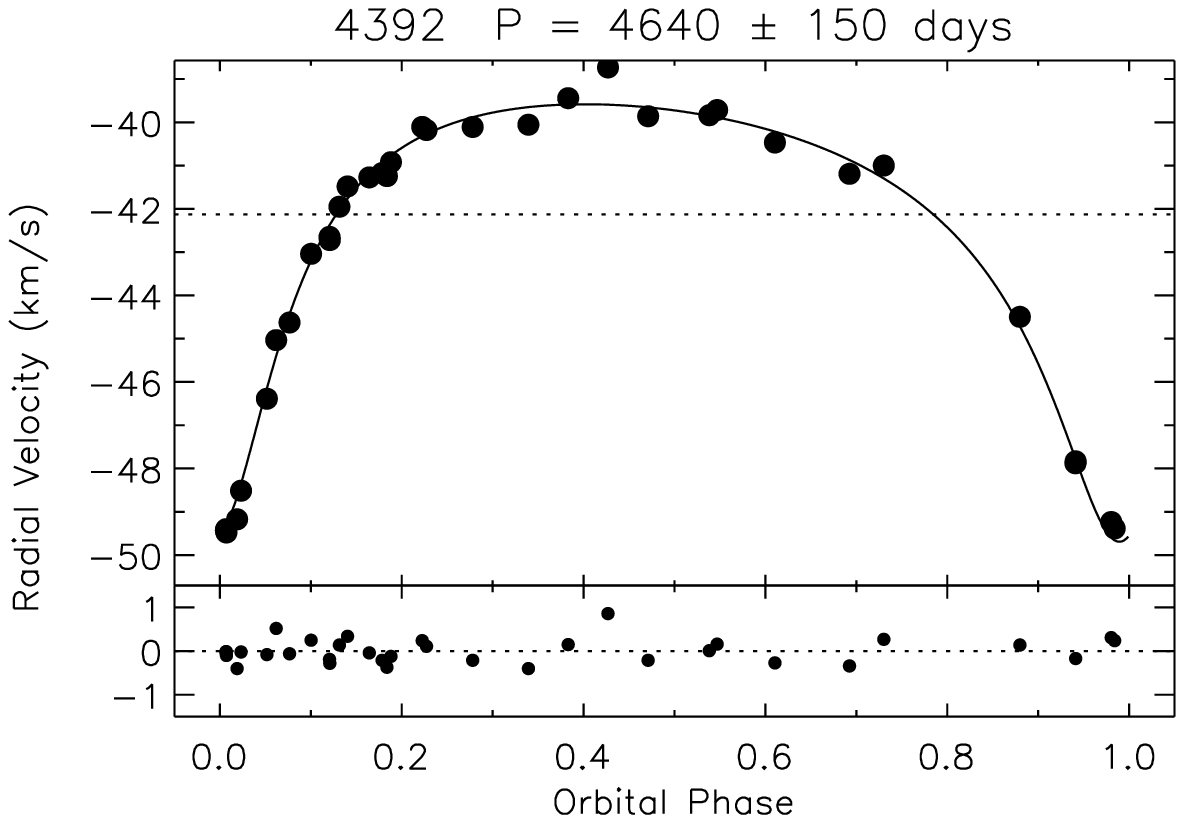,width=0.3\linewidth} & \epsfig{file=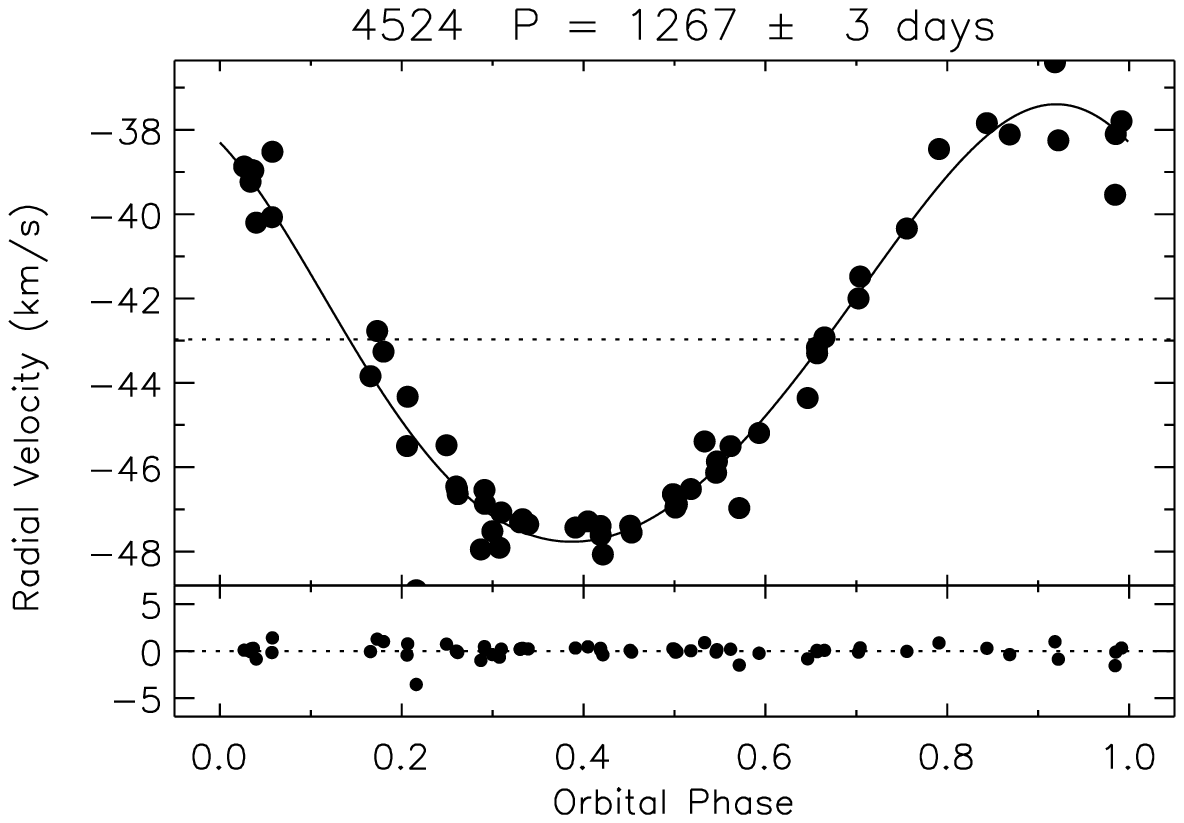,width=0.3\linewidth} & \epsfig{file=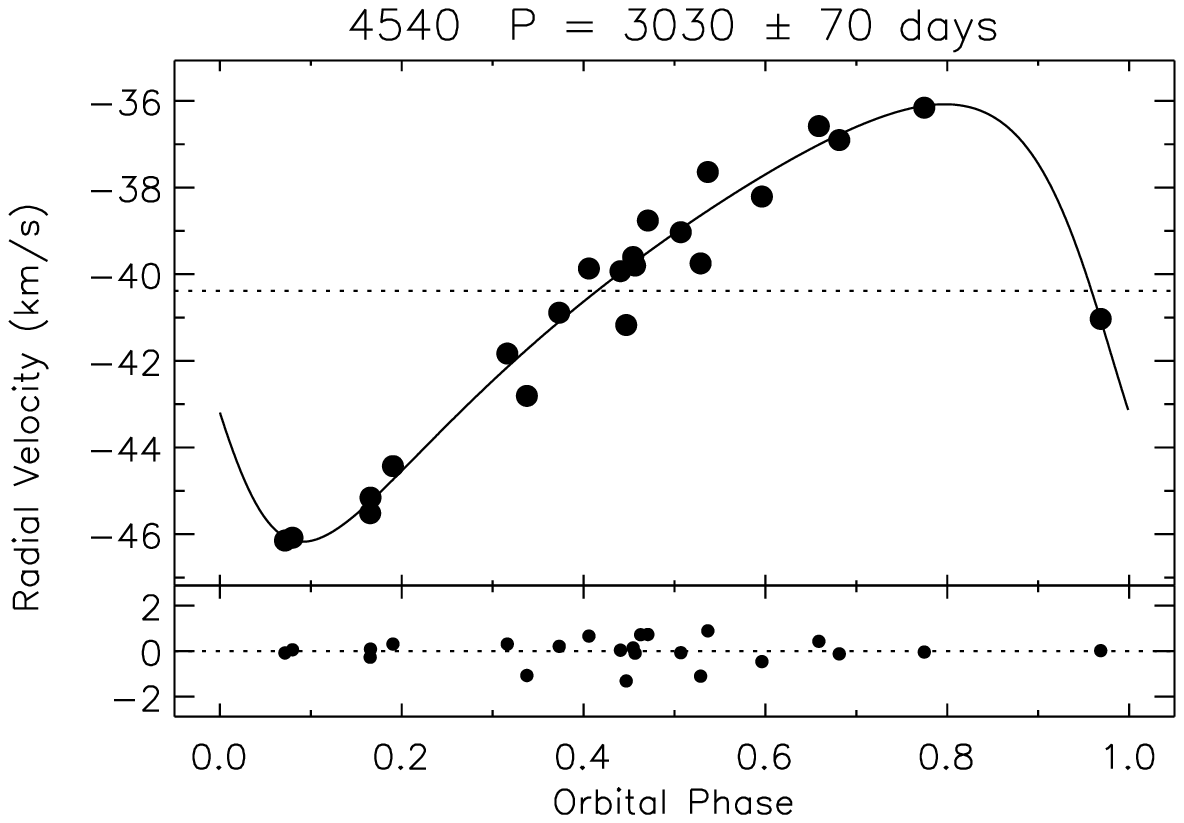,width=0.3\linewidth} \\
\epsfig{file=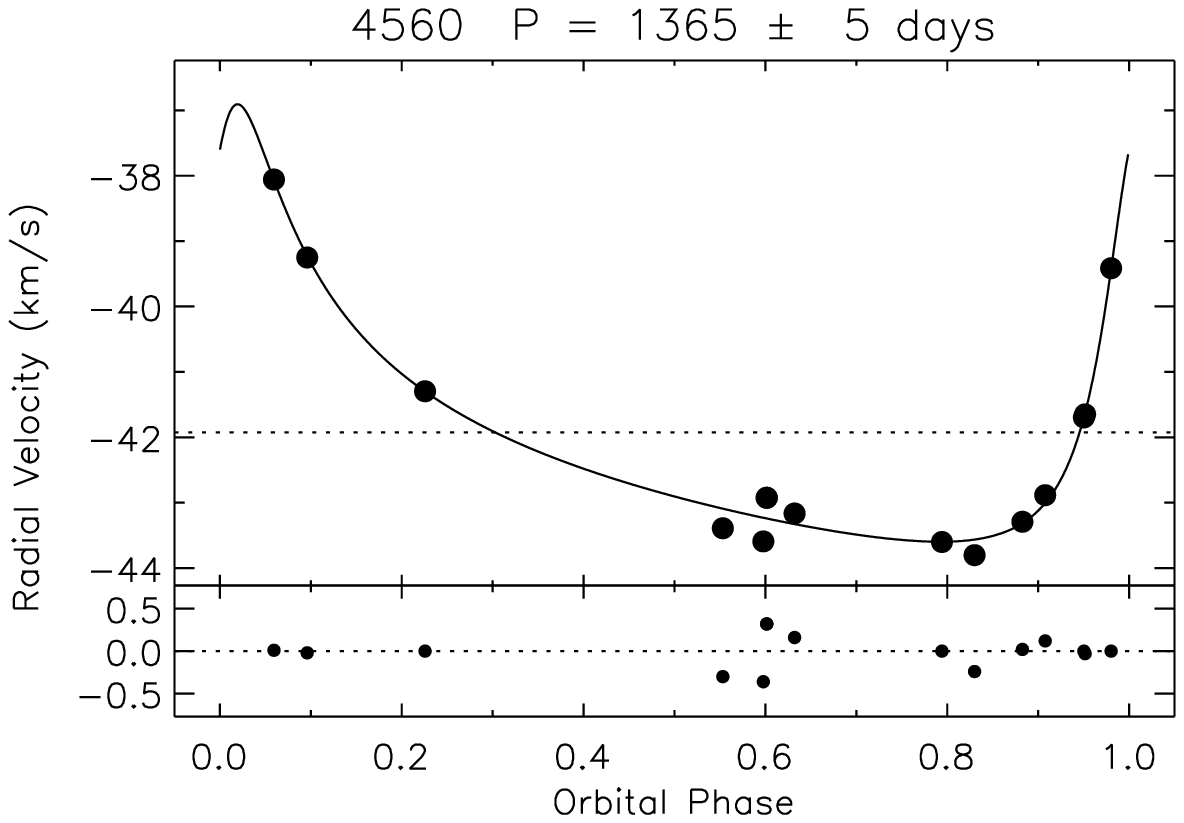,width=0.3\linewidth} & \epsfig{file=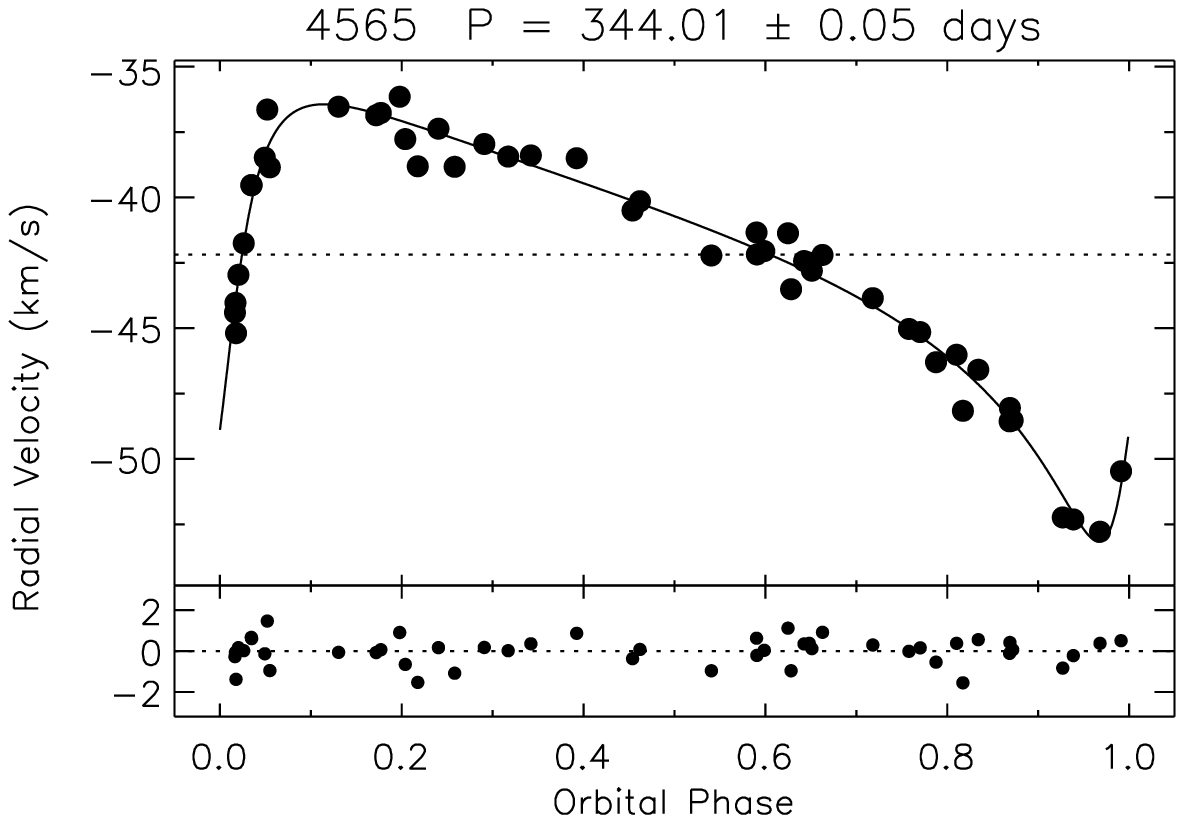,width=0.3\linewidth} & \epsfig{file=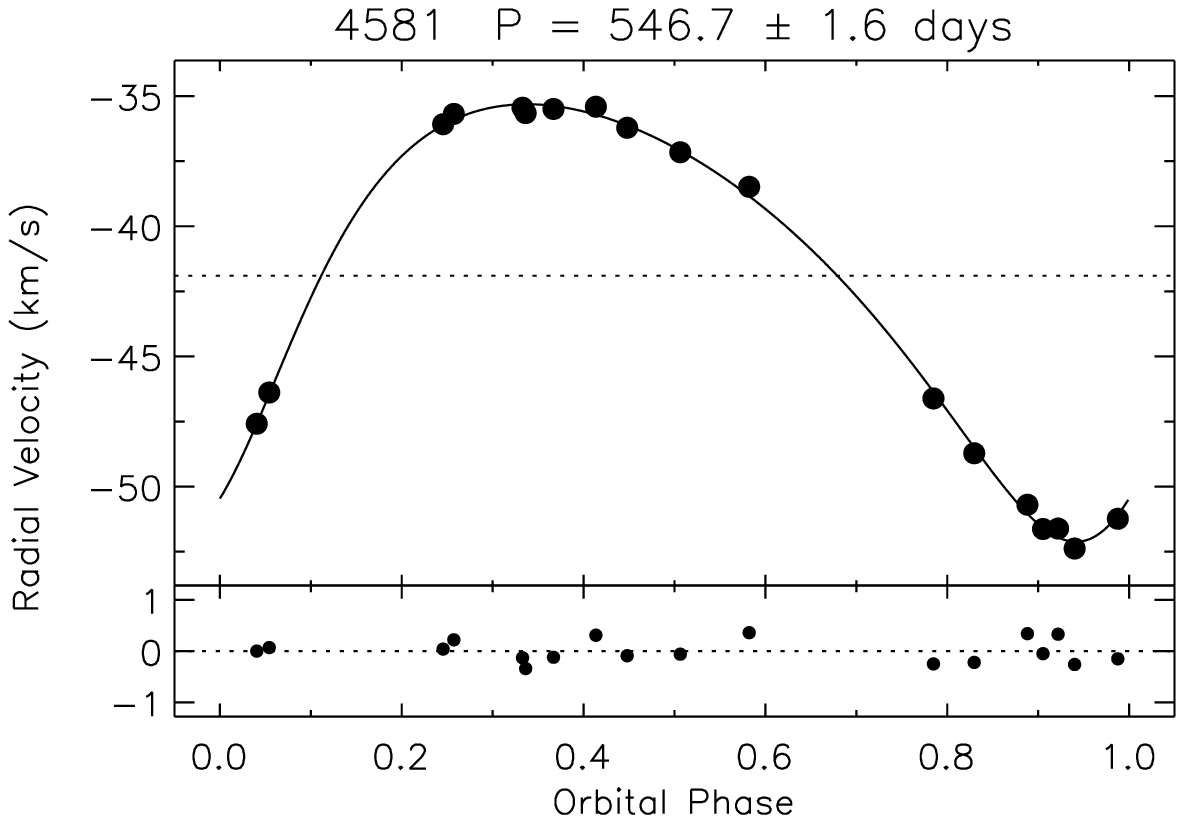,width=0.3\linewidth} \\
\epsfig{file=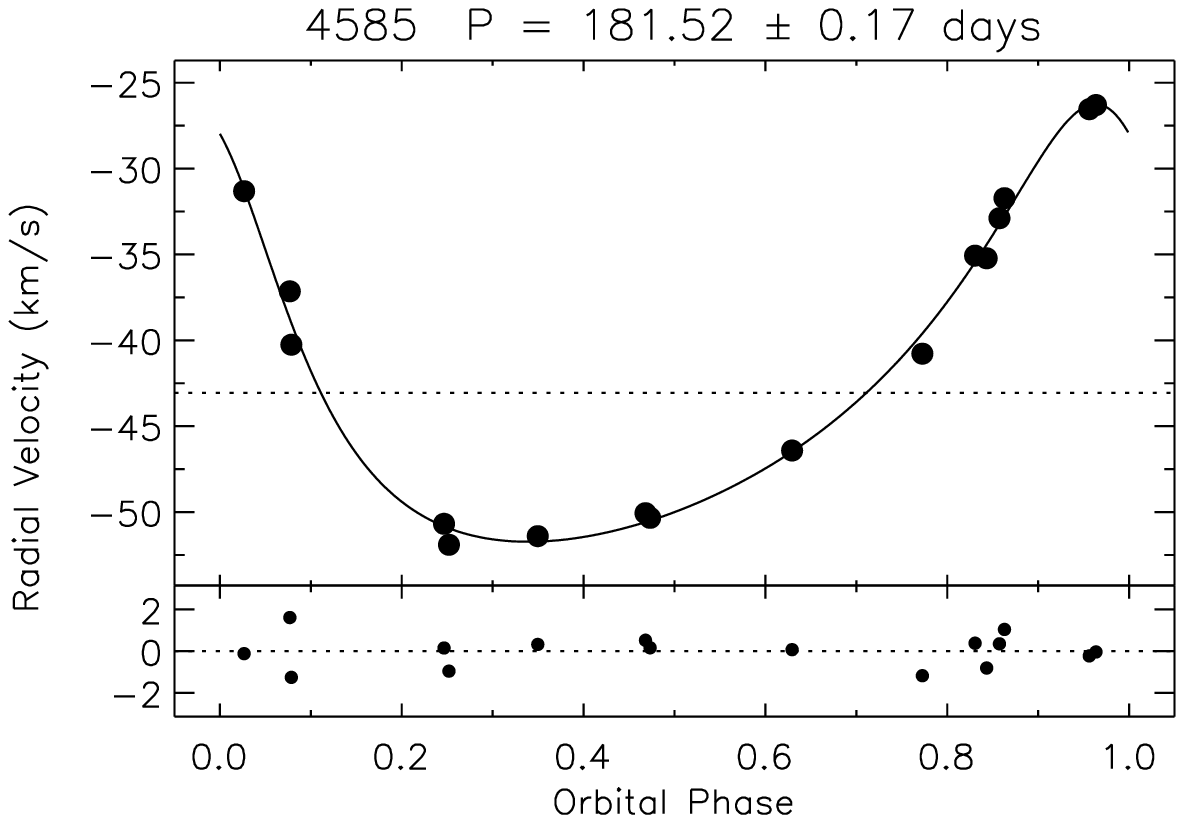,width=0.3\linewidth} & \epsfig{file=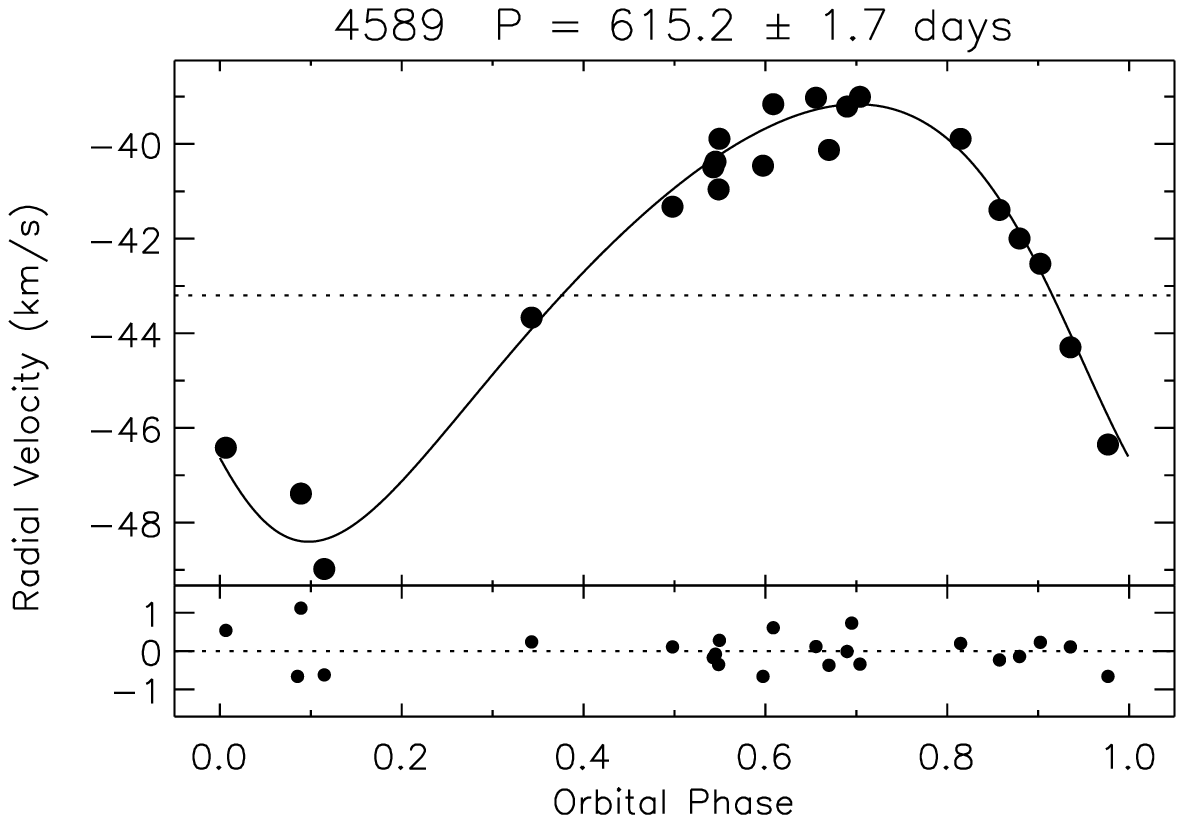,width=0.3\linewidth} & \epsfig{file=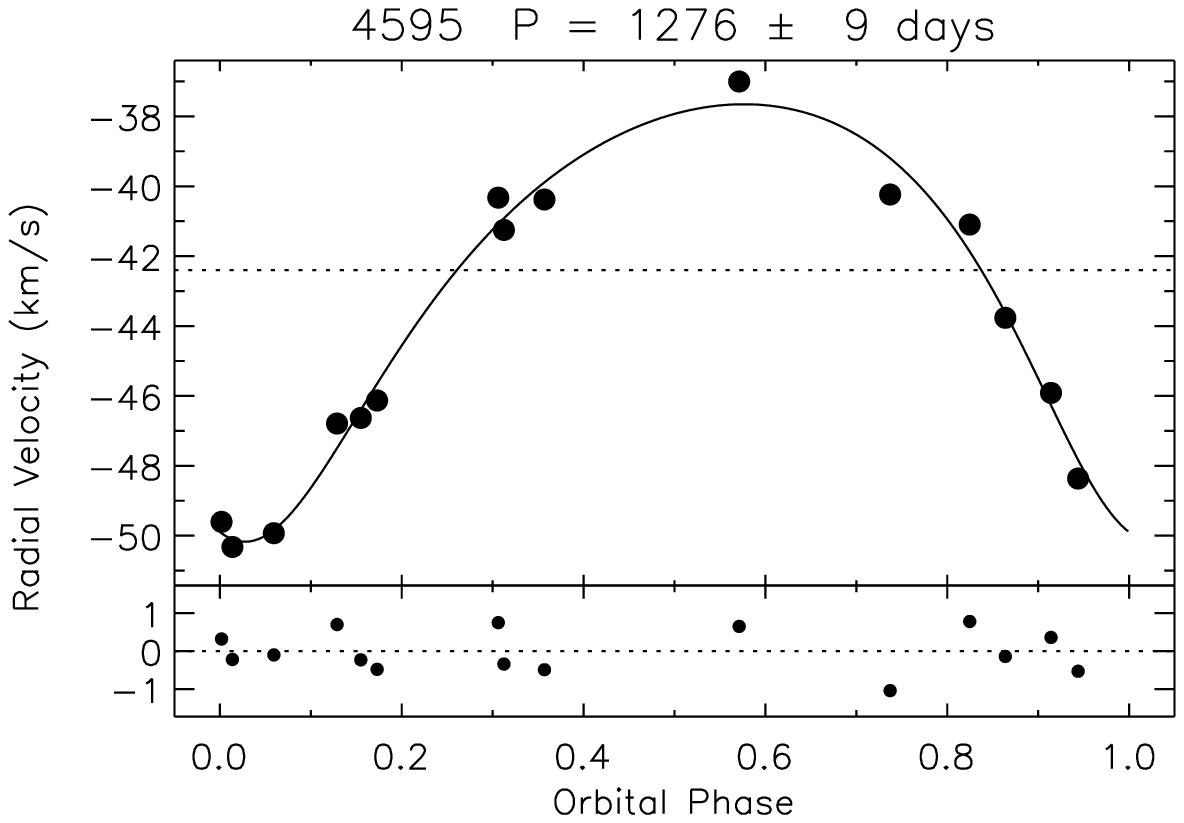,width=0.3\linewidth} \\
\epsfig{file=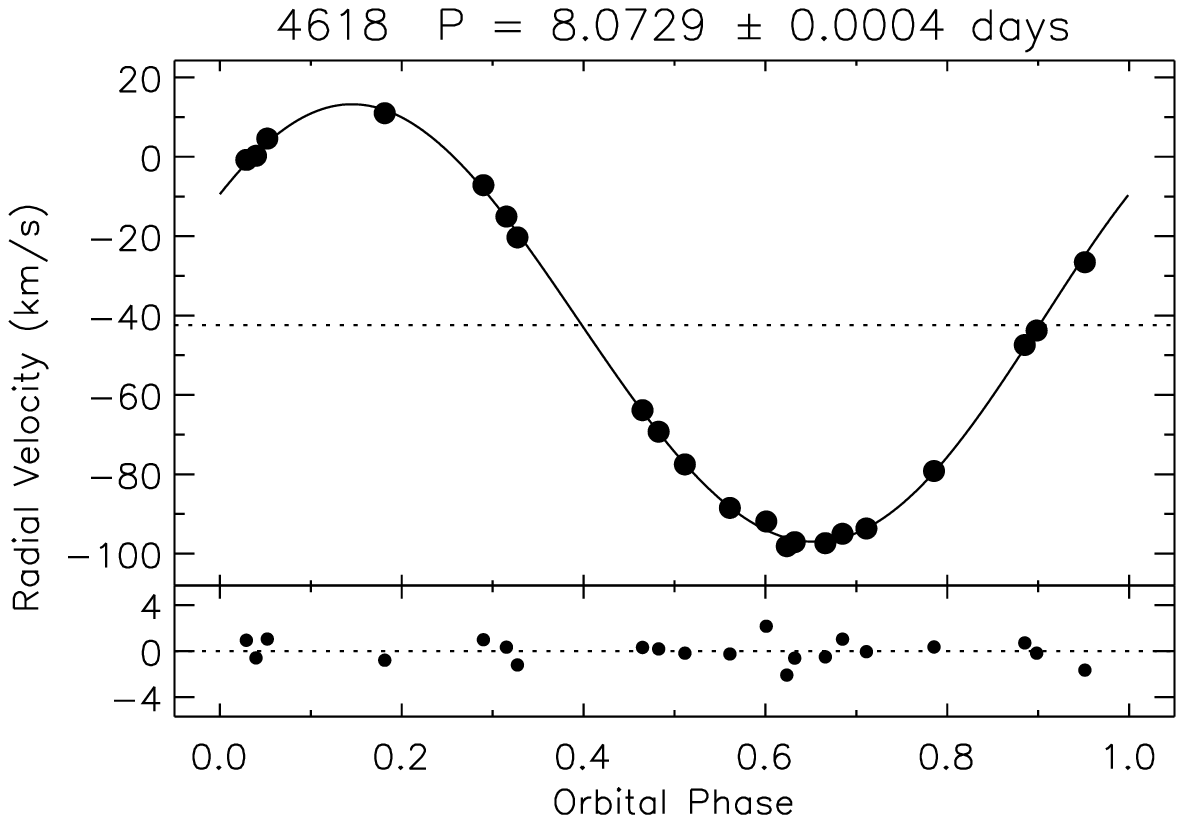,width=0.3\linewidth} & \epsfig{file=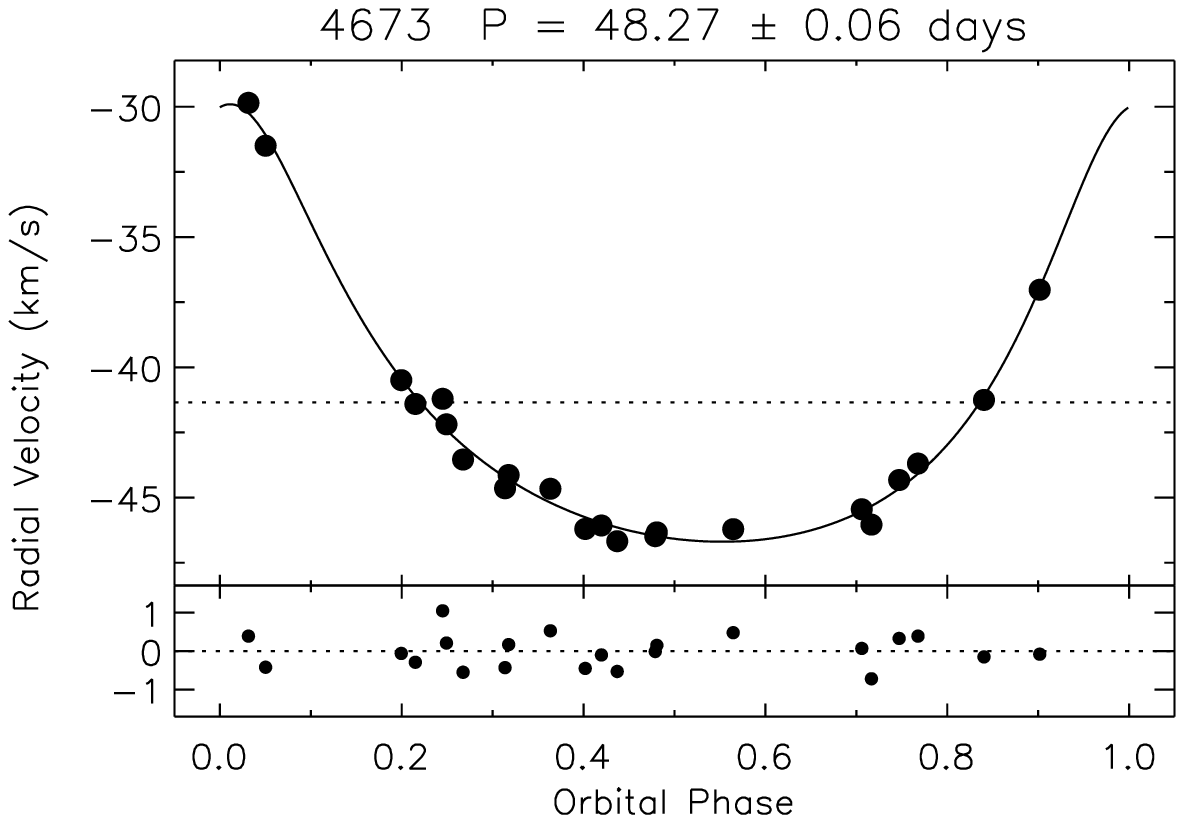,width=0.3\linewidth} & \epsfig{file=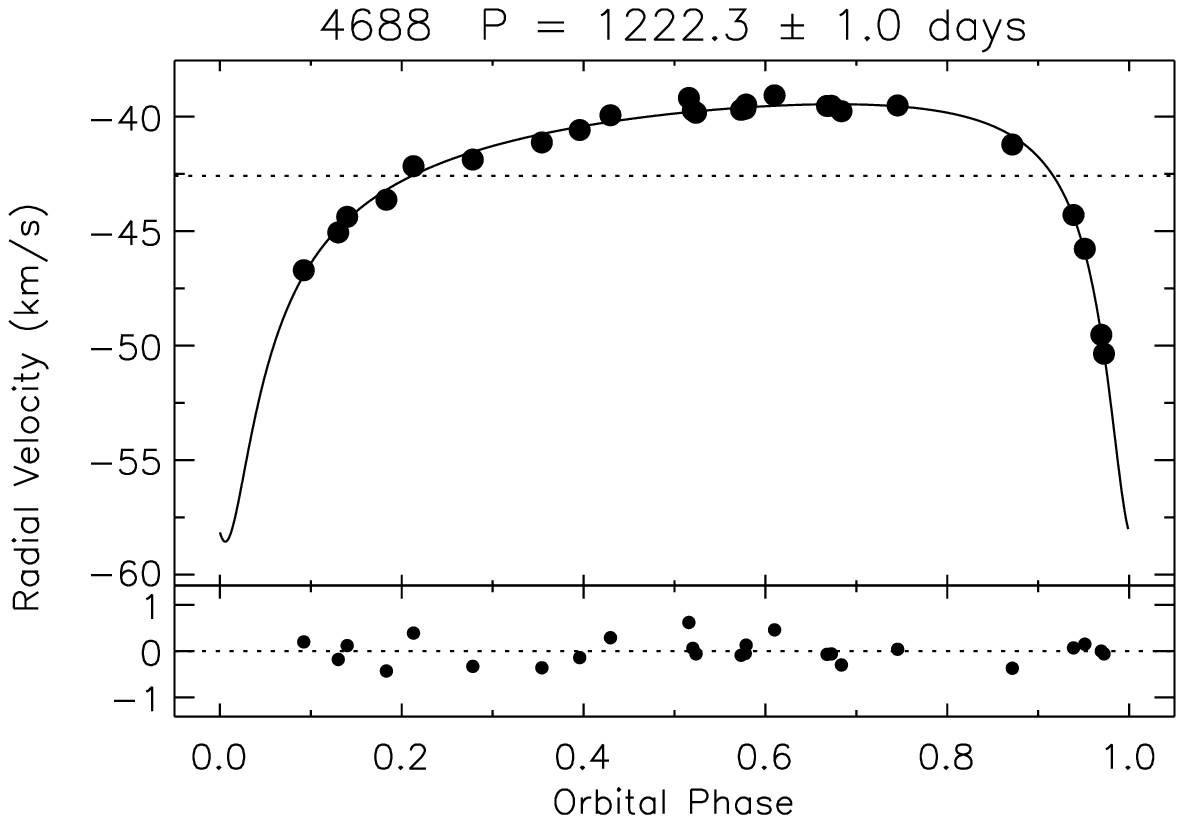,width=0.3\linewidth} \\
\epsfig{file=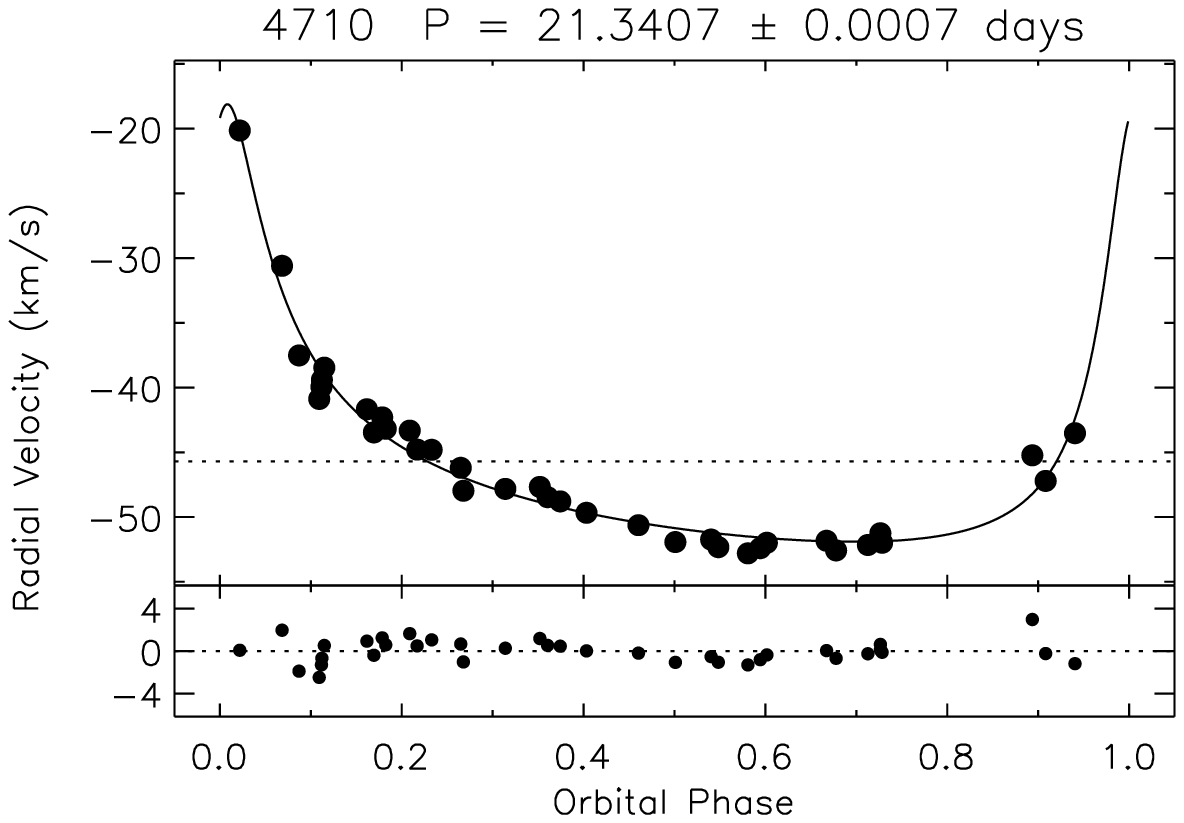,width=0.3\linewidth} & \epsfig{file=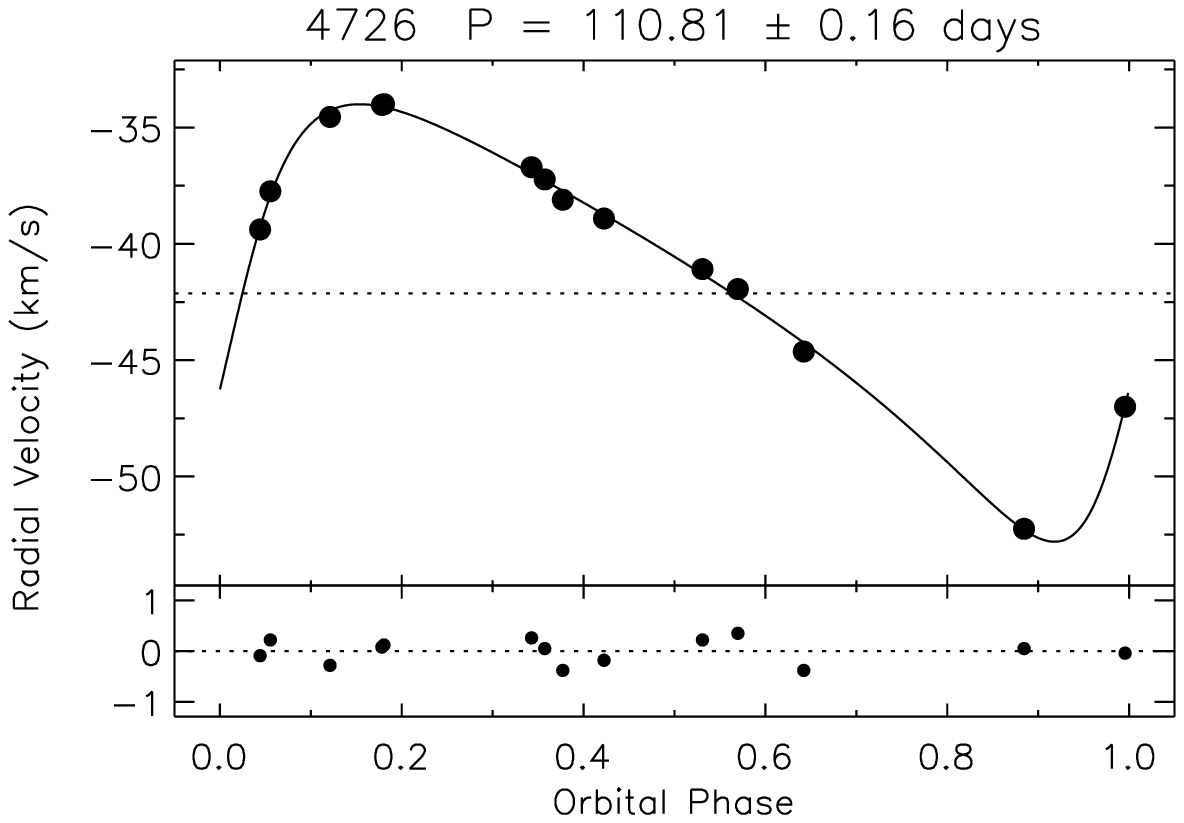,width=0.3\linewidth} & \epsfig{file=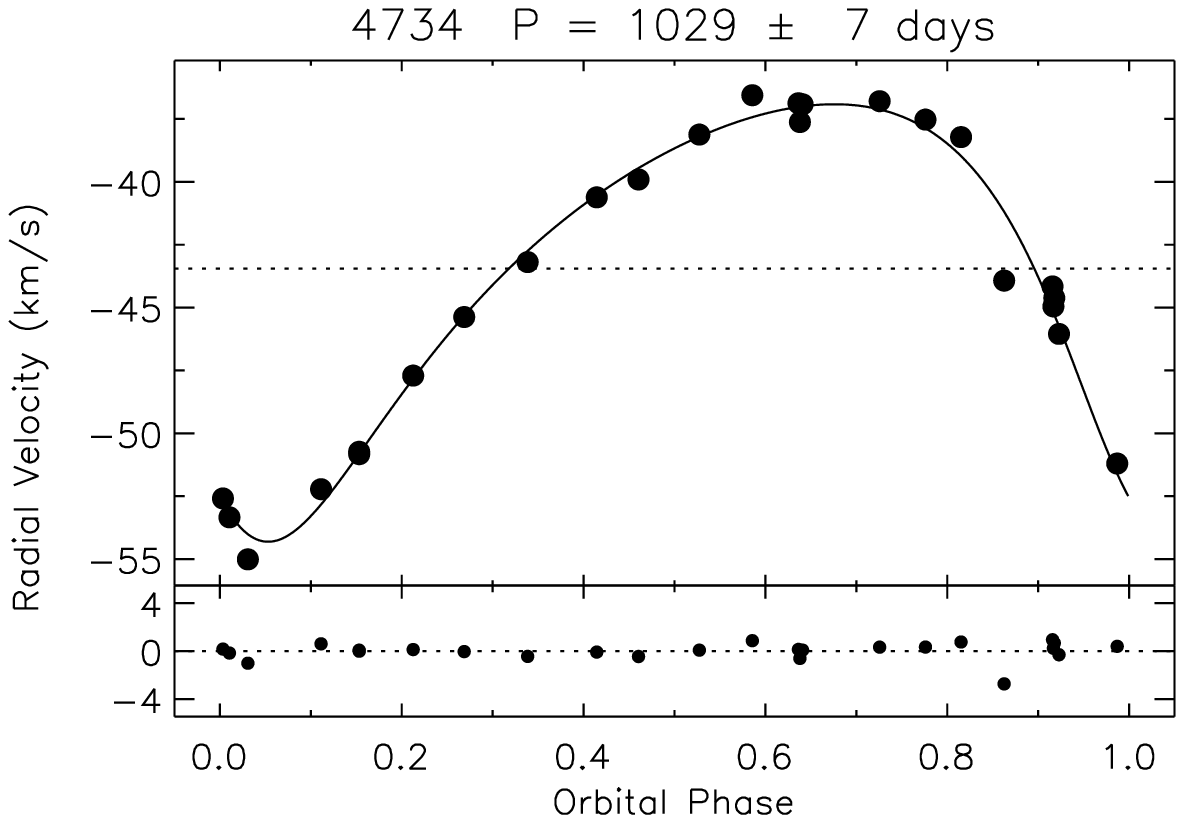,width=0.3\linewidth} \\
\epsfig{file=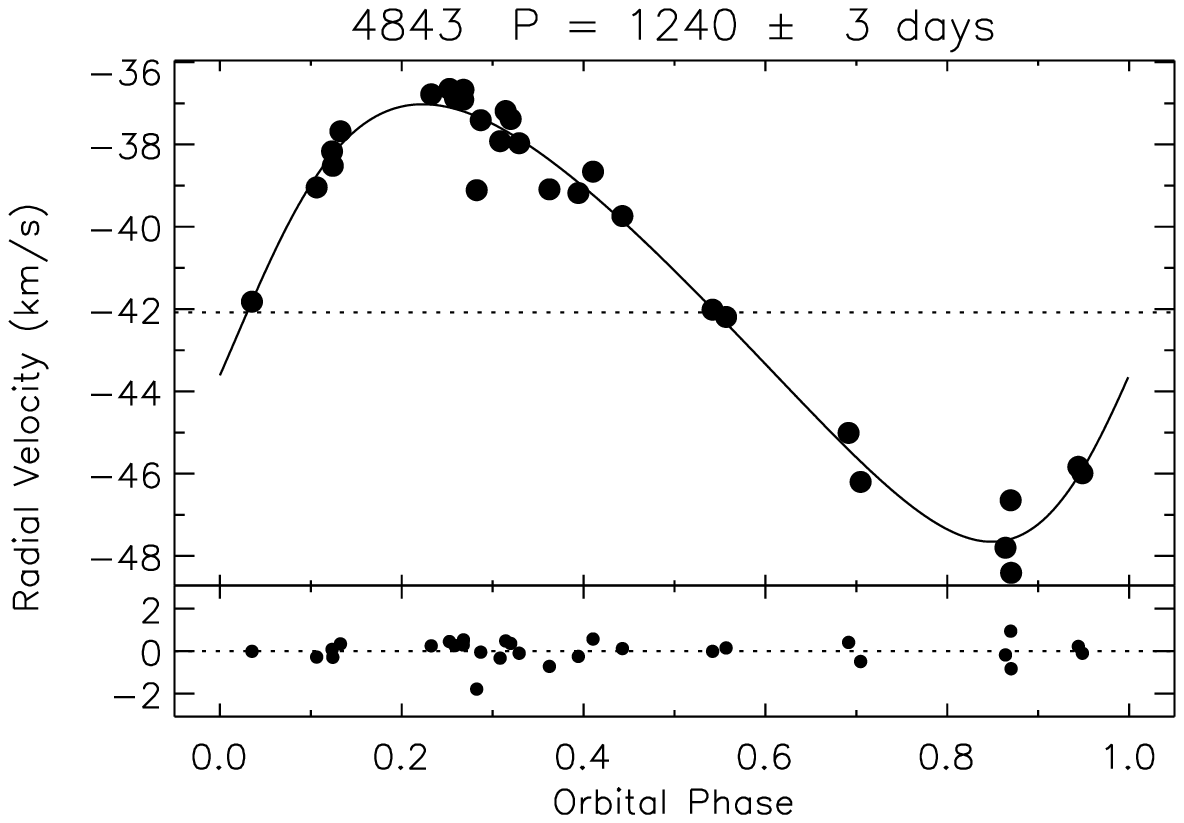,width=0.3\linewidth} & \epsfig{file=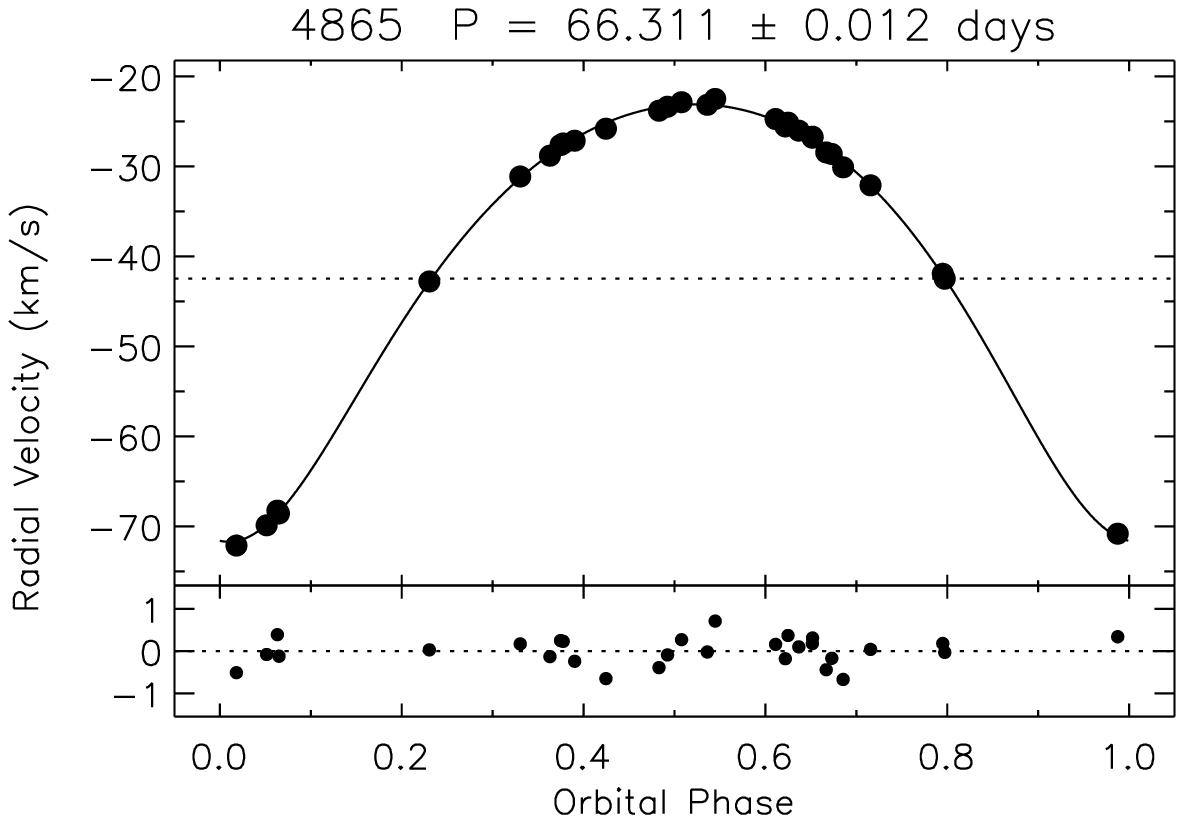,width=0.3\linewidth} & \epsfig{file=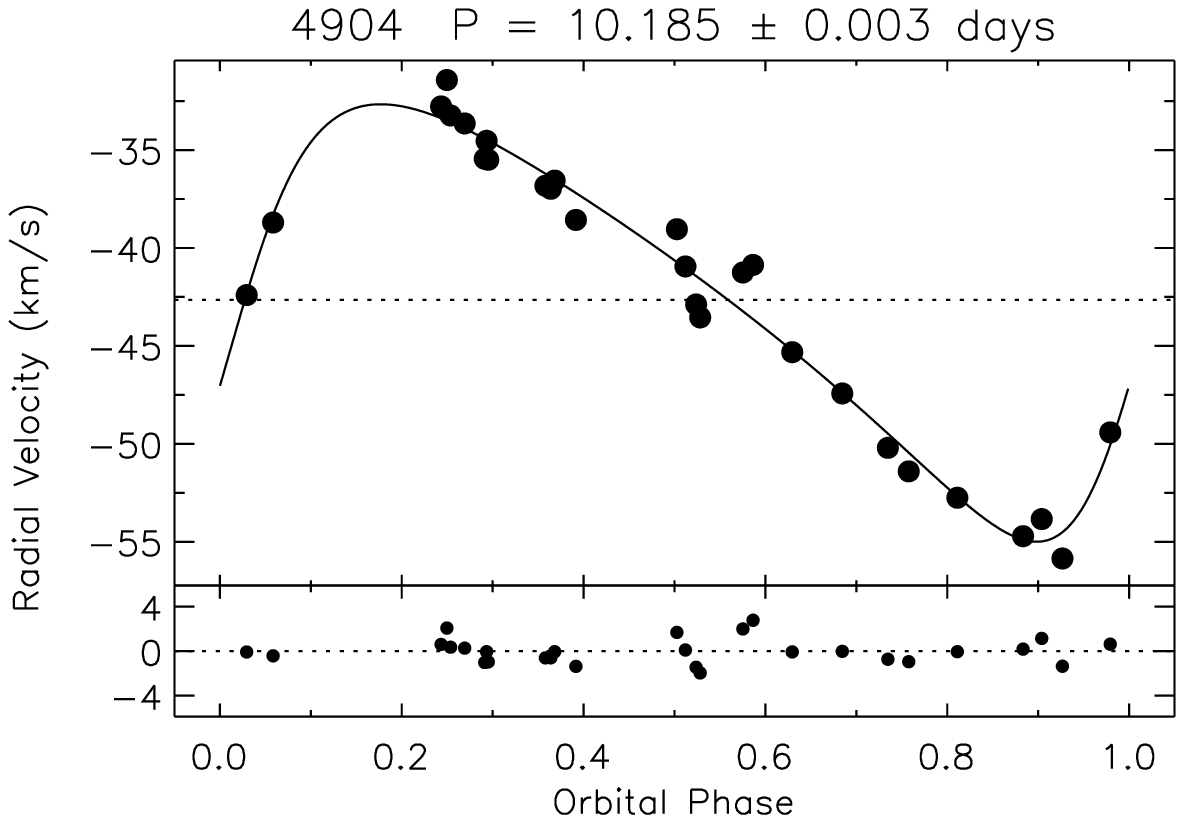,width=0.3\linewidth} \\
\epsfig{file=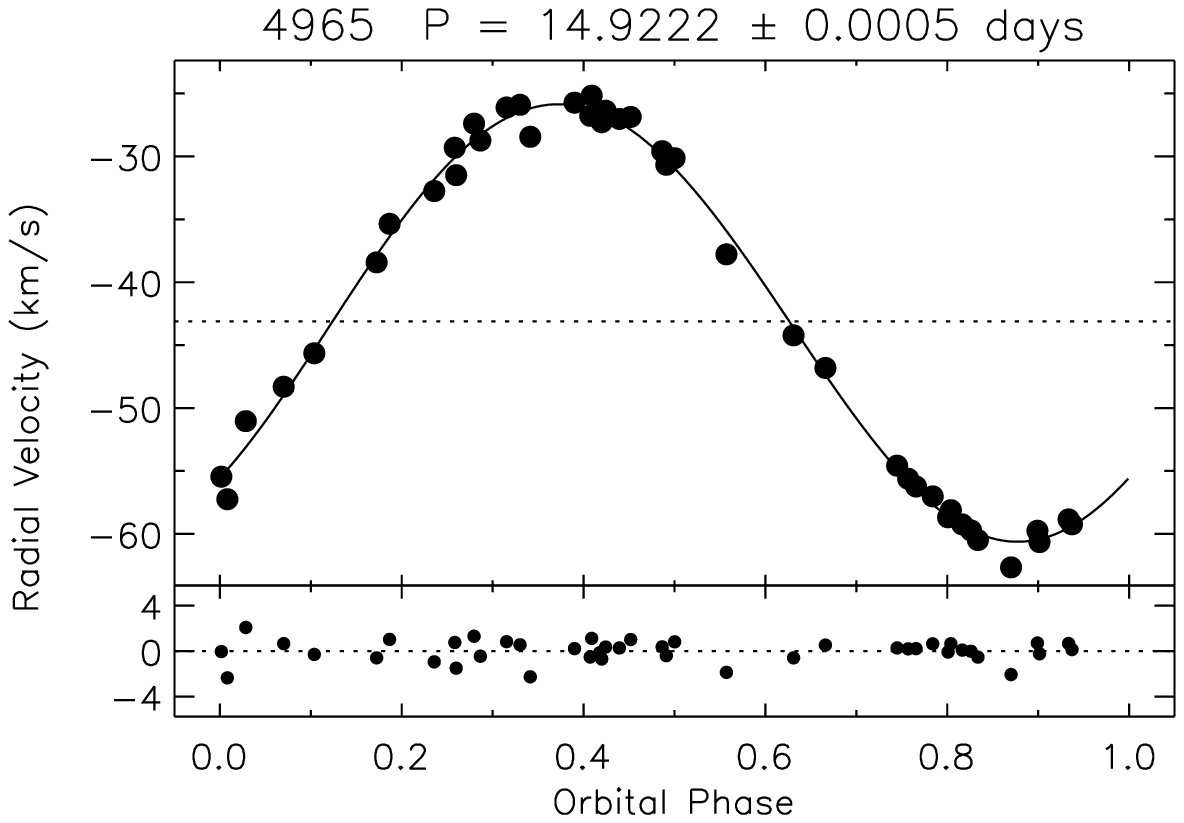,width=0.3\linewidth} & \epsfig{file=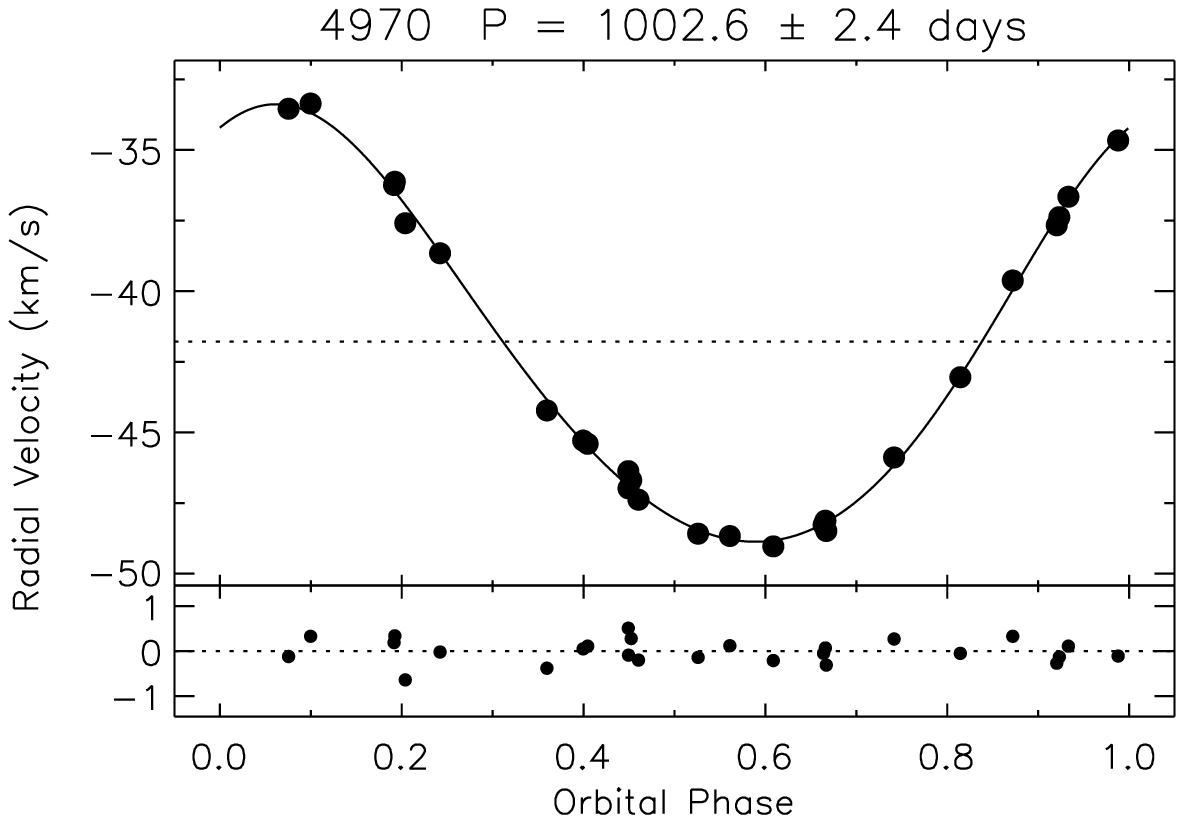,width=0.3\linewidth} & \epsfig{file=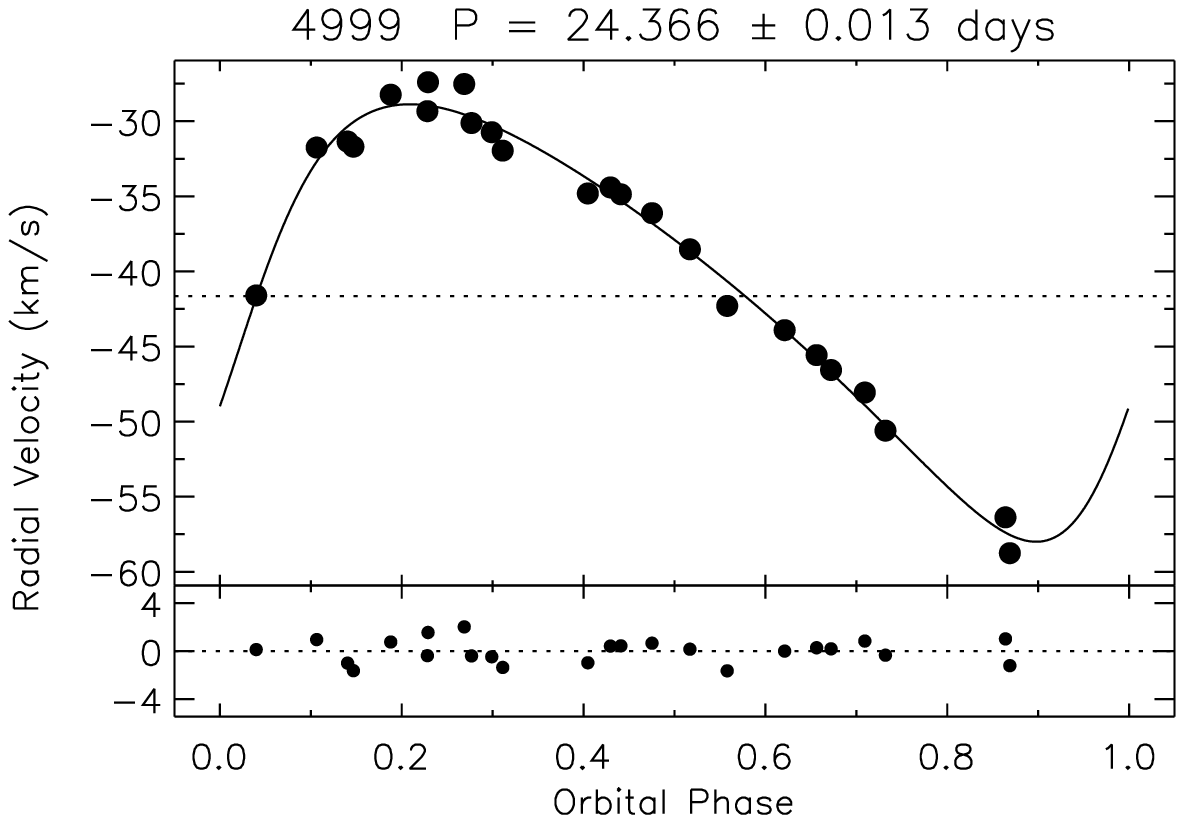,width=0.3\linewidth} \\
\epsfig{file=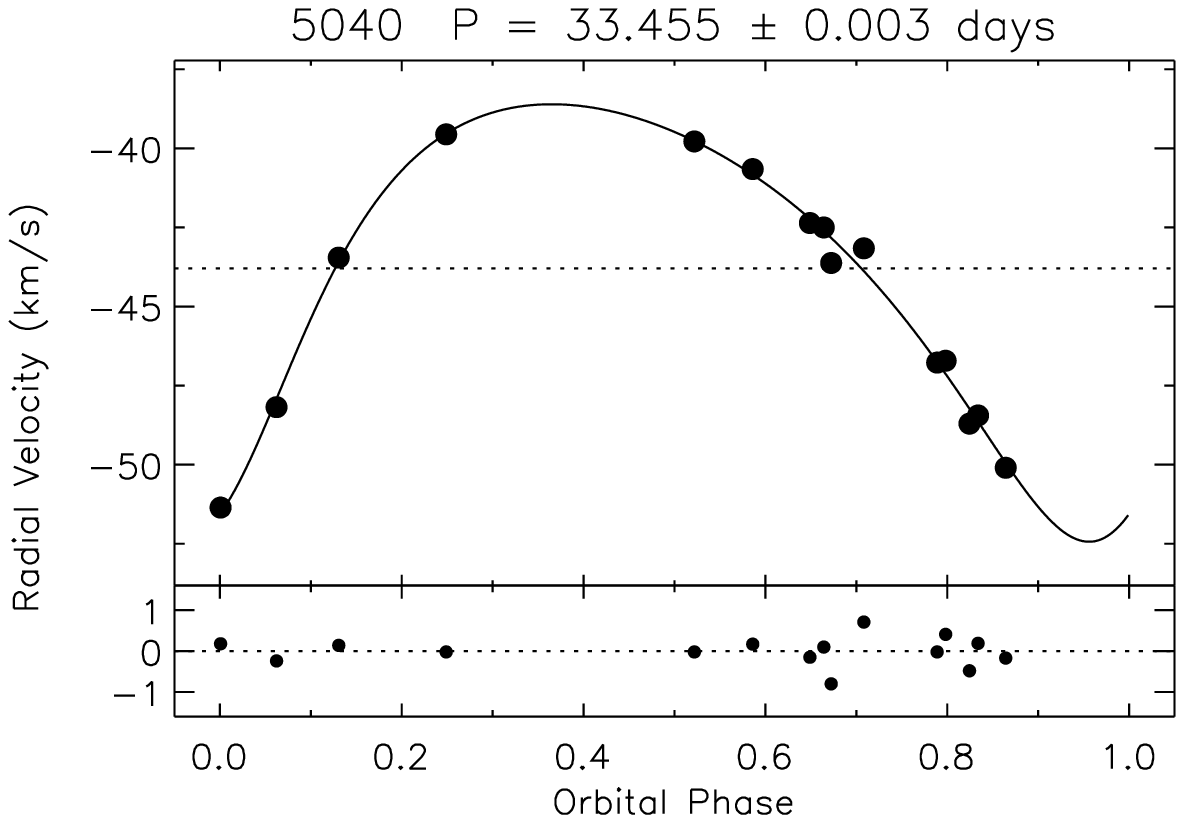,width=0.3\linewidth} & \epsfig{file=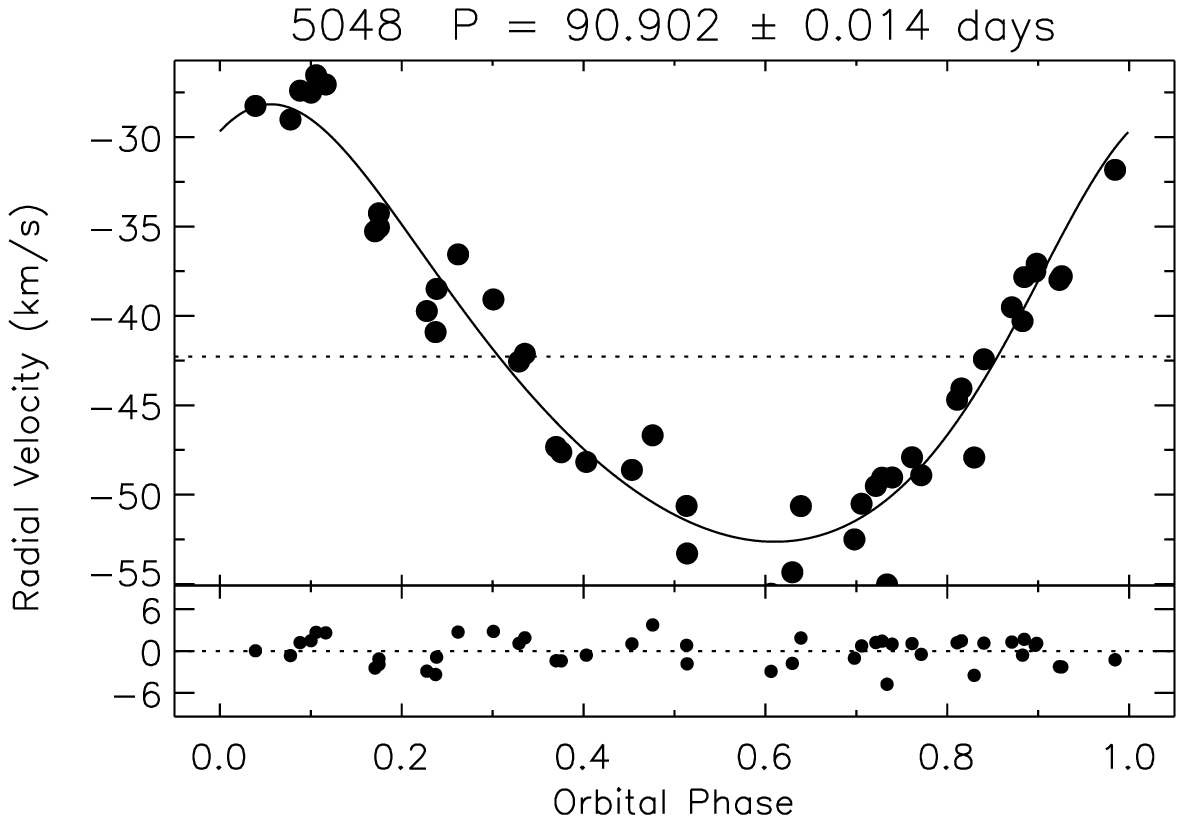,width=0.3\linewidth} & \epsfig{file=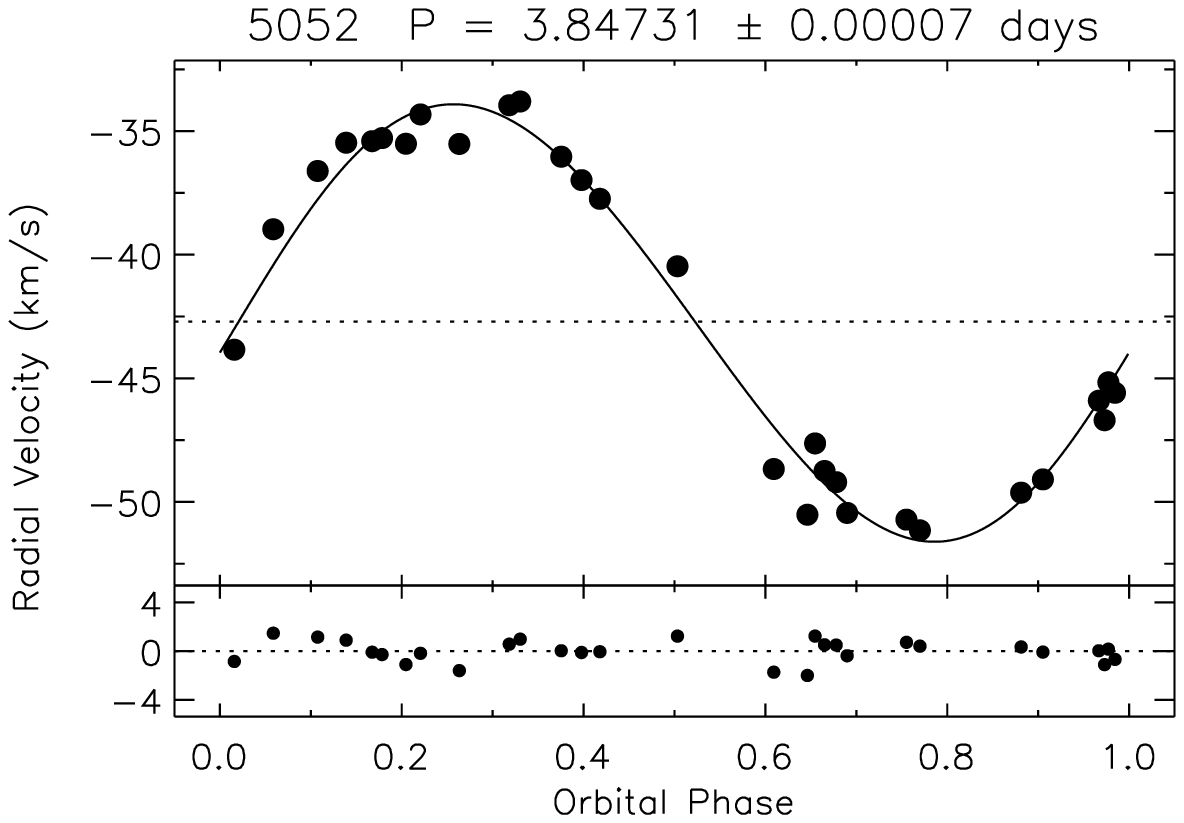,width=0.3\linewidth} \\
\epsfig{file=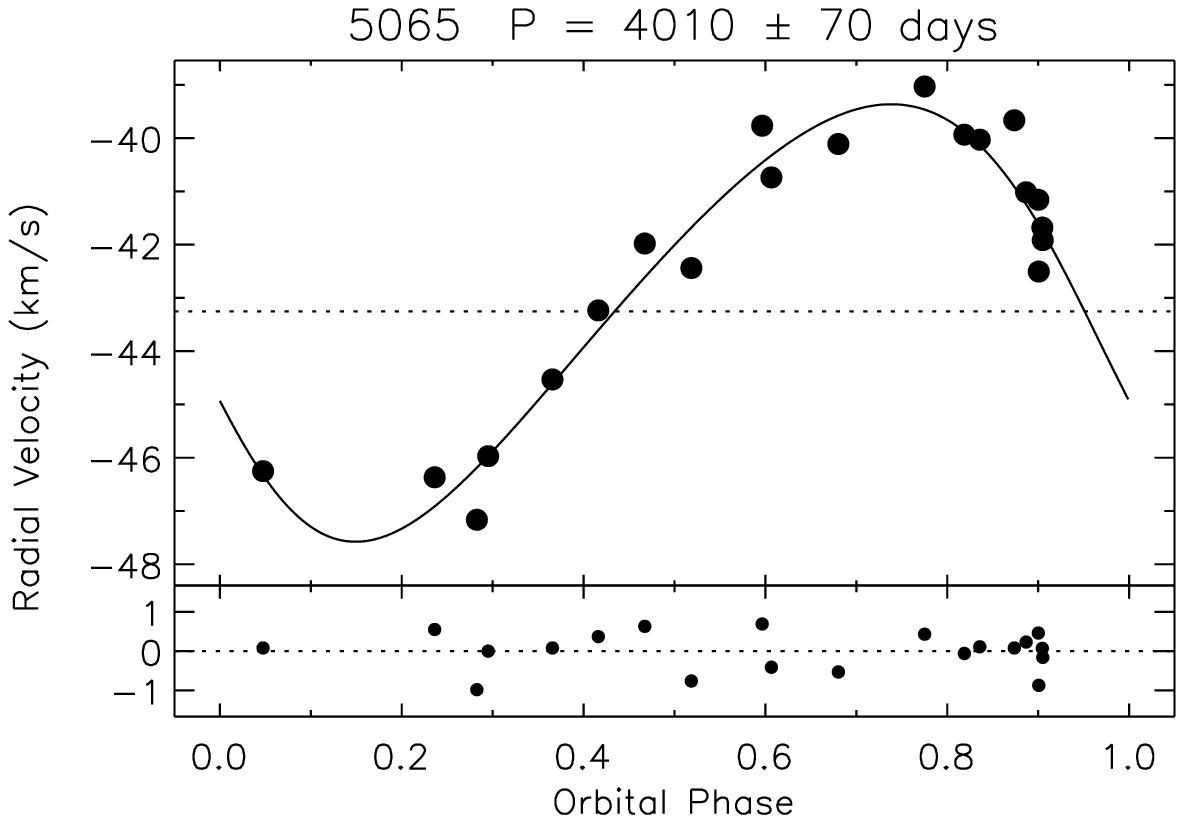,width=0.3\linewidth} & \epsfig{file=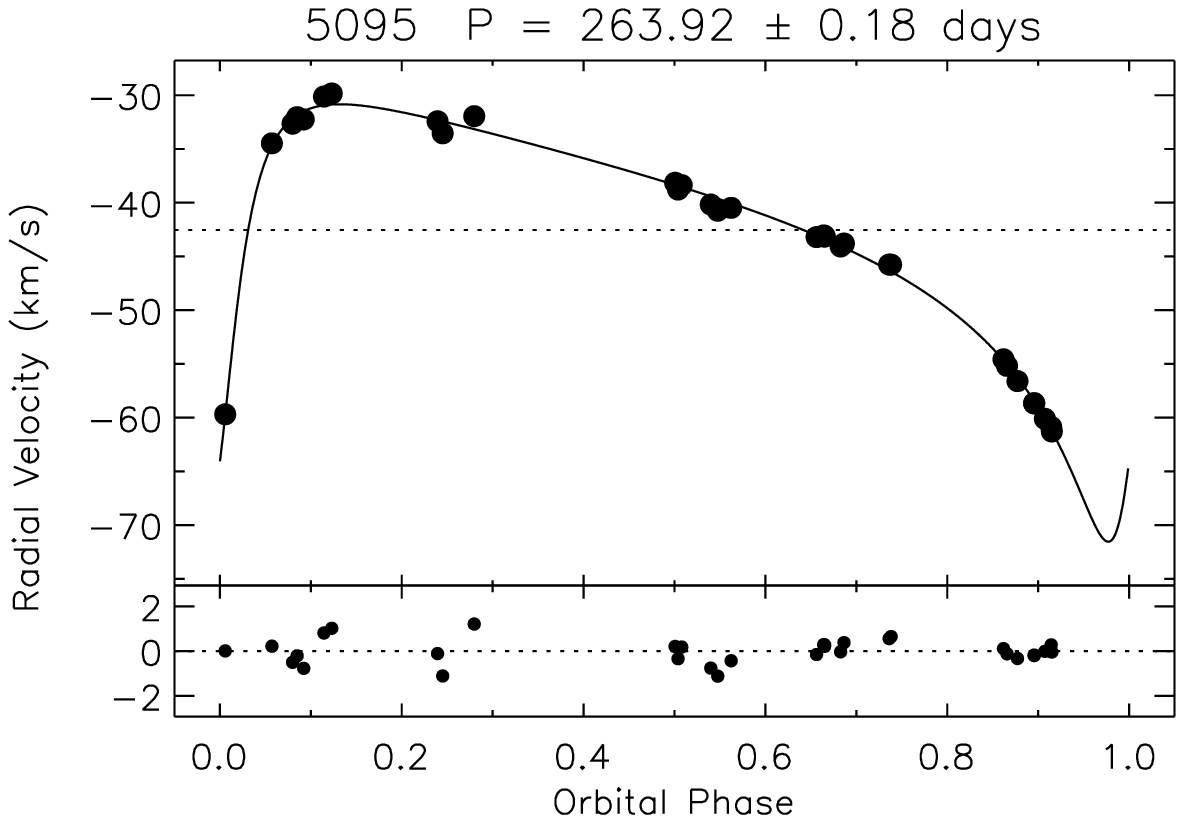,width=0.3\linewidth} & \epsfig{file=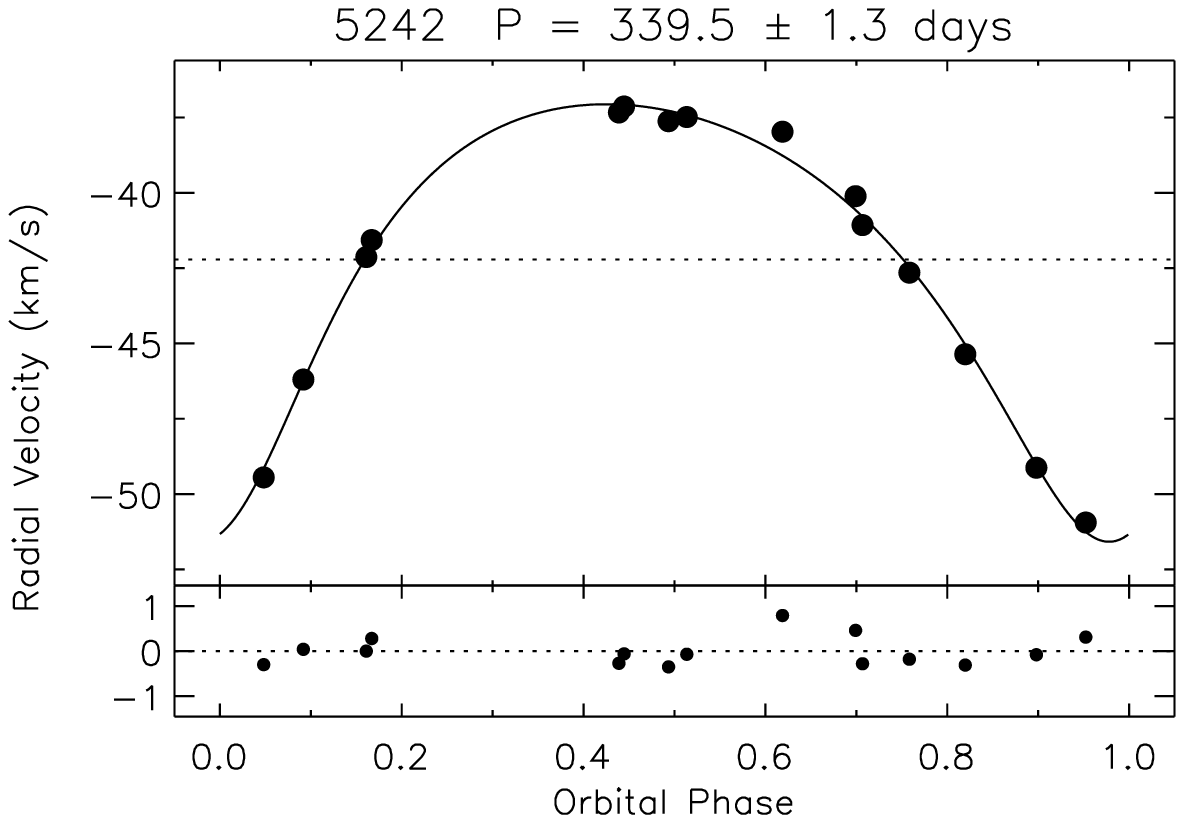,width=0.3\linewidth} \\
\epsfig{file=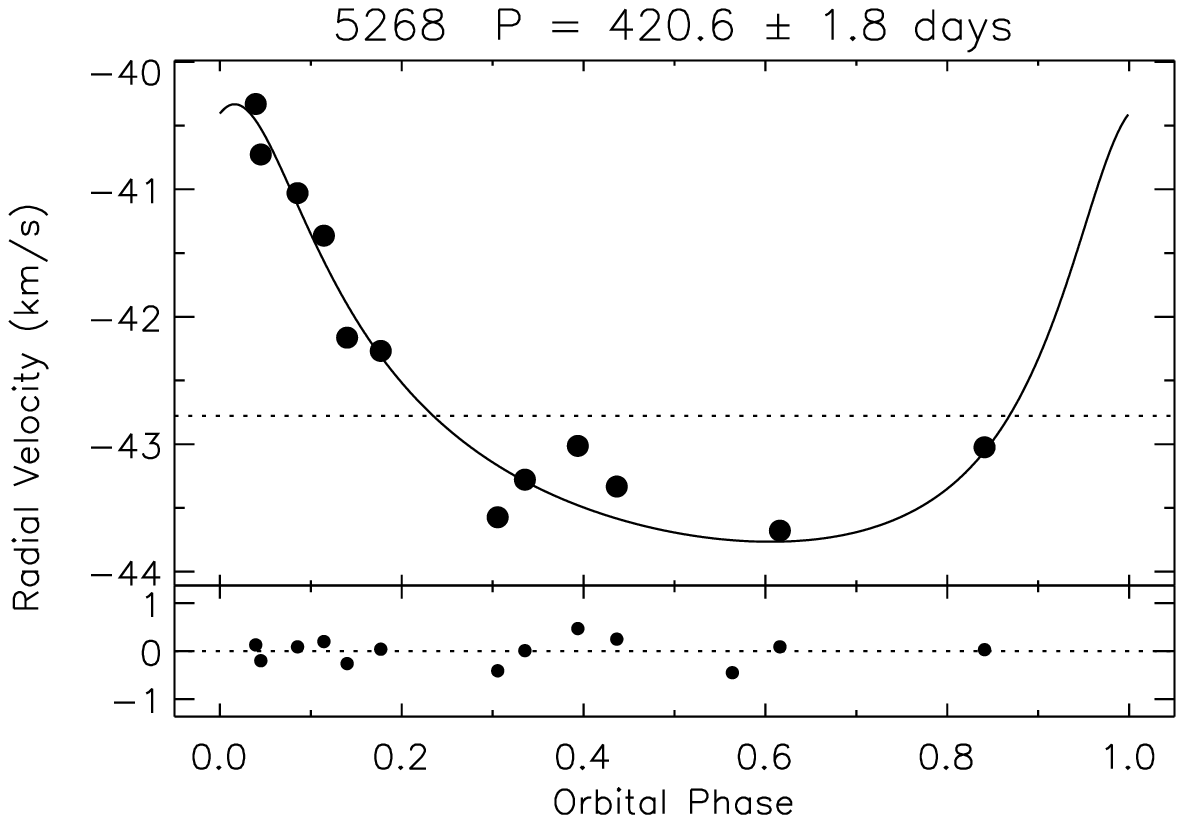,width=0.3\linewidth} & \epsfig{file=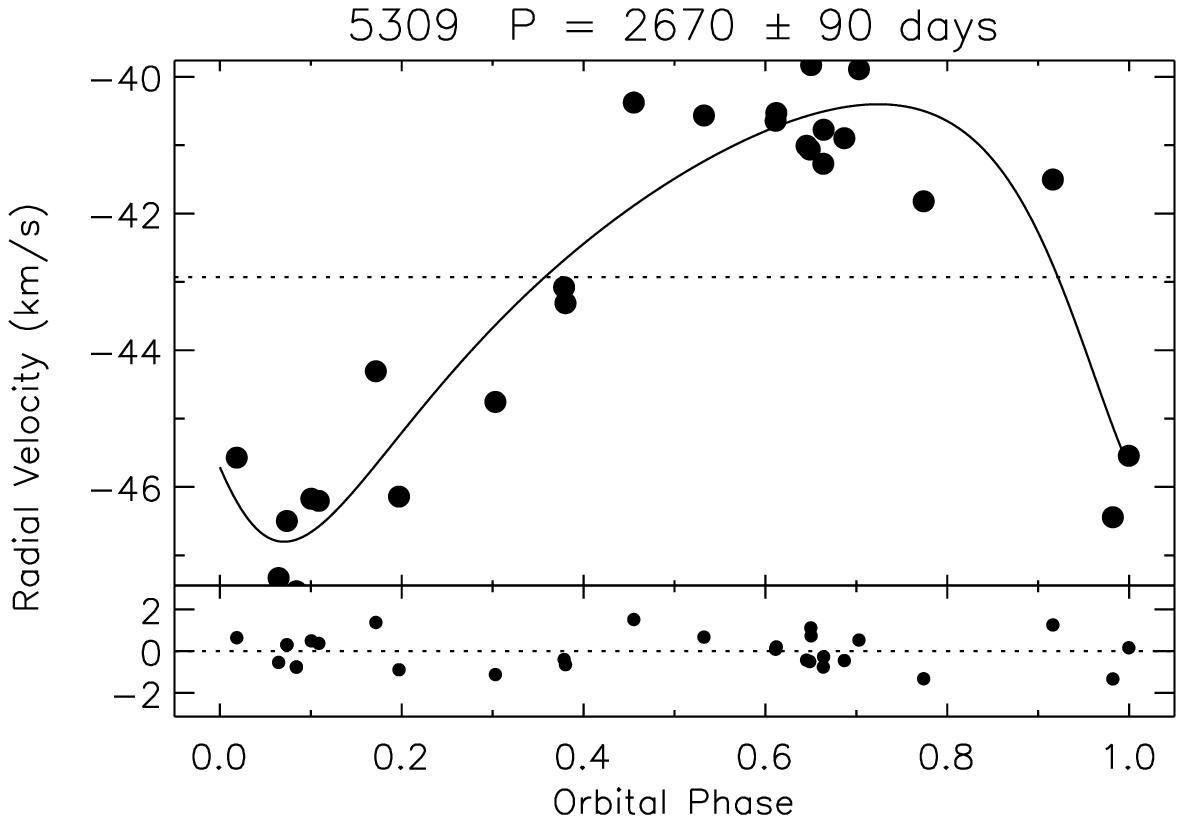,width=0.3\linewidth} & \epsfig{file=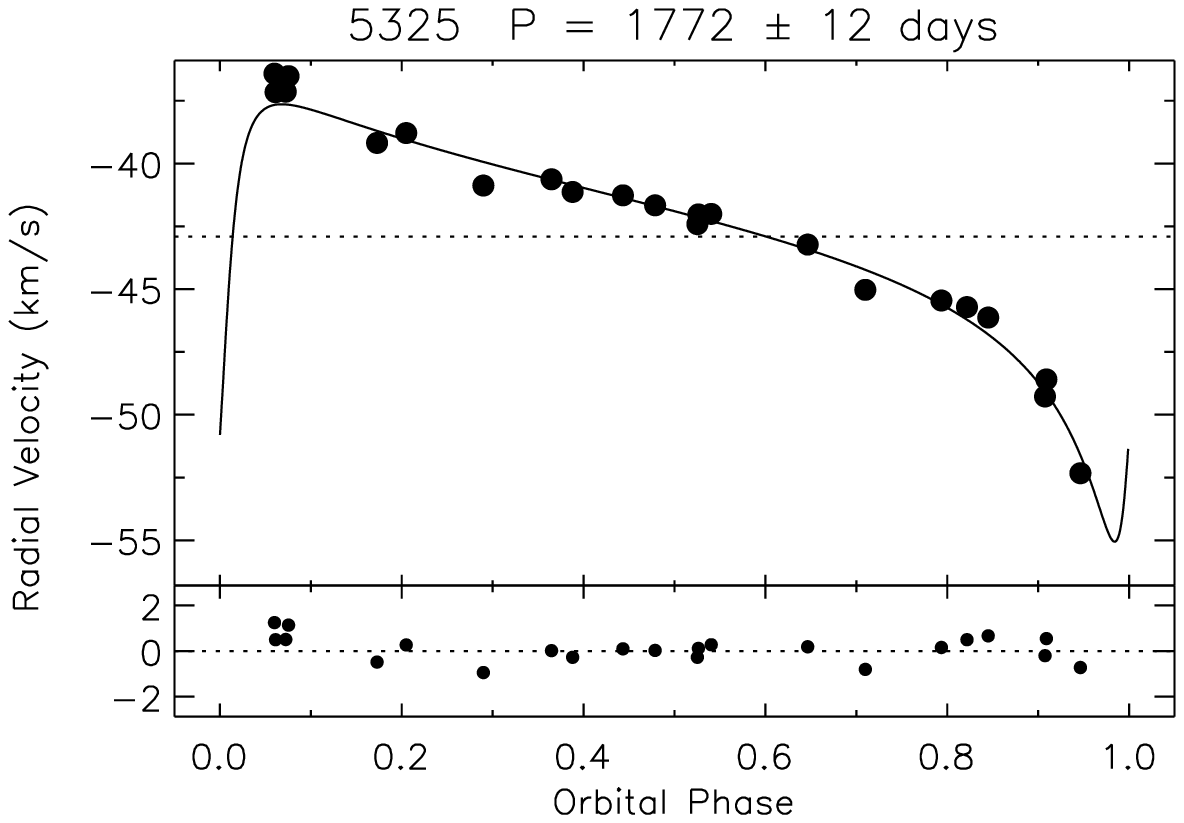,width=0.3\linewidth} \\
\epsfig{file=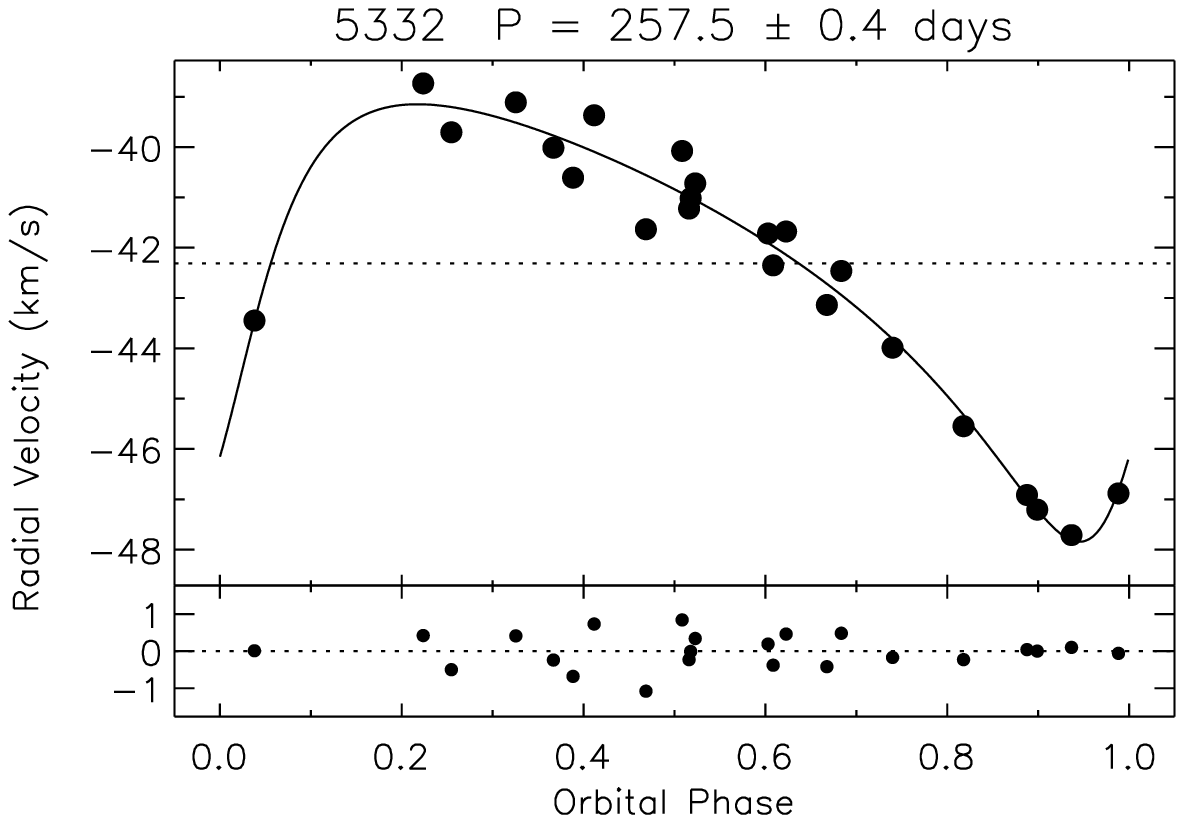,width=0.3\linewidth} & \epsfig{file=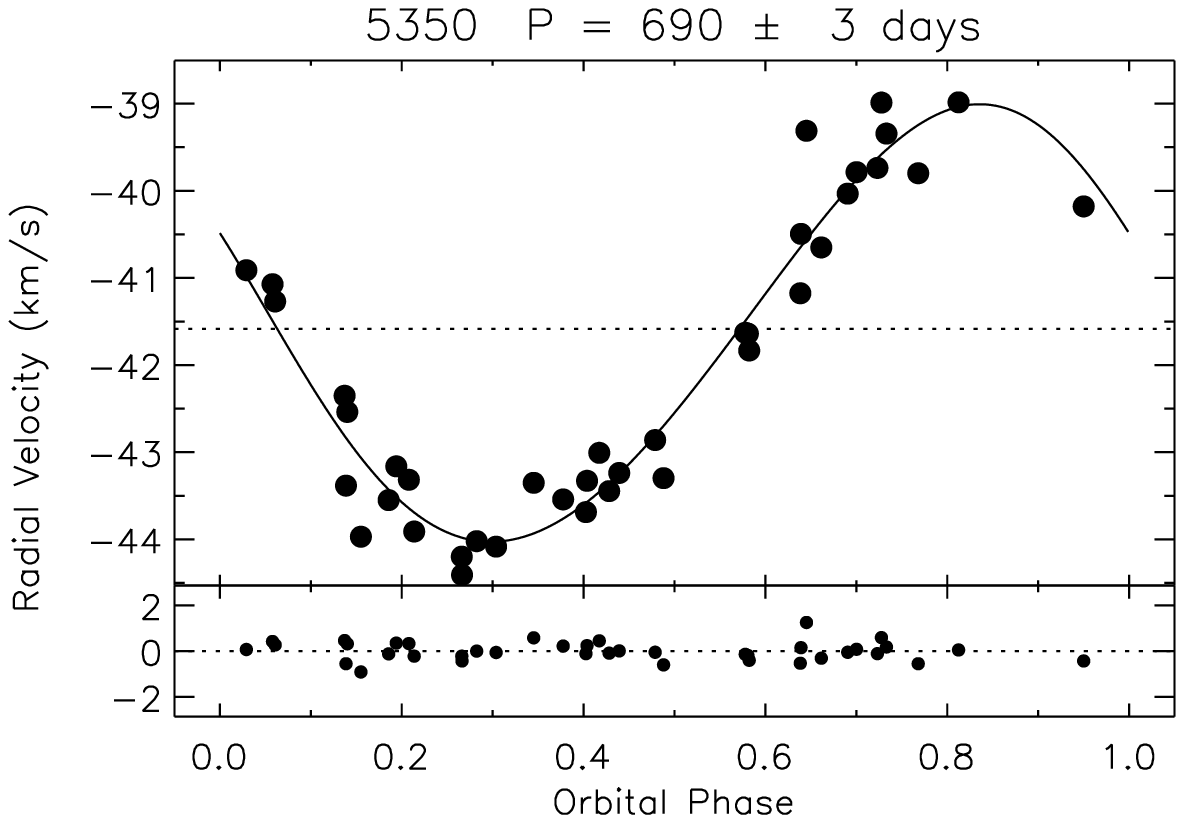,width=0.3\linewidth} & \epsfig{file=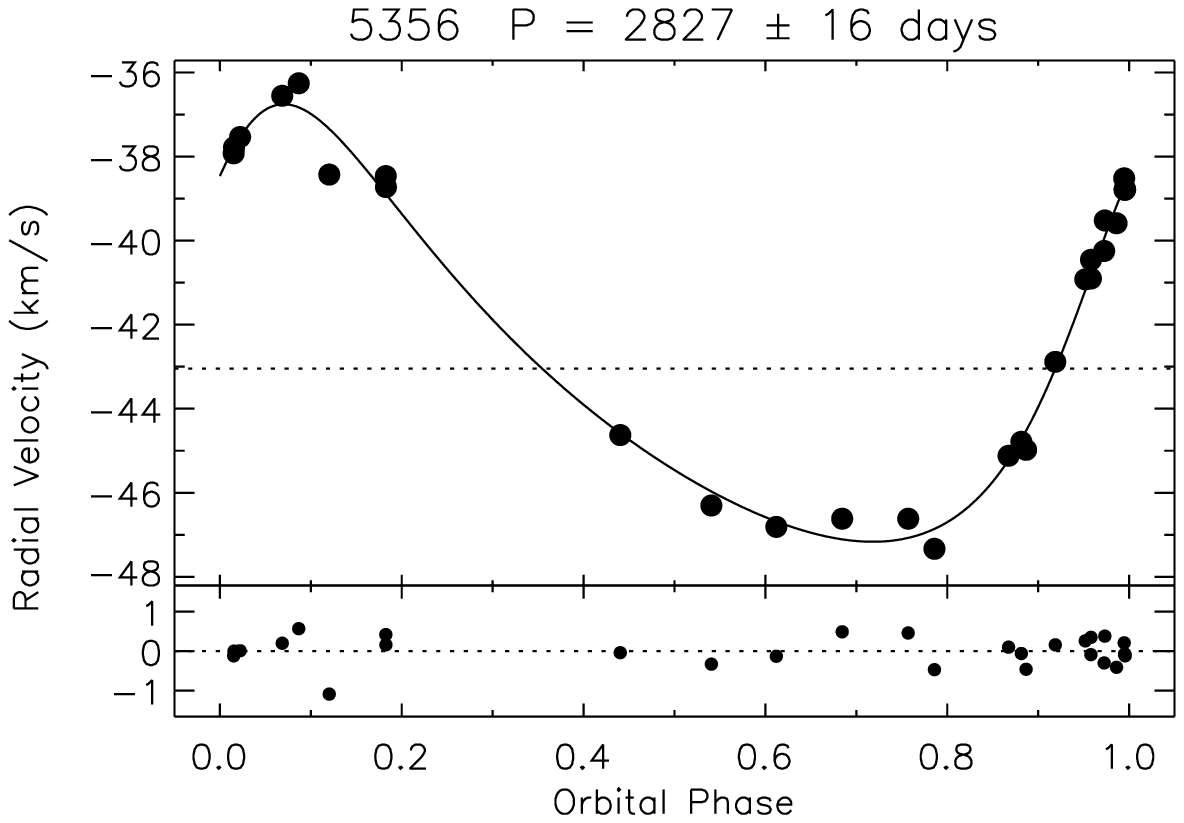,width=0.3\linewidth} \\
\epsfig{file=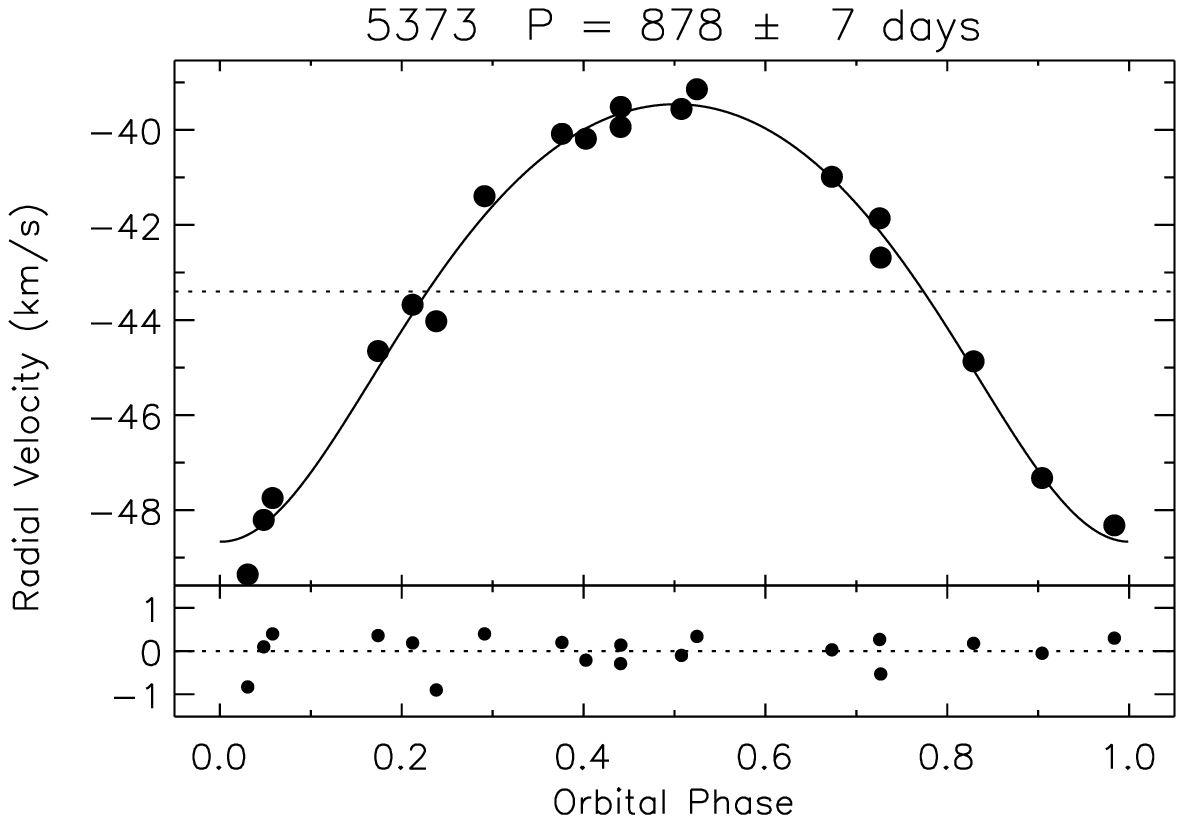,width=0.3\linewidth} & \epsfig{file=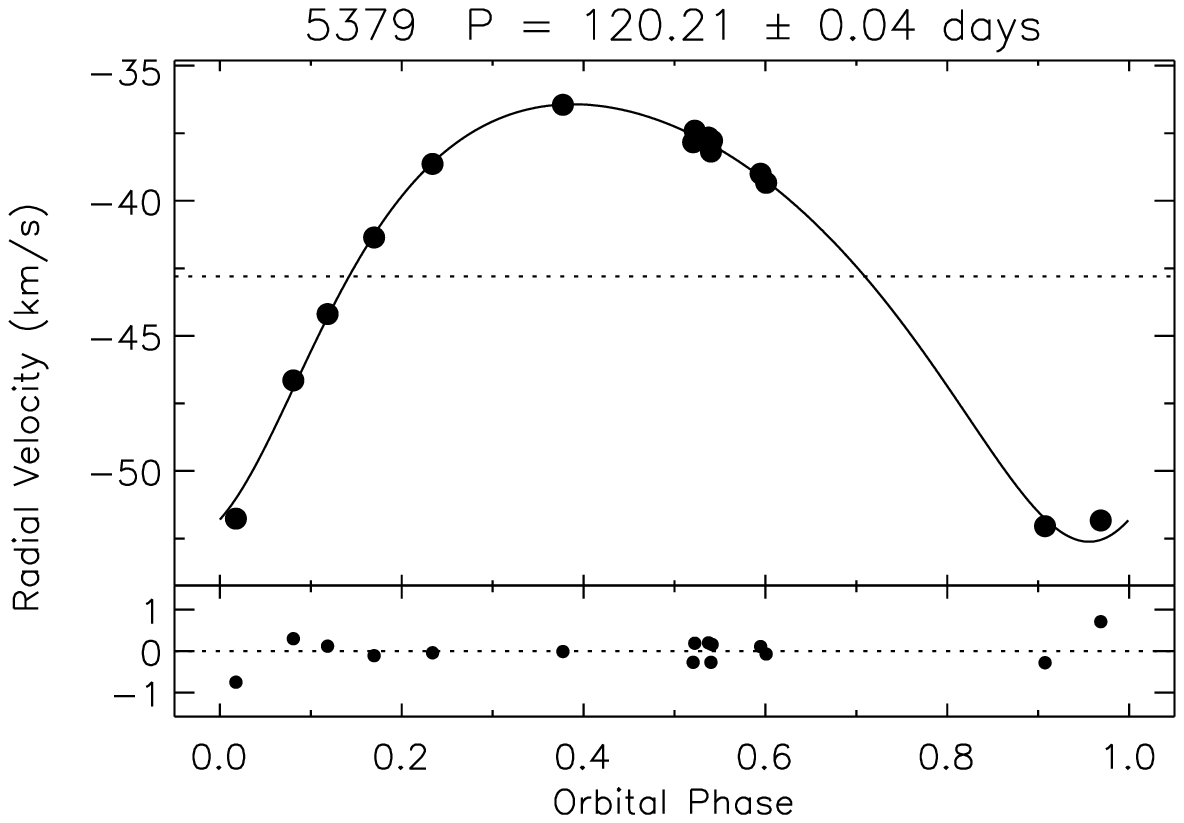,width=0.3\linewidth} & \epsfig{file=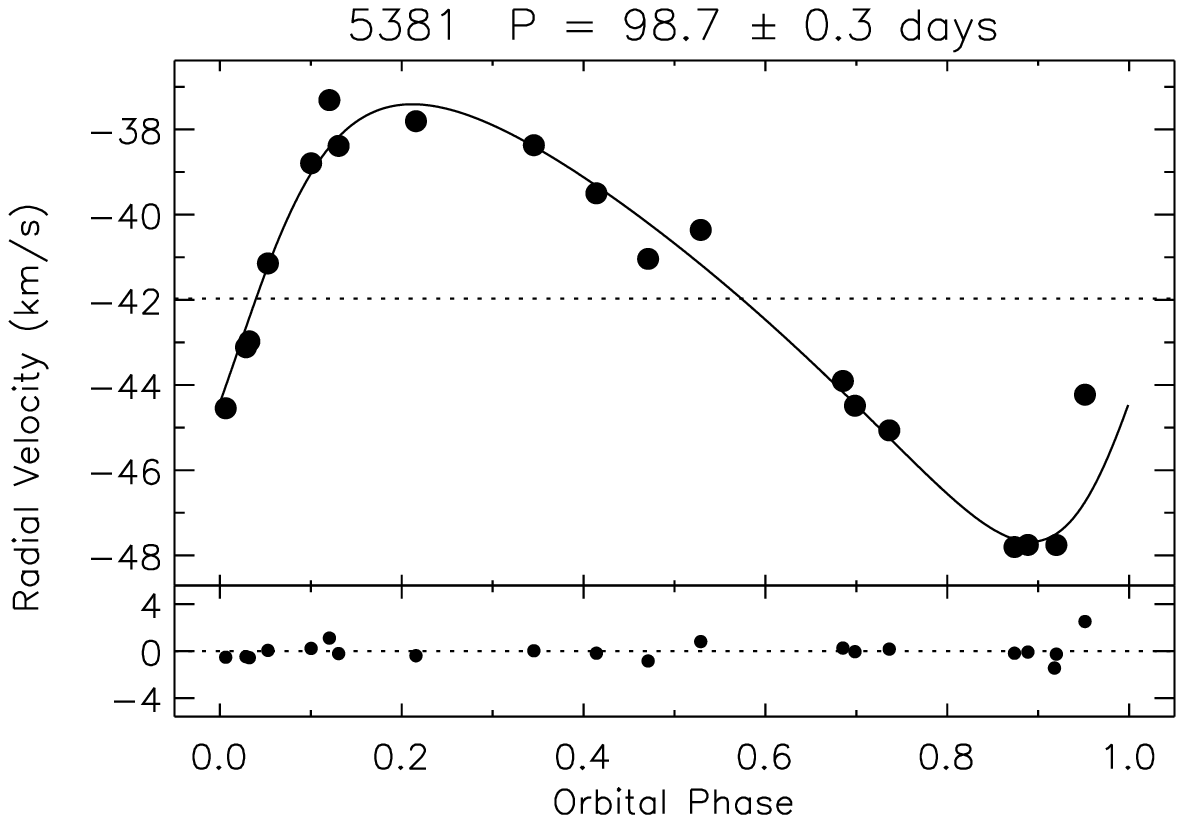,width=0.3\linewidth} \\
\epsfig{file=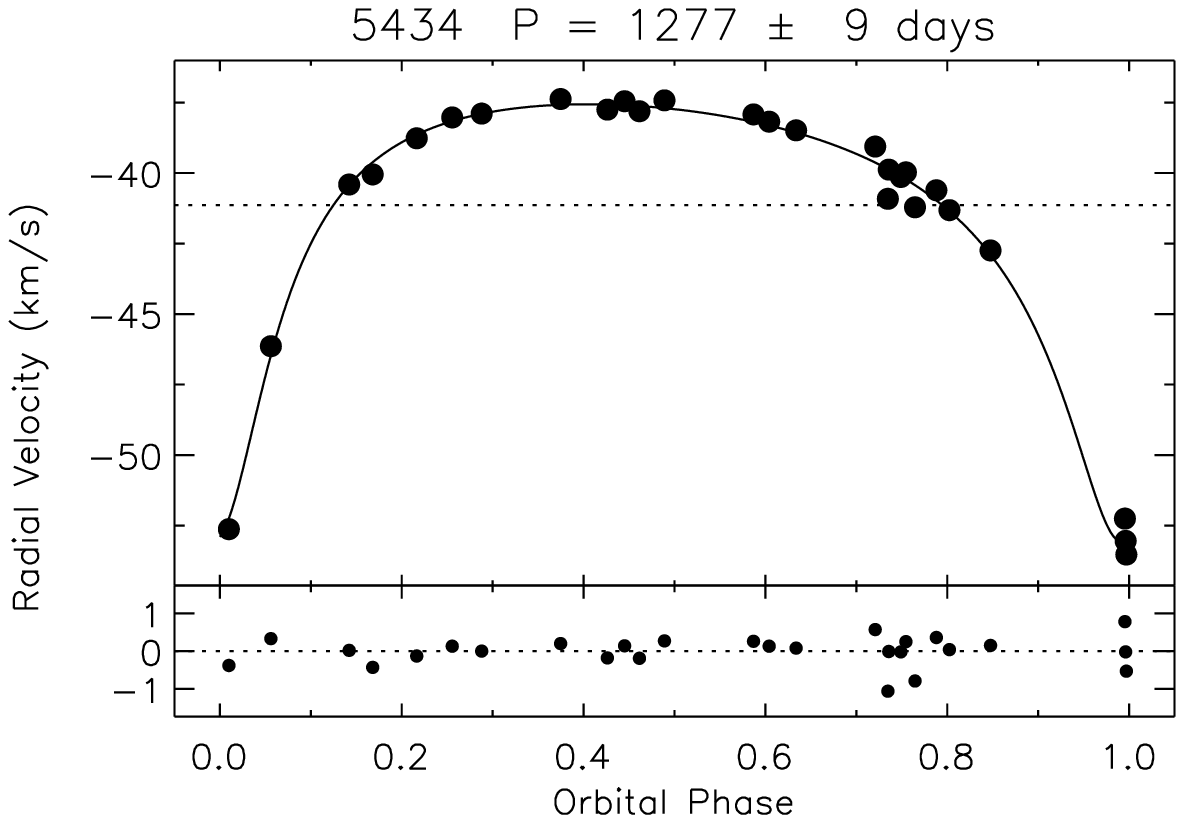,width=0.3\linewidth} & \epsfig{file=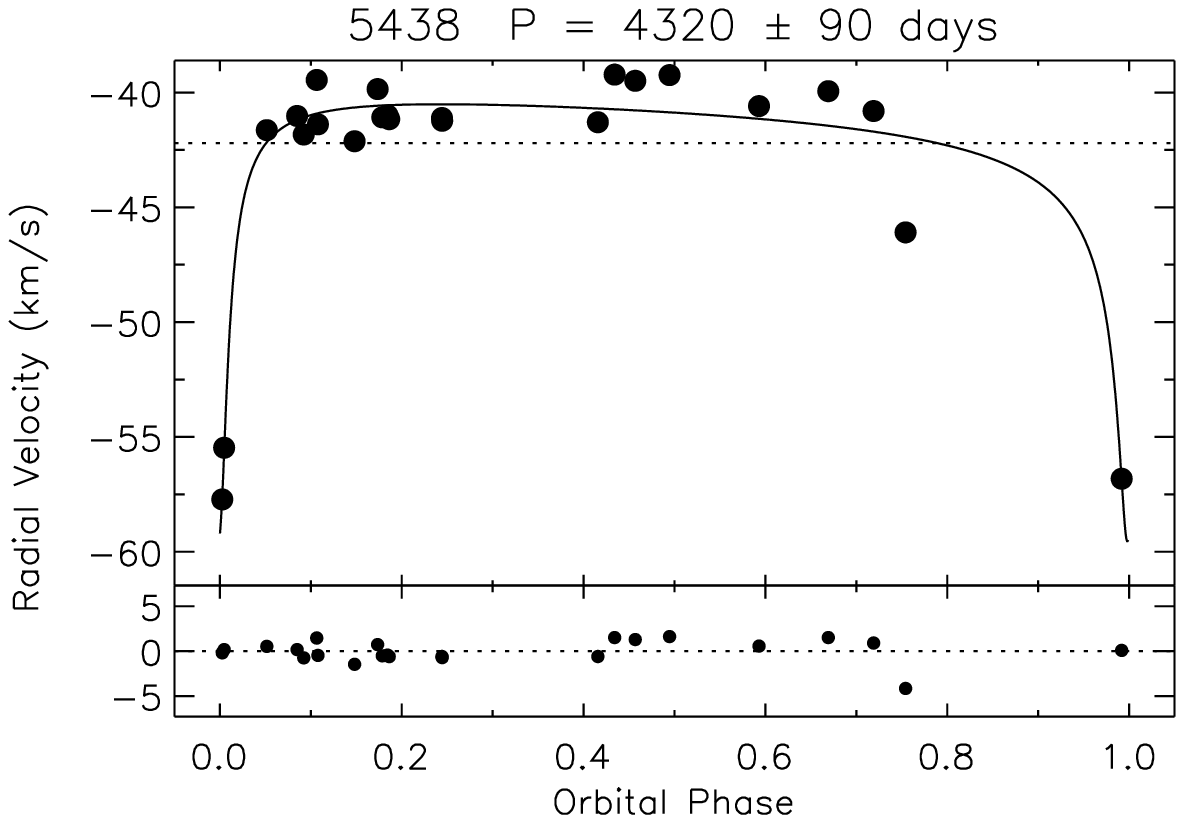,width=0.3\linewidth} & \epsfig{file=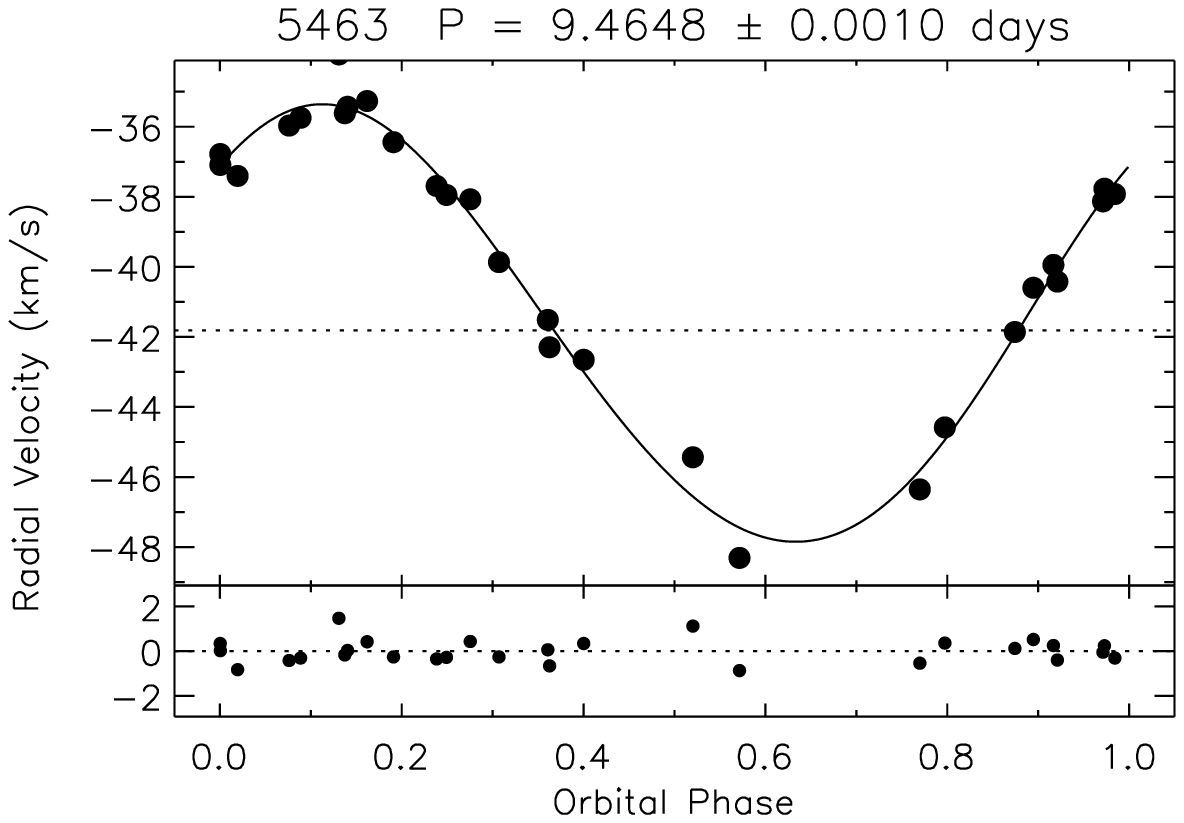,width=0.3\linewidth} \\
\epsfig{file=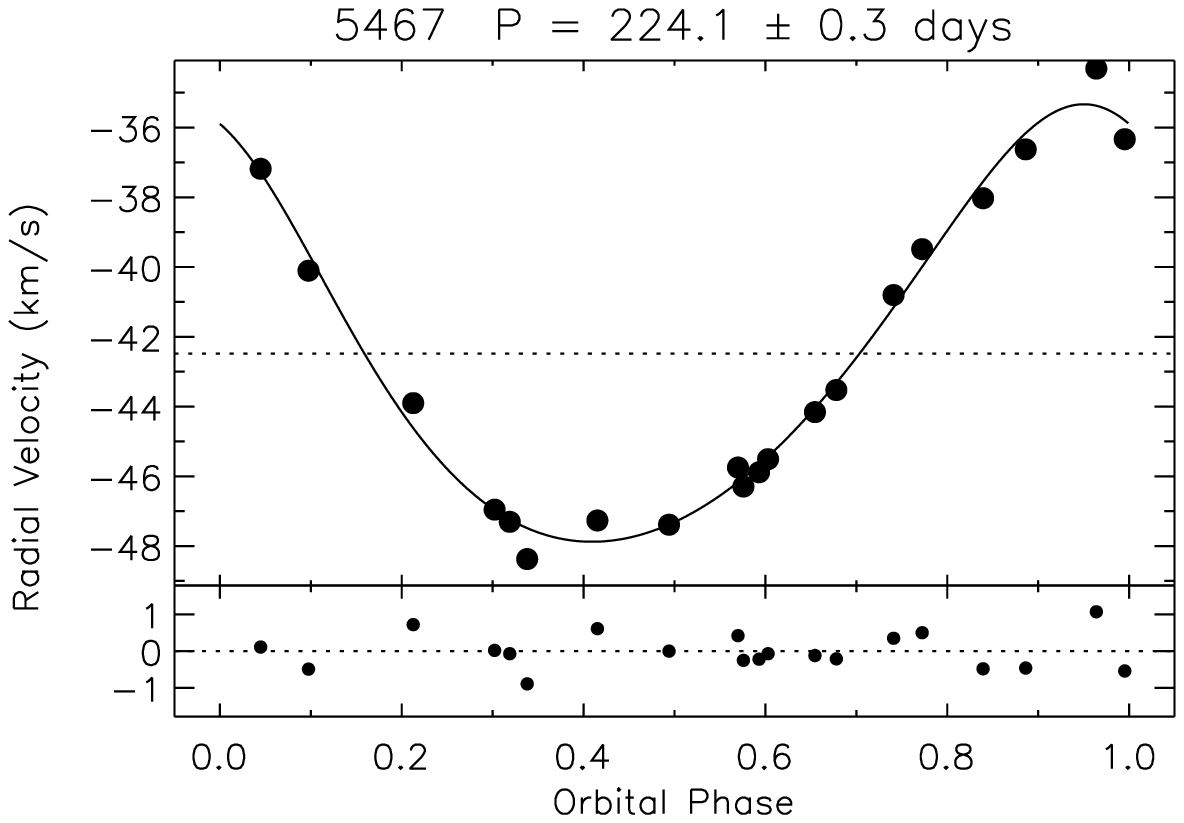,width=0.3\linewidth} & \epsfig{file=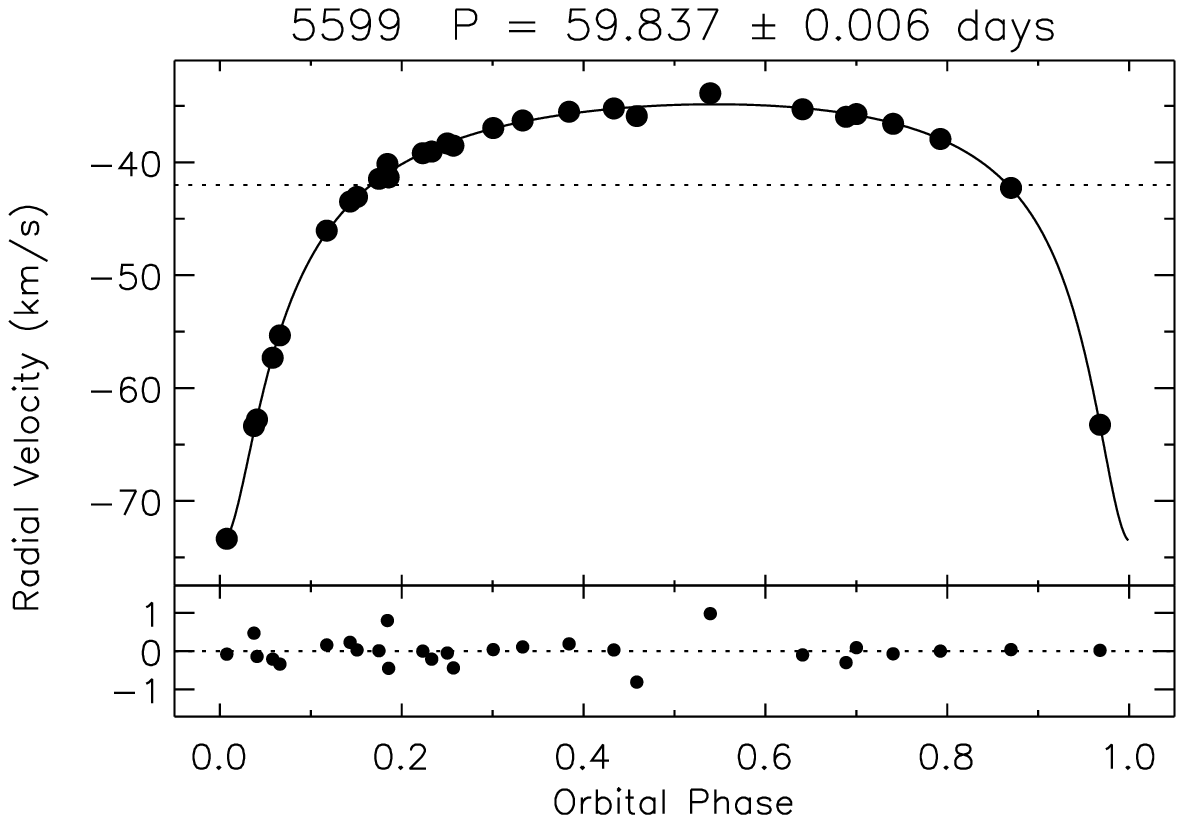,width=0.3\linewidth} & \epsfig{file=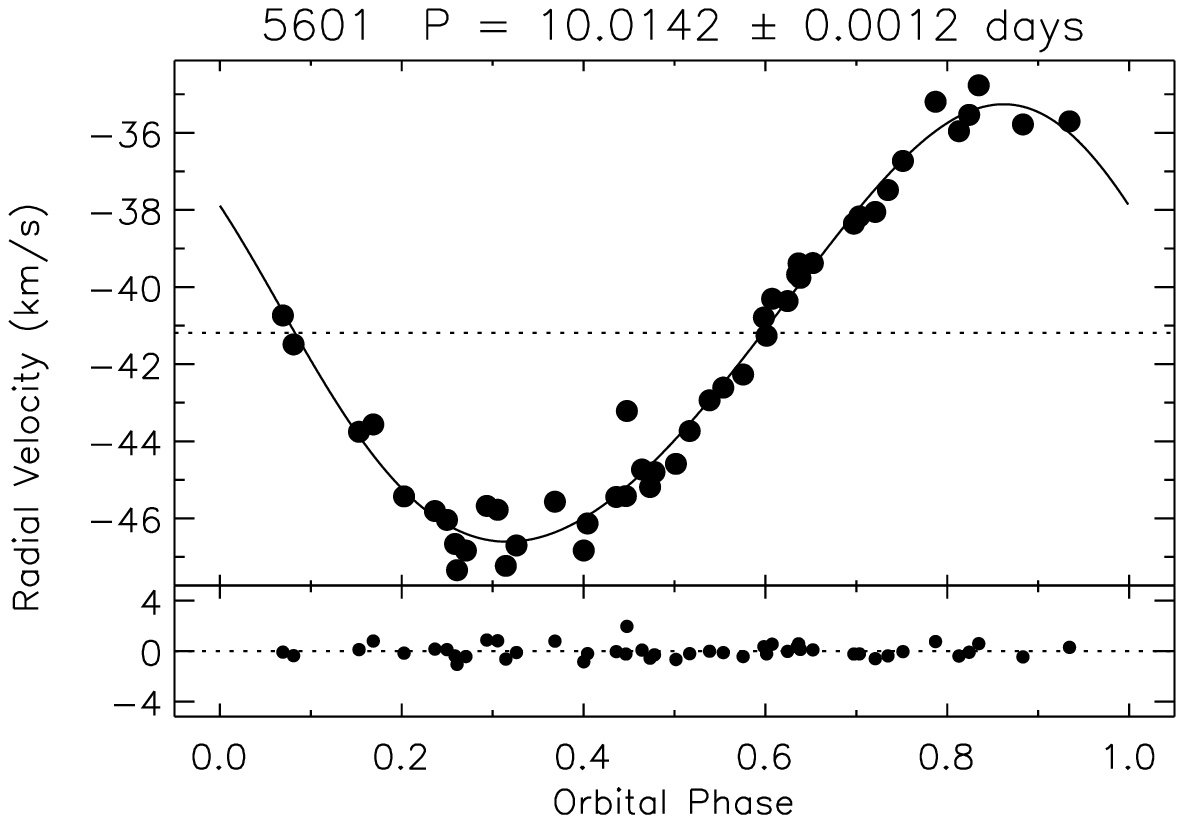,width=0.3\linewidth} \\
\epsfig{file=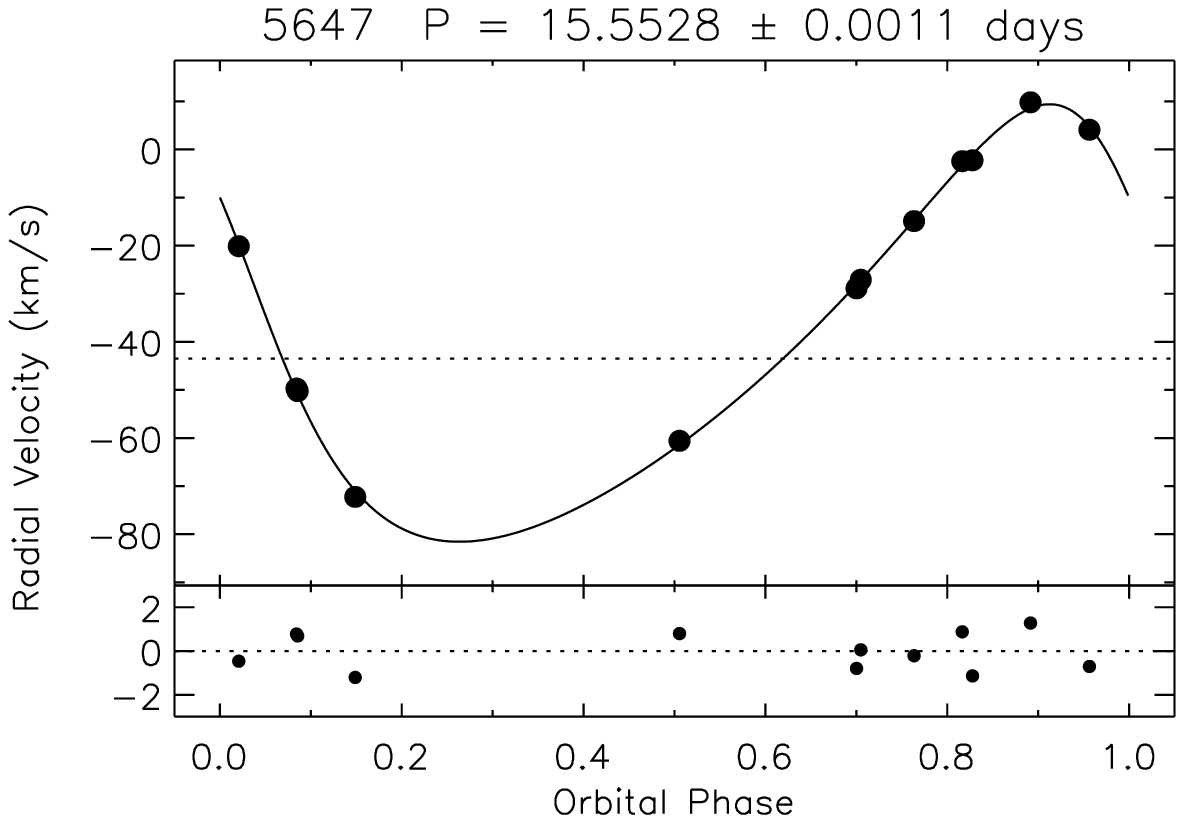,width=0.3\linewidth} & \epsfig{file=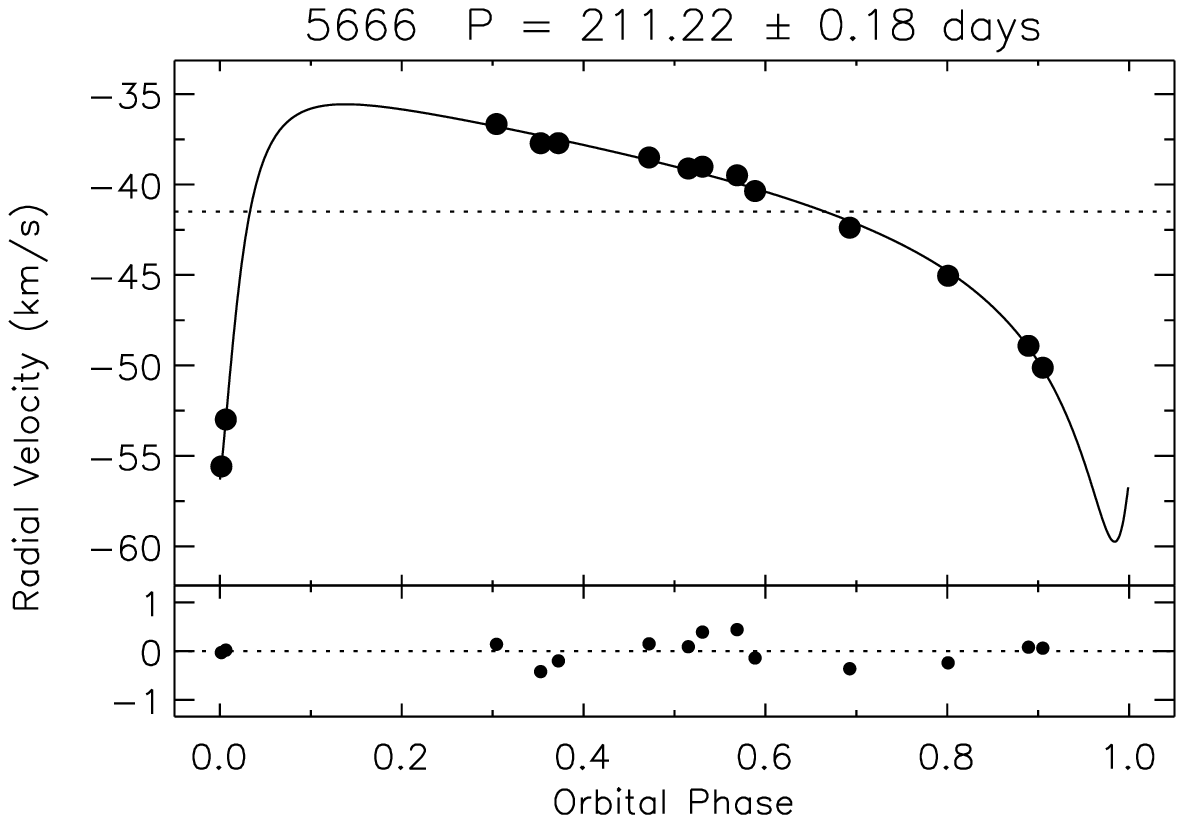,width=0.3\linewidth} & \epsfig{file=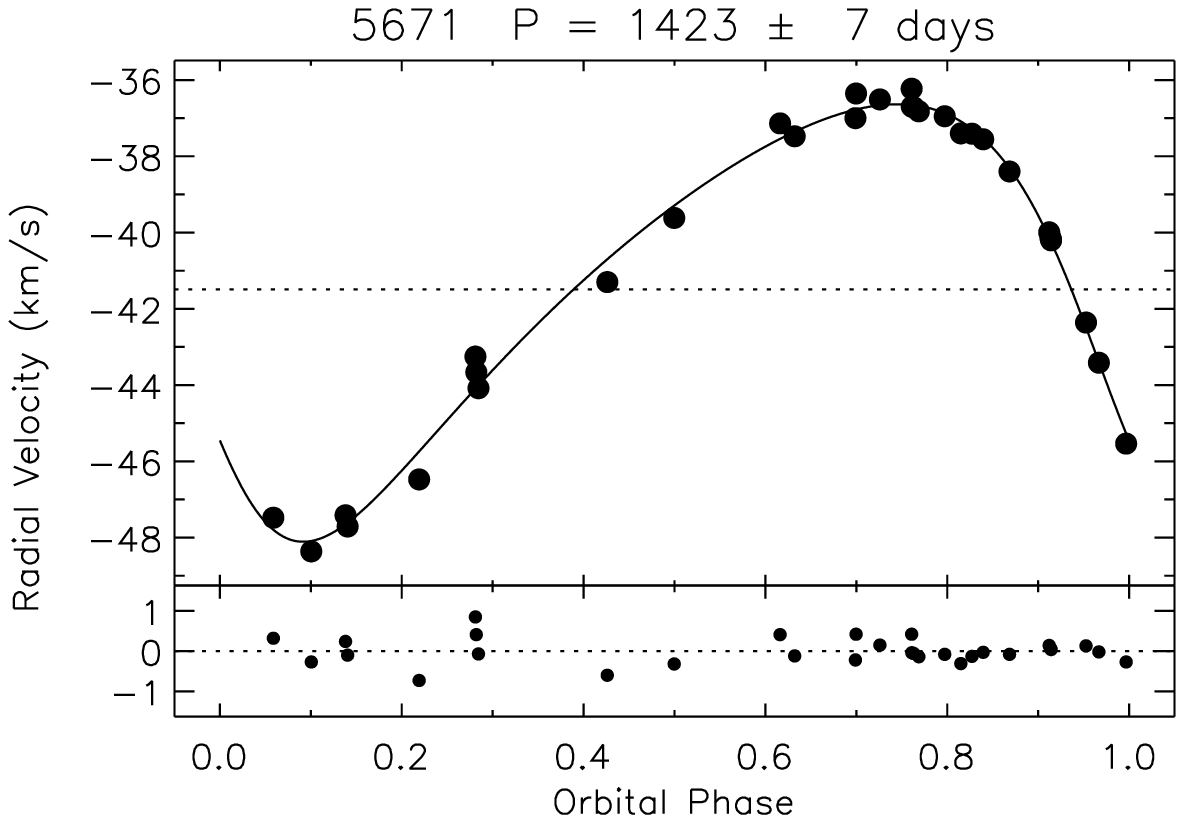,width=0.3\linewidth} \\
\epsfig{file=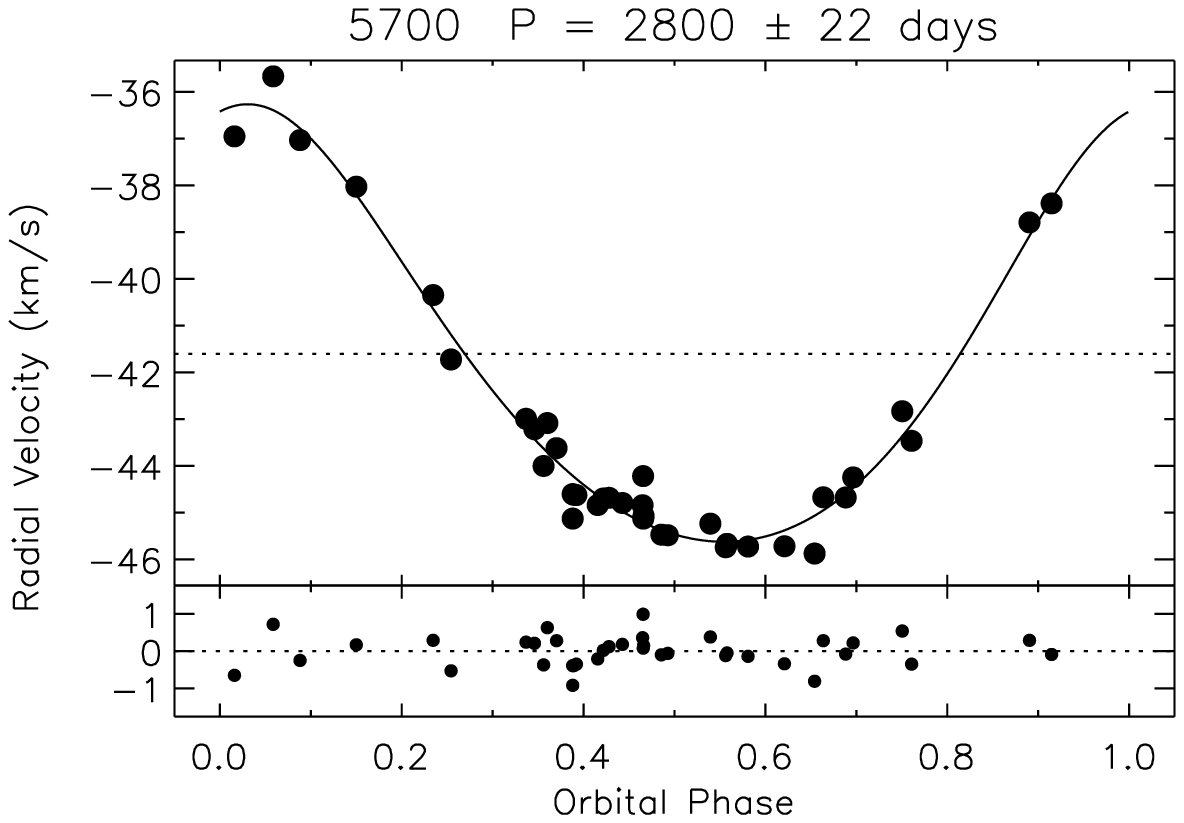,width=0.3\linewidth} & \epsfig{file=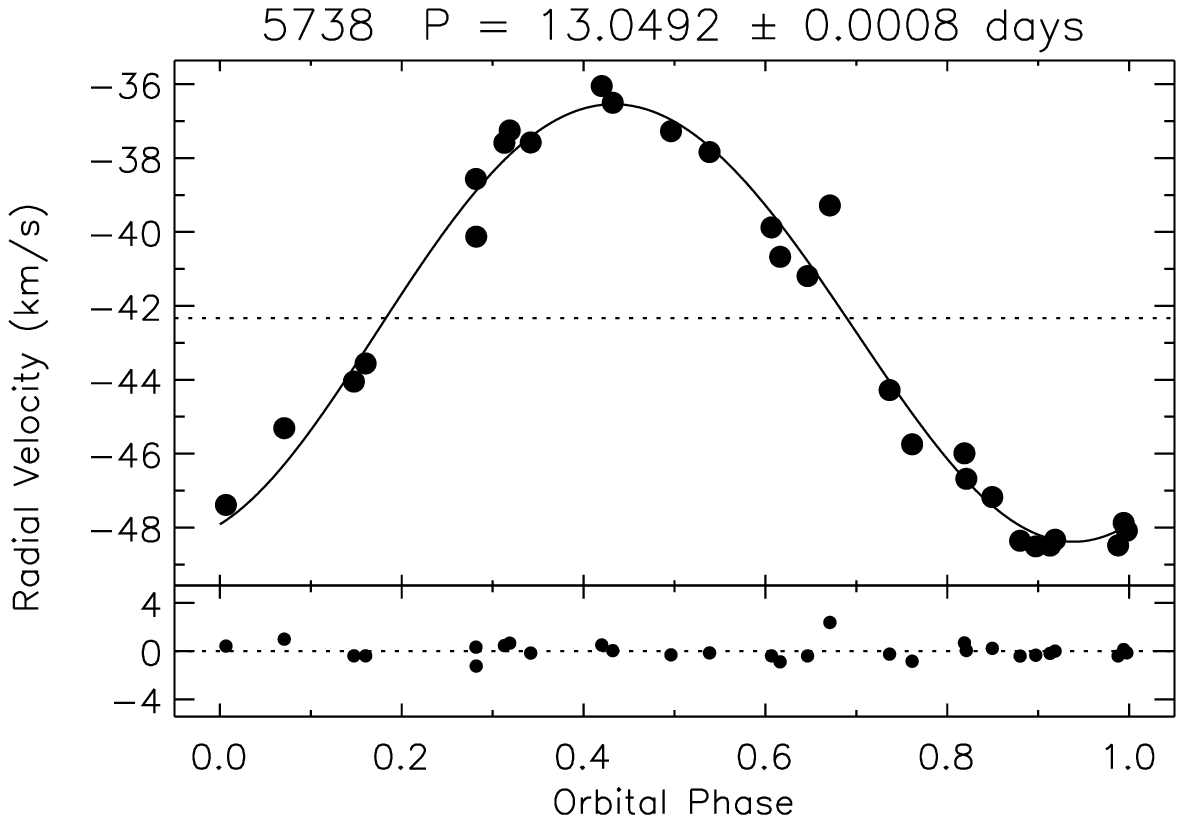,width=0.3\linewidth} & \epsfig{file=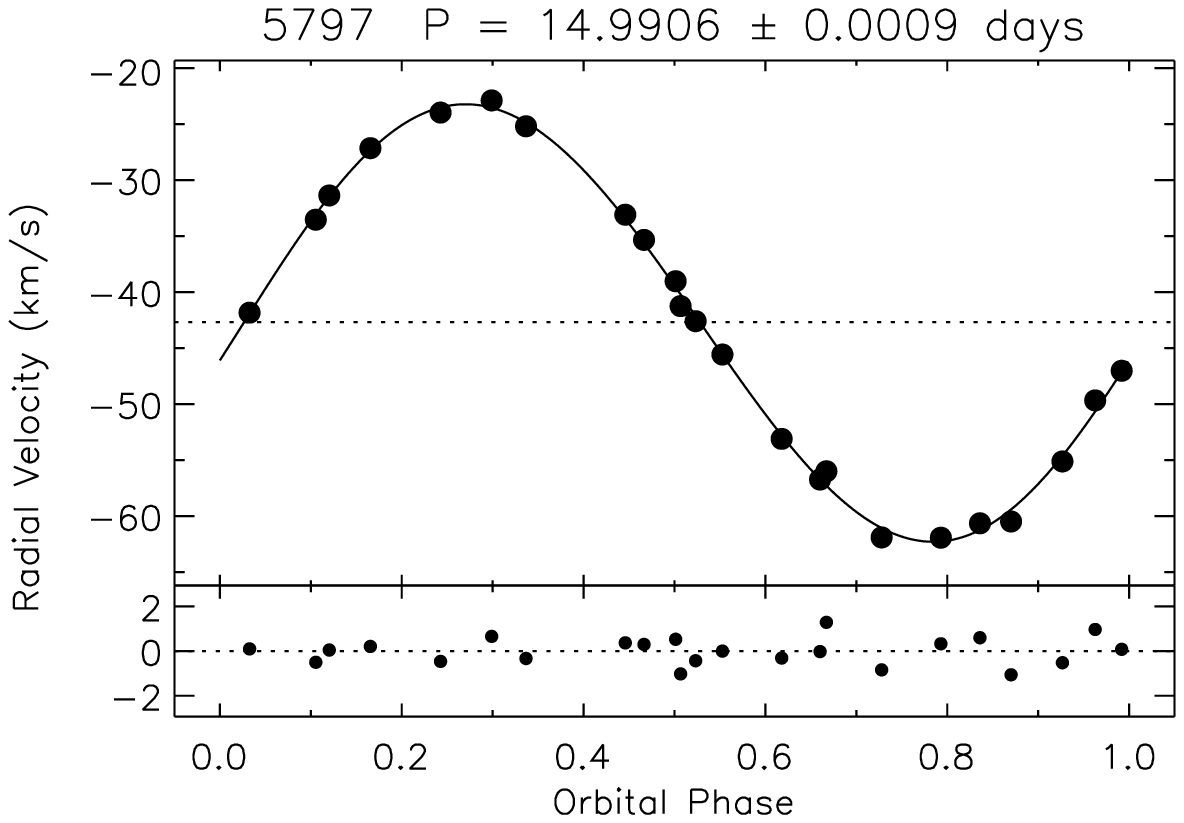,width=0.3\linewidth} \\
\epsfig{file=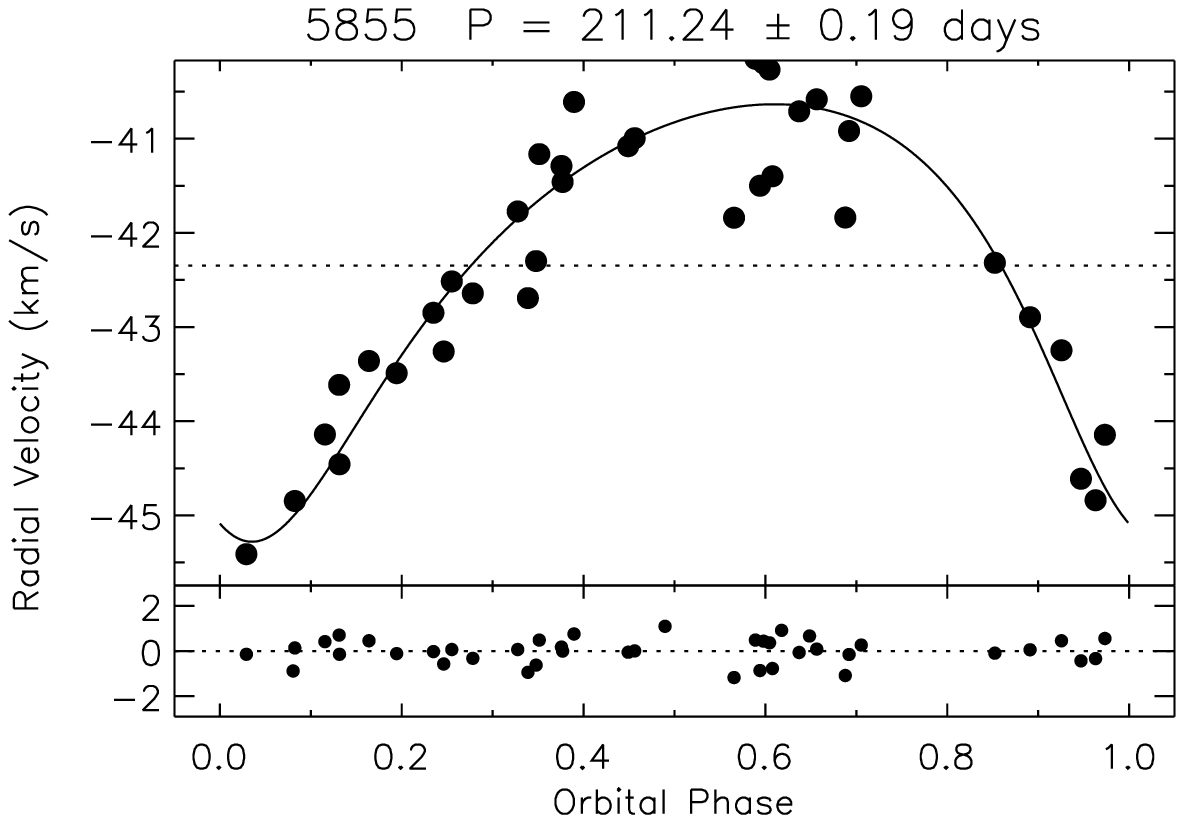,width=0.3\linewidth} & \epsfig{file=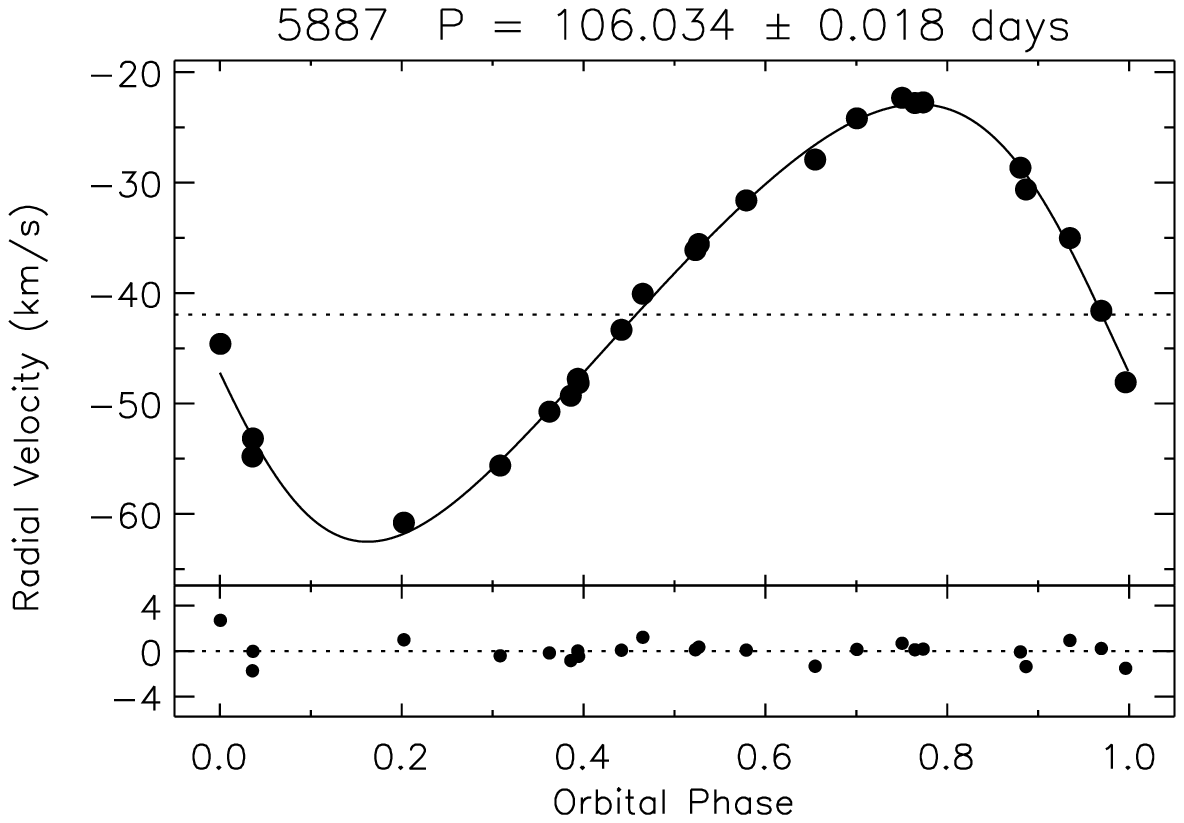,width=0.3\linewidth} & \epsfig{file=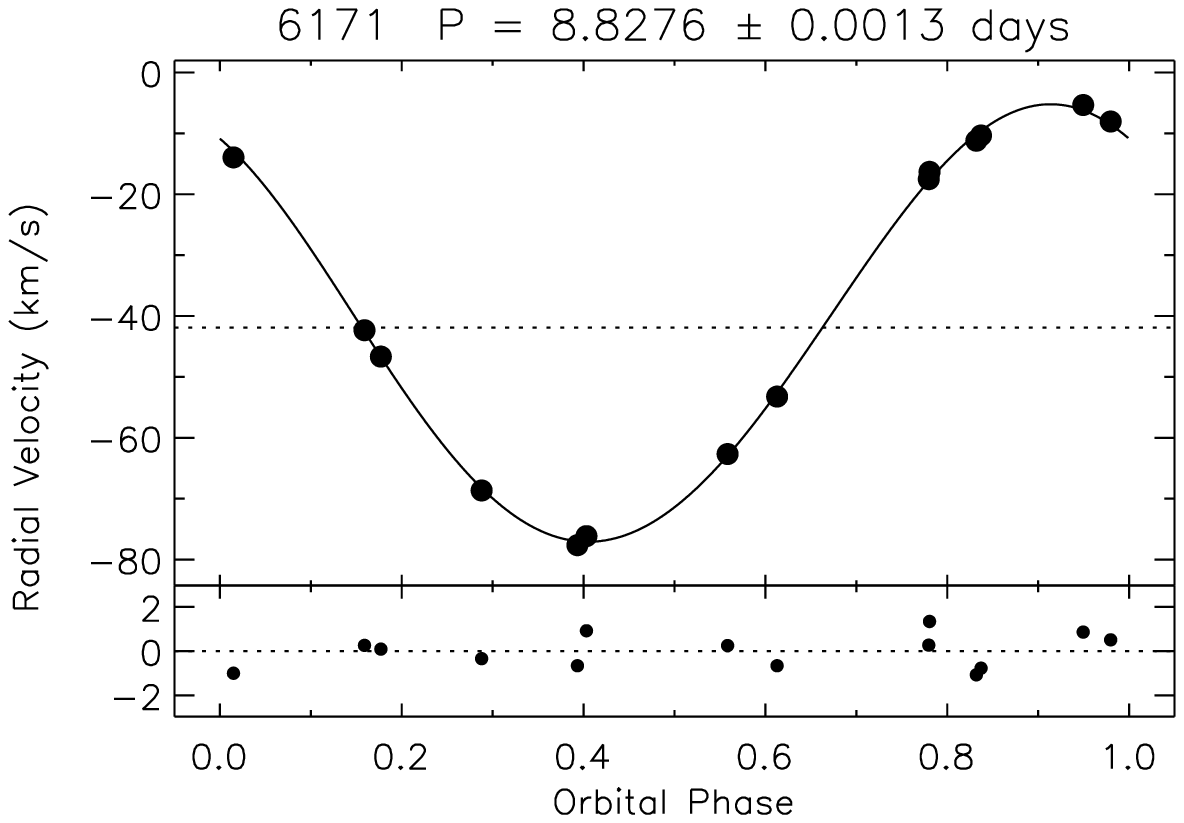,width=0.3\linewidth} \\
\epsfig{file=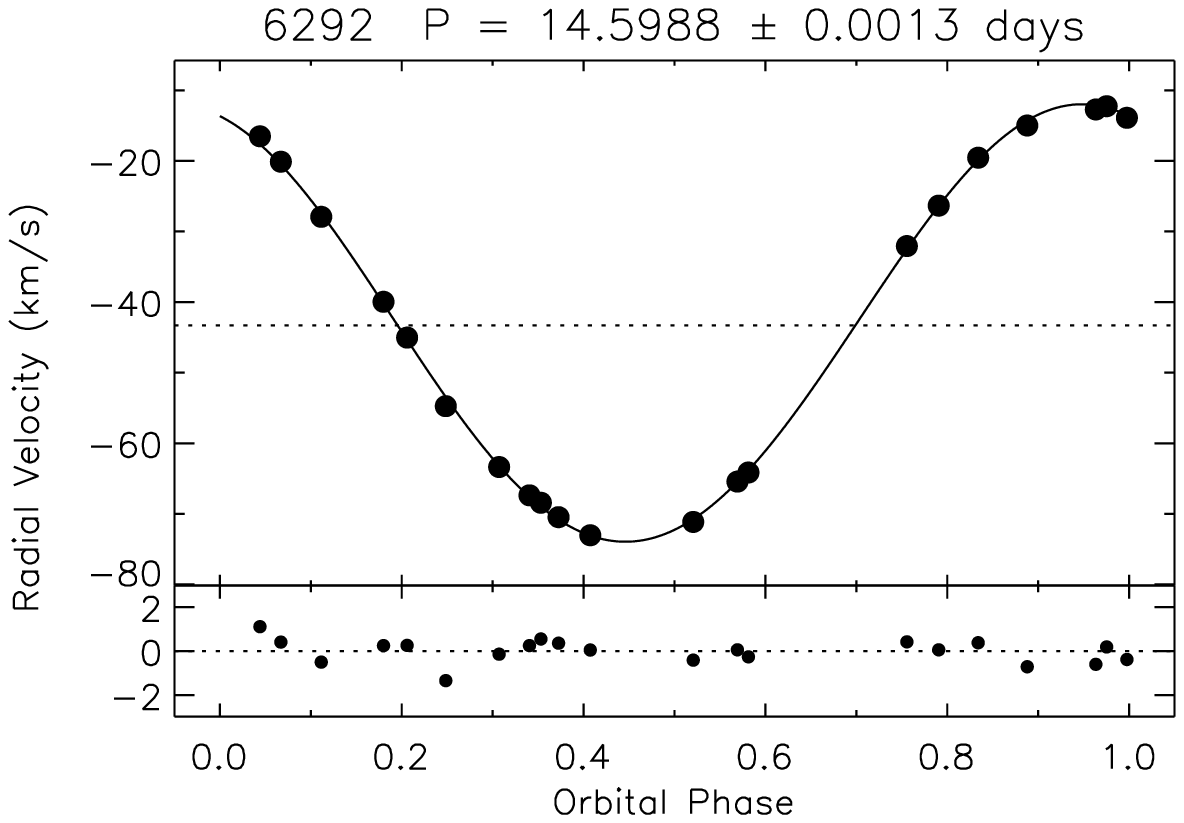,width=0.3\linewidth} & \epsfig{file=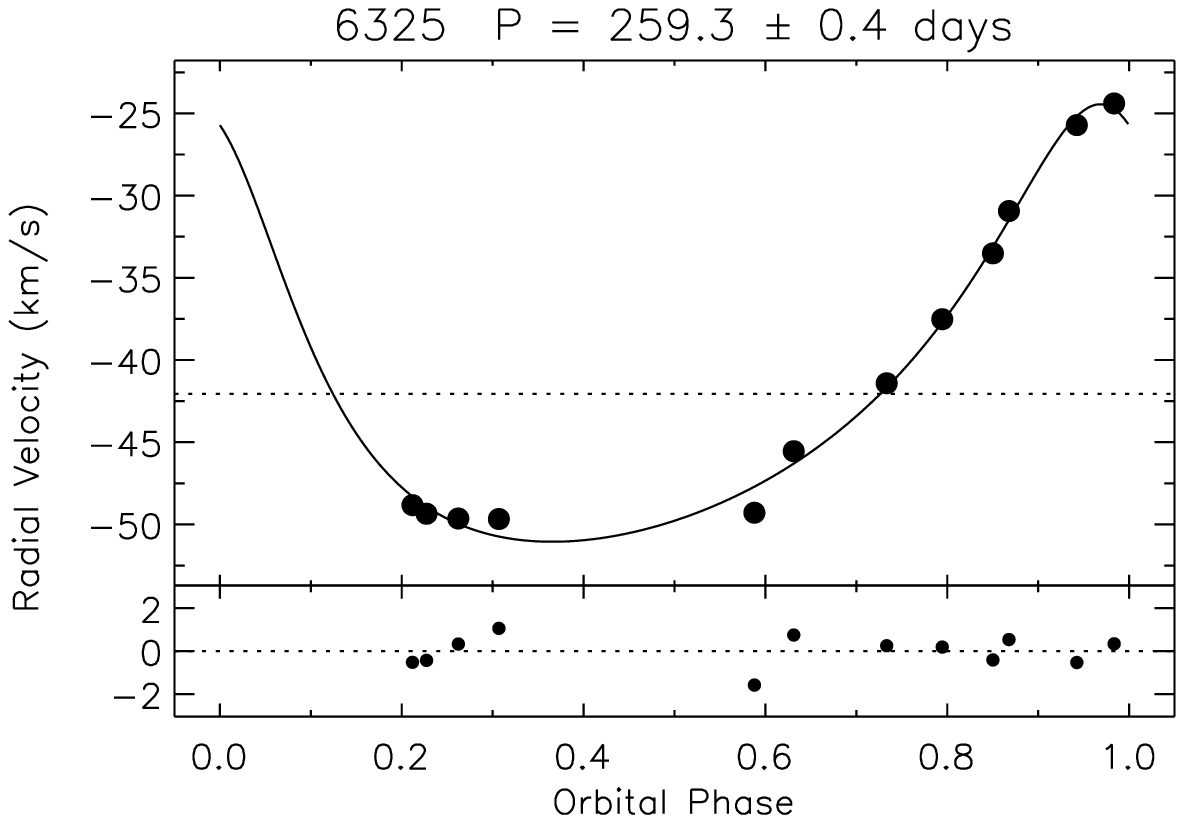,width=0.3\linewidth} & \epsfig{file=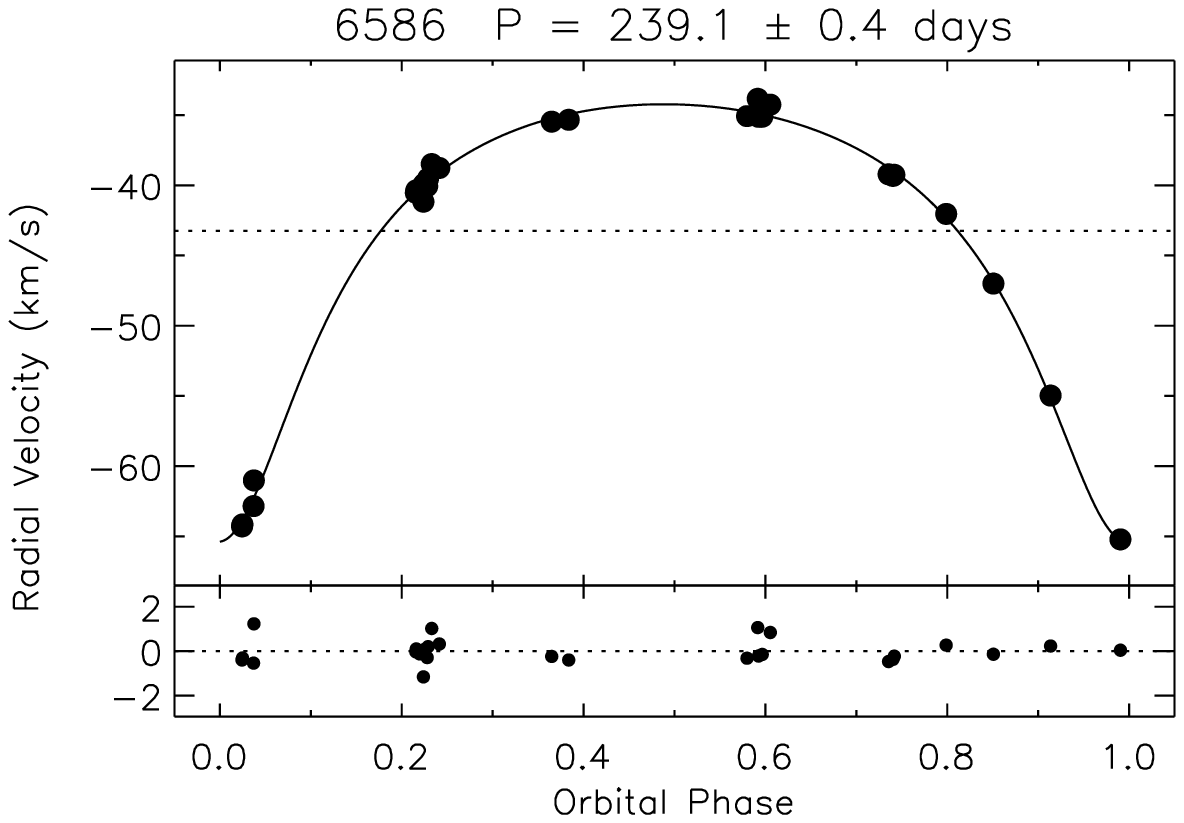,width=0.3\linewidth} \\
\epsfig{file=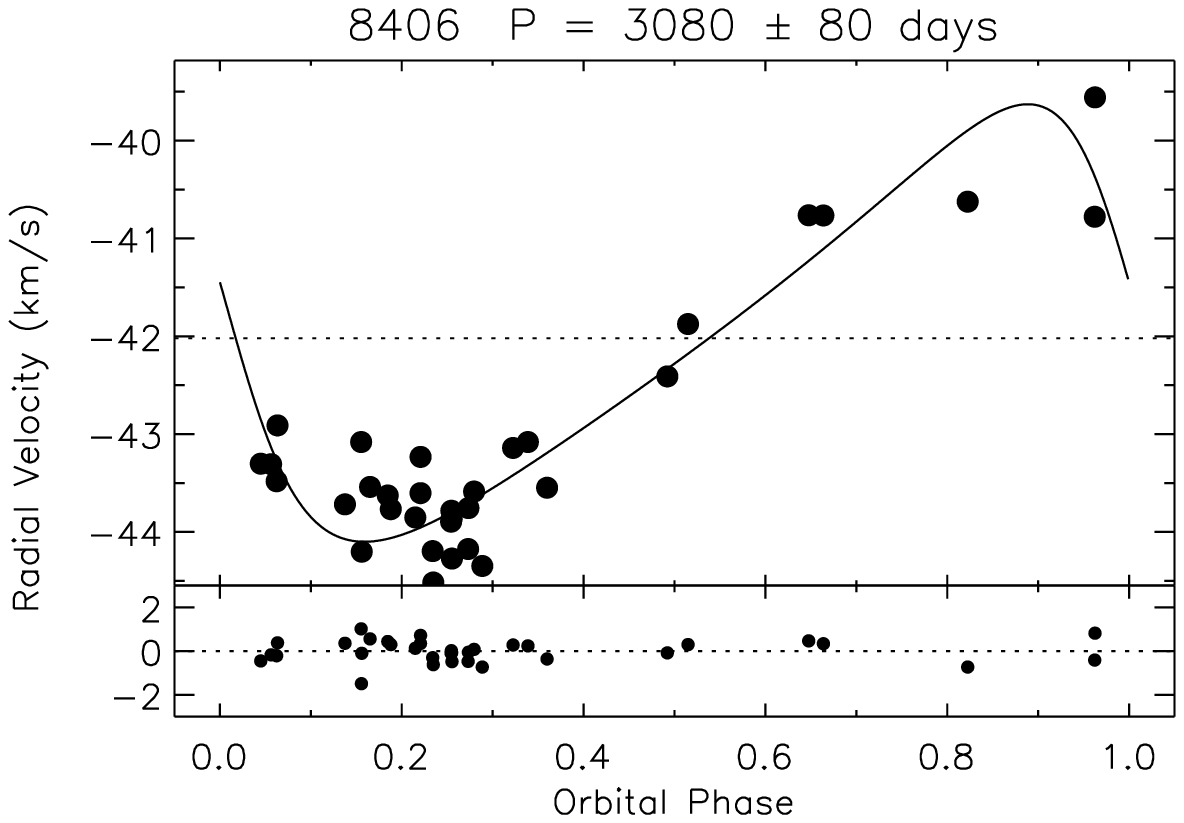,width=0.3\linewidth} & & \\
\end{longtable}
\end{center}
Fig. 1. --- \footnotesize NGC 188 SB1 orbit plots.  For each binary, we plot RV against orbital phase, showing the data points with black dots and the orbital fit to the data with the solid line; the dotted line marks the $\gamma$-velocity.  Beneath each orbit plot, we show the residuals from the fit.  Above each plot, we give the binary ID and orbital period.\normalsize

\addtocounter{figure}{+1}
\addtocounter{table}{-1}

\begin{deluxetable}{l r c r r r r r r r c c}
\tabletypesize{\scriptsize}
\tablewidth{0pt}
\tablecaption{Orbital Parameters For NGC 188 Single-Lined Binaries\label{SB1tab}}
\tablehead{\colhead{ID} & \colhead{P} & \colhead{Orbital} & \colhead{$\gamma$} & \colhead{K} & \colhead{e} & \colhead{$\omega$} & \colhead{T$_\circ$} & \colhead{a$\sin$ i} & \colhead{f(m)} & \colhead{$\sigma$} & \colhead{N} \\
\colhead{} & \colhead{(days)} & \colhead{Cycles} & \colhead{(\kms)} & \colhead{(\kms)} & \colhead{} & \colhead{(deg)} & \colhead{(HJD-2400000 d)} & \colhead{(10$^6$ km)} & \colhead{(\Msolar)} & \colhead{(\kms)} & \colhead{}}
\startdata
    ~451 &             722 &   5.5 &          -41.04 &            4.35 &            0.34 &             235 &           51010 &            40.6 &          5.1e-3 &  0.37 &   28 \\
         &         $\pm$ 4 &       &      $\pm$ 0.10 &      $\pm$ 0.17 &      $\pm$ 0.03 &         $\pm$ 6 &        $\pm$ 12 &       $\pm$ 1.6 &    $\pm$ 0.6e-3 &       &      \\
    ~880 &          20.292 &  42.1 &          -41.71 &            15.4 &           0.213 &             238 &        50715.29 &            4.21 &          7.2e-3 &  0.77 &   31 \\
         &     $\pm$ 0.004 &       &      $\pm$ 0.24 &       $\pm$ 0.3 &     $\pm$ 0.016 &         $\pm$ 4 &      $\pm$ 0.21 &      $\pm$ 0.08 &    $\pm$ 0.4e-3 &       &      \\
    1888 &            2240 &   1.8 &          -41.44 &             6.4 &            0.21 &               2 &           52510 &             192 &          5.7e-2 &  1.11 &   40 \\
         &        $\pm$ 30 &       &      $\pm$ 0.22 &       $\pm$ 0.4 &      $\pm$ 0.04 &        $\pm$ 13 &        $\pm$ 80 &        $\pm$ 12 &    $\pm$ 1.0e-2 &       &      \\
    2679 &            1033 &   3.8 &           -42.3 &             6.2 &            0.07 &              50 &           52500 &              88 &          2.5e-2 &  1.22 &   30 \\
         &         $\pm$ 8 &       &       $\pm$ 0.3 &       $\pm$ 0.4 &      $\pm$ 0.05 &        $\pm$ 50 &       $\pm$ 130 &         $\pm$ 5 &    $\pm$ 0.5e-2 &       &      \\
    3171 &          123.01 &   8.3 &          -42.56 &            8.40 &           0.129 &             275 &           50827 &           14.09 &          7.4e-3 &  0.51 &   28 \\
         &      $\pm$ 0.17 &       &      $\pm$ 0.13 &      $\pm$ 0.14 &     $\pm$ 0.020 &        $\pm$ 10 &         $\pm$ 3 &      $\pm$ 0.24 &    $\pm$ 0.4e-3 &       &      \\
3732$^a$ &             893 &   4.4 &          -42.60 &             4.1 &            0.74 &              46 &           52181 &              34 &          2.0e-3 &  0.48 &   25 \\
         &         $\pm$ 3 &       &      $\pm$ 0.17 &       $\pm$ 1.0 &      $\pm$ 0.11 &         $\pm$ 8 &        $\pm$ 10 &        $\pm$ 10 &    $\pm$ 1.6e-3 &       &      \\
    3871 &            1960 &   1.9 &          -42.87 &            2.91 &            0.54 &              10 &           52630 &              66 &          3.0e-3 &  0.84 &   33 \\
         &        $\pm$ 50 &       &      $\pm$ 0.18 &      $\pm$ 0.23 &      $\pm$ 0.07 &        $\pm$ 12 &        $\pm$ 60 &         $\pm$ 6 &    $\pm$ 0.7e-3 &       &      \\
    4080 &          32.197 &  31.7 &          -42.45 &            20.7 &           0.069 &             264 &         50884.6 &            9.16 &         2.96e-2 &  1.05 &   21 \\
         &     $\pm$ 0.013 &       &      $\pm$ 0.24 &       $\pm$ 0.3 &     $\pm$ 0.015 &        $\pm$ 18 &       $\pm$ 1.6 &      $\pm$ 0.15 &    $\pm$ 1.4e-3 &       &      \\
    4147 &            1194 &   3.1 &          -42.30 &            3.62 &            0.14 &             321 &           52120 &            58.9 &          5.7e-3 &  0.32 &   19 \\
         &         $\pm$ 7 &       &      $\pm$ 0.09 &      $\pm$ 0.13 &      $\pm$ 0.04 &        $\pm$ 14 &        $\pm$ 50 &       $\pm$ 2.1 &    $\pm$ 0.6e-3 &       &      \\
    4289 &         11.4877 & 116.2 &          -42.62 &            40.8 &           0.012 &             200 &         50744.5 &            6.44 &         8.07e-2 &  1.11 &   25 \\
         &    $\pm$ 0.0009 &       &      $\pm$ 0.23 &       $\pm$ 0.3 &     $\pm$ 0.010 &        $\pm$ 50 &       $\pm$ 1.7 &      $\pm$ 0.05 &    $\pm$ 2.0e-3 &       &      \\
    4333 &            1001 &   3.5 &           -42.7 &             4.9 &            0.50 &             207 &           50825 &              58 &          7.7e-3 &  1.37 &   31 \\
         &        $\pm$ 18 &       &       $\pm$ 0.3 &       $\pm$ 0.4 &      $\pm$ 0.06 &        $\pm$ 10 &        $\pm$ 20 &         $\pm$ 6 &    $\pm$ 2.2e-3 &       &      \\
    4348 &            1168 &   3.2 &           -40.7 &             6.7 &            0.09 &             220 &           51750 &             107 &          3.6e-2 &  1.34 &   27 \\
         &         $\pm$ 8 &       &       $\pm$ 0.3 &       $\pm$ 0.4 &      $\pm$ 0.05 &        $\pm$ 40 &       $\pm$ 130 &         $\pm$ 7 &    $\pm$ 0.7e-2 &       &      \\
    4369 &             441 &   2.0 &          -41.91 &            6.42 &            0.38 &              53 &           50625 &            36.0 &          9.5e-3 &  0.66 &   20 \\
         &         $\pm$ 3 &       &      $\pm$ 0.16 &      $\pm$ 0.22 &      $\pm$ 0.03 &         $\pm$ 6 &         $\pm$ 6 &       $\pm$ 1.4 &    $\pm$ 1.0e-3 &       &      \\
    4386 &         108.393 &  11.7 &          -42.97 &            20.2 &           0.785 &           133.7 &        51125.56 &            18.6 &          2.2e-2 &  0.54 &   32 \\
         &     $\pm$ 0.018 &       &      $\pm$ 0.12 &       $\pm$ 1.1 &     $\pm$ 0.013 &       $\pm$ 0.9 &      $\pm$ 0.23 &       $\pm$ 1.2 &    $\pm$ 0.4e-2 &       &      \\
    4390 &            1313 &   3.0 &          -42.40 &             5.5 &            0.37 &             143 &           51203 &              92 &          1.8e-2 &  0.60 &   32 \\
         &         $\pm$ 7 &       &      $\pm$ 0.23 &       $\pm$ 0.9 &      $\pm$ 0.08 &         $\pm$ 8 &        $\pm$ 23 &        $\pm$ 15 &    $\pm$ 0.9e-2 &       &      \\
    4392 &            4640 &   0.9 &          -42.13 &            5.06 &           0.511 &             194 &           55530 &             277 &          3.9e-2 &  0.30 &   34 \\
         &       $\pm$ 150 &       &      $\pm$ 0.07 &      $\pm$ 0.09 &     $\pm$ 0.015 &         $\pm$ 3 &       $\pm$ 160 &         $\pm$ 6 &    $\pm$ 0.3e-2 &       &      \\
    4524 &            1267 &   7.6 &          -42.97 &            5.19 &            0.09 &              34 &           49020 &              90 &         1.81e-2 &  0.76 &   61 \\
         &         $\pm$ 3 &       &      $\pm$ 0.11 &      $\pm$ 0.16 &      $\pm$ 0.03 &        $\pm$ 18 &        $\pm$ 70 &         $\pm$ 3 &    $\pm$ 1.6e-3 &       &      \\
    4540 &            3030 &   1.5 &           -40.4 &             5.0 &            0.36 &             114 &           53140 &             196 &          3.3e-2 &  0.65 &   23 \\
         &        $\pm$ 70 &       &       $\pm$ 0.3 &       $\pm$ 0.4 &      $\pm$ 0.07 &        $\pm$ 10 &       $\pm$ 100 &        $\pm$ 15 &    $\pm$ 0.7e-2 &       &      \\
    4560 &            1365 &   2.9 &          -41.93 &             3.3 &            0.63 &             322 &           53413 &              49 &          2.5e-3 &  0.24 &   15 \\
         &         $\pm$ 5 &       &      $\pm$ 0.07 &       $\pm$ 0.5 &      $\pm$ 0.07 &         $\pm$ 6 &         $\pm$ 7 &         $\pm$ 9 &    $\pm$ 1.2e-3 &       &      \\
    4565 &          344.01 &  19.8 &          -42.18 &            8.36 &           0.641 &           240.8 &         47601.9 &            30.4 &          9.4e-3 &  0.71 &   48 \\
         &      $\pm$ 0.05 &       &      $\pm$ 0.11 &      $\pm$ 0.23 &     $\pm$ 0.013 &       $\pm$ 2.5 &       $\pm$ 0.9 &       $\pm$ 0.9 &    $\pm$ 0.8e-3 &       &      \\
    4581 &           546.7 &   2.6 &          -41.90 &            8.40 &           0.269 &             217 &           51222 &            60.8 &         3.00e-2 &  0.27 &   18 \\
         &       $\pm$ 1.6 &       &      $\pm$ 0.08 &      $\pm$ 0.09 &     $\pm$ 0.015 &         $\pm$ 3 &         $\pm$ 3 &       $\pm$ 0.7 &    $\pm$ 1.0e-3 &       &      \\
    4585 &          181.52 &  11.4 &           -43.1 &            12.7 &           0.370 &              30 &         52110.4 &            29.5 &          3.1e-2 &  0.95 &   16 \\
         &      $\pm$ 0.17 &       &       $\pm$ 0.3 &       $\pm$ 0.4 &     $\pm$ 0.023 &         $\pm$ 6 &       $\pm$ 2.1 &       $\pm$ 0.9 &    $\pm$ 0.3e-2 &       &      \\
    4589 &           615.2 &   6.9 &          -43.20 &            4.62 &            0.21 &             128 &           52125 &            38.3 &          5.9e-3 &  0.54 &   23 \\
         &       $\pm$ 1.7 &       &      $\pm$ 0.16 &      $\pm$ 0.17 &      $\pm$ 0.04 &        $\pm$ 12 &        $\pm$ 19 &       $\pm$ 1.5 &    $\pm$ 0.7e-3 &       &      \\
    4595 &            1276 &   2.4 &          -42.40 &             6.3 &            0.25 &             163 &           51720 &             106 &          2.9e-2 &  0.70 &   15 \\
         &         $\pm$ 9 &       &      $\pm$ 0.22 &       $\pm$ 0.3 &      $\pm$ 0.04 &        $\pm$ 10 &        $\pm$ 30 &         $\pm$ 5 &    $\pm$ 0.4e-2 &       &      \\
    4618 &          8.0729 & 104.4 &           -42.4 &            55.1 &           0.017 &             306 &         50819.5 &            6.12 &         1.40e-1 &  1.15 &   21 \\
         &    $\pm$ 0.0004 &       &       $\pm$ 0.3 &       $\pm$ 0.4 &     $\pm$ 0.007 &        $\pm$ 24 &       $\pm$ 0.5 &      $\pm$ 0.04 &    $\pm$ 0.3e-2 &       &      \\
    4673 &           48.27 &  22.4 &          -41.35 &             8.4 &            0.37 &             351 &         50836.6 &            5.18 &          2.4e-3 &  0.49 &   22 \\
         &      $\pm$ 0.06 &       &      $\pm$ 0.12 &       $\pm$ 0.3 &      $\pm$ 0.03 &         $\pm$ 4 &       $\pm$ 0.6 &      $\pm$ 0.19 &    $\pm$ 0.3e-3 &       &      \\
4688$^a$ &          1222.3 &   3.0 &          -42.59 &             9.6 &            0.70 &             163 &           52238 &             114 &          4.0e-2 &  0.30 &   25 \\
         &       $\pm$ 1.0 &       &      $\pm$ 0.18 &       $\pm$ 1.7 &      $\pm$ 0.04 &         $\pm$ 3 &         $\pm$ 5 &        $\pm$ 21 &    $\pm$ 2.1e-2 &       &      \\
    4710 &         21.3407 & 203.7 &           -45.7 &            16.9 &           0.675 &             340 &        52412.34 &             3.7 &          4.3e-3 &  1.18 &   37 \\
         &    $\pm$ 0.0007 &       &       $\pm$ 0.3 &       $\pm$ 1.5 &     $\pm$ 0.024 &         $\pm$ 3 &      $\pm$ 0.16 &       $\pm$ 0.3 &    $\pm$ 1.1e-3 &       &      \\
    4726 &          110.81 &   8.0 &          -42.13 &            9.40 &           0.446 &           252.4 &         51017.3 &            12.8 &          6.9e-3 &  0.30 &   14 \\
         &      $\pm$ 0.16 &       &      $\pm$ 0.10 &      $\pm$ 0.20 &     $\pm$ 0.016 &       $\pm$ 2.2 &       $\pm$ 0.5 &       $\pm$ 0.3 &    $\pm$ 0.5e-3 &       &      \\
    4734 &            1029 &   1.8 &          -43.45 &             8.7 &           0.312 &             143 &           51730 &             117 &          6.0e-2 &  0.83 &   25 \\
         &         $\pm$ 7 &       &      $\pm$ 0.17 &       $\pm$ 0.3 &     $\pm$ 0.024 &         $\pm$ 6 &        $\pm$ 15 &         $\pm$ 4 &    $\pm$ 0.5e-2 &       &      \\
    4843 &            1240 &   9.1 &          -42.08 &            5.32 &            0.21 &             256 &           49310 &              89 &         1.81e-2 &  0.58 &   29 \\
         &         $\pm$ 3 &       &      $\pm$ 0.13 &      $\pm$ 0.16 &      $\pm$ 0.03 &        $\pm$ 10 &        $\pm$ 30 &         $\pm$ 3 &    $\pm$ 1.7e-3 &       &      \\
    4865 &          66.311 &  19.2 &          -42.46 &           24.31 &           0.204 &           174.7 &         51076.6 &           21.67 &         9.26e-2 &  0.36 &   29 \\
         &     $\pm$ 0.012 &       &      $\pm$ 0.10 &      $\pm$ 0.12 &     $\pm$ 0.007 &       $\pm$ 1.4 &       $\pm$ 0.3 &      $\pm$ 0.11 &    $\pm$ 1.4e-3 &       &      \\
    4904 &          10.185 & 100.5 &           -42.7 &            11.2 &            0.37 &             253 &        50894.60 &            1.45 &         1.18e-3 &  1.26 &   28 \\
         &     $\pm$ 0.003 &       &       $\pm$ 0.3 &       $\pm$ 0.5 &      $\pm$ 0.04 &         $\pm$ 7 &      $\pm$ 0.16 &      $\pm$ 0.07 &    $\pm$ 1.6e-4 &       &      \\
    4965 &         14.9222 & 257.3 &          -43.11 &           17.38 &           0.010 &             220 &           52346 &            3.57 &          8.1e-3 &  1.02 &   43 \\
         &    $\pm$ 0.0005 &       &      $\pm$ 0.17 &      $\pm$ 0.20 &     $\pm$ 0.014 &        $\pm$ 70 &         $\pm$ 3 &      $\pm$ 0.04 &    $\pm$ 0.3e-3 &       &      \\
    4970 &          1002.6 &   2.7 &          -41.78 &            7.74 &           0.095 &             333 &           51256 &           106.3 &         4.76e-2 &  0.29 &   26 \\
         &       $\pm$ 2.4 &       &      $\pm$ 0.06 &      $\pm$ 0.09 &     $\pm$ 0.013 &         $\pm$ 6 &        $\pm$ 17 &       $\pm$ 1.2 &    $\pm$ 1.7e-3 &       &      \\
    4999 &          24.366 &  30.4 &           -41.6 &            14.6 &            0.32 &             248 &         50867.6 &            4.62 &          6.6e-3 &  1.10 &   24 \\
         &     $\pm$ 0.013 &       &       $\pm$ 0.4 &       $\pm$ 0.6 &      $\pm$ 0.04 &         $\pm$ 5 &       $\pm$ 0.4 &      $\pm$ 0.21 &    $\pm$ 0.9e-3 &       &      \\
    5040 &          33.455 & 104.3 &          -43.79 &            6.91 &            0.29 &             209 &         51356.4 &            3.05 &         1.01e-3 &  0.45 &   15 \\
         &     $\pm$ 0.003 &       &      $\pm$ 0.16 &      $\pm$ 0.24 &      $\pm$ 0.03 &         $\pm$ 6 &       $\pm$ 0.5 &      $\pm$ 0.11 &    $\pm$ 1.1e-4 &       &      \\
    5048 &          90.902 &  97.9 &           -42.3 &            12.2 &            0.18 &             331 &           49095 &            15.1 &         1.65e-2 &  2.08 &   46 \\
         &     $\pm$ 0.014 &       &       $\pm$ 0.3 &       $\pm$ 0.5 &      $\pm$ 0.04 &        $\pm$ 11 &         $\pm$ 3 &       $\pm$ 0.6 &    $\pm$ 2.0e-3 &       &      \\
    5052 &         3.84731 & 923.0 &          -42.71 &             8.8 &            0.05 &             260 &         51570.2 &           0.468 &          2.8e-4 &  1.01 &   29 \\
         &   $\pm$ 0.00007 &       &      $\pm$ 0.19 &       $\pm$ 0.3 &      $\pm$ 0.03 &        $\pm$ 50 &       $\pm$ 0.5 &     $\pm$ 0.015 &    $\pm$ 0.3e-4 &       &      \\
    5065 &            4010 &   1.0 &          -43.25 &             4.1 &            0.15 &             111 &           51000 &             224 &          2.8e-2 &  0.57 &   20 \\
         &        $\pm$ 70 &       &      $\pm$ 0.16 &       $\pm$ 0.3 &      $\pm$ 0.05 &        $\pm$ 23 &       $\pm$ 250 &        $\pm$ 15 &    $\pm$ 0.5e-2 &       &      \\
    5095 &          263.92 &  12.9 &          -42.54 &            20.4 &           0.673 &           230.8 &         51475.3 &            54.7 &          9.3e-2 &  0.59 &   32 \\
         &      $\pm$ 0.18 &       &      $\pm$ 0.13 &       $\pm$ 0.8 &     $\pm$ 0.019 &       $\pm$ 1.8 &       $\pm$ 0.6 &       $\pm$ 2.5 &    $\pm$ 1.1e-2 &       &      \\
    5242 &           339.5 &   3.4 &          -42.22 &            7.26 &            0.30 &             195 &           51144 &            32.3 &         1.17e-2 &  0.41 &   15 \\
         &       $\pm$ 1.3 &       &      $\pm$ 0.12 &      $\pm$ 0.20 &      $\pm$ 0.03 &         $\pm$ 5 &         $\pm$ 4 &       $\pm$ 0.9 &    $\pm$ 1.0e-3 &       &      \\
5268$^a$ &           420.6 &   7.7 &          -42.78 &             1.7 &            0.44 &             343 &           52950 &             8.9 &          1.6e-4 &  0.34 &   13 \\
         &       $\pm$ 1.8 &       &      $\pm$ 0.17 &       $\pm$ 0.4 &      $\pm$ 0.08 &        $\pm$ 21 &        $\pm$ 30 &       $\pm$ 2.1 &    $\pm$ 1.1e-4 &       &      \\
    5309 &            2670 &   1.5 &          -42.93 &             3.2 &            0.32 &             131 &           51650 &             112 &          7.7e-3 &  0.91 &   27 \\
         &        $\pm$ 90 &       &      $\pm$ 0.25 &       $\pm$ 0.3 &      $\pm$ 0.10 &        $\pm$ 18 &       $\pm$ 120 &        $\pm$ 10 &    $\pm$ 1.8e-3 &       &      \\
    5325 &            1772 &   1.9 &          -42.91 &             8.7 &            0.77 &           239.3 &           53588 &             126 &          3.1e-2 &  0.66 &   22 \\
         &        $\pm$ 12 &       &      $\pm$ 0.23 &       $\pm$ 0.8 &      $\pm$ 0.03 &       $\pm$ 2.3 &        $\pm$ 24 &        $\pm$ 15 &    $\pm$ 0.9e-2 &       &      \\
    5332 &           257.5 &  12.7 &          -42.31 &             4.3 &            0.45 &             232 &           52560 &            13.8 &          1.6e-3 &  0.52 &   23 \\
         &       $\pm$ 0.4 &       &      $\pm$ 0.14 &       $\pm$ 0.3 &      $\pm$ 0.05 &         $\pm$ 7 &         $\pm$ 3 &       $\pm$ 1.0 &    $\pm$ 0.3e-3 &       &      \\
    5350 &             690 &  10.2 &          -41.58 &            2.51 &            0.07 &              70 &           51210 &            23.8 &         1.12e-3 &  0.44 &   39 \\
         &         $\pm$ 3 &       &      $\pm$ 0.08 &      $\pm$ 0.13 &      $\pm$ 0.05 &        $\pm$ 40 &        $\pm$ 70 &       $\pm$ 1.3 &    $\pm$ 1.8e-4 &       &      \\
    5356 &            2827 &   1.4 &          -43.04 &            5.21 &           0.310 &             312 &           50930 &             192 &          3.6e-2 &  0.40 &   28 \\
         &        $\pm$ 16 &       &      $\pm$ 0.12 &      $\pm$ 0.12 &     $\pm$ 0.020 &         $\pm$ 6 &        $\pm$ 30 &         $\pm$ 5 &    $\pm$ 0.3e-2 &       &      \\
    5373 &             878 &   3.5 &          -43.40 &            4.60 &            0.14 &             180 &           51730 &            55.0 &          8.6e-3 &  0.46 &   19 \\
         &         $\pm$ 7 &       &      $\pm$ 0.11 &      $\pm$ 0.16 &      $\pm$ 0.03 &        $\pm$ 16 &        $\pm$ 40 &       $\pm$ 1.9 &    $\pm$ 0.9e-3 &       &      \\
    5379 &          120.21 &  35.1 &          -42.80 &            8.10 &            0.24 &             206 &         52715.7 &            13.0 &          6.1e-3 &  0.41 &   15 \\
         &      $\pm$ 0.04 &       &      $\pm$ 0.14 &      $\pm$ 0.17 &      $\pm$ 0.03 &         $\pm$ 6 &       $\pm$ 1.7 &       $\pm$ 0.3 &    $\pm$ 0.4e-3 &       &      \\
    5381 &            98.7 &  35.3 &           -42.0 &             5.1 &            0.31 &             249 &           51008 &             6.6 &         1.19e-3 &  0.94 &   20 \\
         &       $\pm$ 0.3 &       &       $\pm$ 0.3 &       $\pm$ 0.3 &      $\pm$ 0.06 &        $\pm$ 13 &         $\pm$ 3 &       $\pm$ 0.4 &    $\pm$ 2.3e-4 &       &      \\
    5434 &            1277 &   2.4 &          -41.14 &             7.8 &           0.551 &             192 &           52449 &             114 &          3.6e-2 &  0.43 &   27 \\
         &         $\pm$ 9 &       &      $\pm$ 0.11 &       $\pm$ 0.3 &     $\pm$ 0.018 &         $\pm$ 3 &        $\pm$ 12 &         $\pm$ 4 &    $\pm$ 0.4e-2 &       &      \\
    5438 &            4320 &   2.4 &           -42.2 &             9.5 &            0.86 &             196 &           49560 &             290 &          5.0e-2 &  1.44 &   23 \\
         &        $\pm$ 90 &       &       $\pm$ 0.5 &       $\pm$ 1.6 &      $\pm$ 0.07 &         $\pm$ 9 &        $\pm$ 90 &        $\pm$ 80 &    $\pm$ 3.0e-2 &       &      \\
    5463 &          9.4648 & 129.4 &          -41.81 &            6.24 &            0.05 &             320 &         50976.5 &            0.81 &         2.38e-4 &  0.59 &   28 \\
         &    $\pm$ 0.0010 &       &      $\pm$ 0.15 &      $\pm$ 0.22 &      $\pm$ 0.03 &        $\pm$ 30 &       $\pm$ 0.9 &      $\pm$ 0.03 &    $\pm$ 2.5e-5 &       &      \\
    5467 &           224.1 &  11.0 &          -42.48 &            6.27 &            0.15 &              25 &           51153 &            19.1 &          5.5e-3 &  0.57 &   20 \\
         &       $\pm$ 0.3 &       &      $\pm$ 0.14 &      $\pm$ 0.19 &      $\pm$ 0.03 &        $\pm$ 10 &         $\pm$ 6 &       $\pm$ 0.6 &    $\pm$ 0.5e-3 &       &      \\
    5599 &          59.837 &  32.5 &          -42.01 &           19.38 &           0.633 &           175.4 &        51879.10 &           12.34 &         2.09e-2 &  0.39 &   28 \\
         &     $\pm$ 0.006 &       &      $\pm$ 0.10 &      $\pm$ 0.22 &     $\pm$ 0.005 &       $\pm$ 0.6 &      $\pm$ 0.09 &      $\pm$ 0.16 &    $\pm$ 0.8e-3 &       &      \\
    5601 &         10.0142 & 132.6 &          -41.19 &            5.67 &            0.09 &              58 &         51162.9 &           0.779 &         1.87e-4 &  0.57 &   47 \\
         &    $\pm$ 0.0012 &       &      $\pm$ 0.10 &      $\pm$ 0.10 &      $\pm$ 0.03 &        $\pm$ 16 &       $\pm$ 0.4 &     $\pm$ 0.020 &    $\pm$ 1.4e-5 &       &      \\
    5647 &         15.5528 &  57.3 &           -43.5 &            45.5 &           0.285 &              55 &        54123.82 &            9.32 &         1.33e-1 &  1.17 &   12 \\
         &    $\pm$ 0.0011 &       &       $\pm$ 0.6 &       $\pm$ 0.7 &     $\pm$ 0.011 &         $\pm$ 3 &      $\pm$ 0.10 &      $\pm$ 0.15 &    $\pm$ 0.6e-2 &       &      \\
    5666 &          211.22 &  13.3 &           -41.5 &            12.1 &           0.712 &             224 &         53321.7 &            24.7 &         1.34e-2 &  0.32 &   14 \\
         &      $\pm$ 0.18 &       &       $\pm$ 0.3 &       $\pm$ 0.6 &     $\pm$ 0.019 &         $\pm$ 5 &       $\pm$ 0.7 &       $\pm$ 1.5 &    $\pm$ 2.2e-3 &       &      \\
    5671 &            1423 &   2.1 &          -41.49 &            5.74 &           0.286 &             123 &           51044 &           107.6 &         2.45e-2 &  0.35 &   30 \\
         &         $\pm$ 7 &       &      $\pm$ 0.08 &      $\pm$ 0.10 &     $\pm$ 0.018 &         $\pm$ 4 &        $\pm$ 12 &       $\pm$ 2.0 &    $\pm$ 1.3e-3 &       &      \\
    5700 &            2800 &   3.7 &          -41.61 &            4.68 &            0.15 &             345 &           52420 &             178 &         2.88e-2 &  0.44 &   38 \\
         &        $\pm$ 22 &       &      $\pm$ 0.11 &      $\pm$ 0.13 &      $\pm$ 0.03 &        $\pm$ 10 &        $\pm$ 80 &         $\pm$ 5 &    $\pm$ 2.5e-3 &       &      \\
    5738 &         13.0492 & 272.3 &          -42.33 &            5.92 &            0.02 &             200 &           51403 &            1.06 &          2.8e-4 &  0.74 &   29 \\
         &    $\pm$ 0.0008 &       &      $\pm$ 0.15 &      $\pm$ 0.20 &      $\pm$ 0.04 &        $\pm$ 80 &         $\pm$ 3 &      $\pm$ 0.04 &    $\pm$ 0.3e-4 &       &      \\
    5797 &         14.9906 &  98.8 &          -42.67 &           19.53 &           0.023 &             260 &         52033.9 &            4.02 &         1.16e-2 &  0.69 &   23 \\
         &    $\pm$ 0.0009 &       &      $\pm$ 0.15 &      $\pm$ 0.23 &     $\pm$ 0.011 &        $\pm$ 30 &       $\pm$ 1.1 &      $\pm$ 0.05 &    $\pm$ 0.4e-3 &       &      \\
    5855 &          211.24 &  47.0 &          -42.35 &            2.32 &            0.29 &             157 &           50203 &             6.5 &          2.4e-4 &  0.59 &   41 \\
         &      $\pm$ 0.19 &       &      $\pm$ 0.10 &      $\pm$ 0.18 &      $\pm$ 0.07 &        $\pm$ 12 &         $\pm$ 6 &       $\pm$ 0.5 &    $\pm$ 0.6e-4 &       &      \\
    5887 &         106.034 &  57.5 &           -41.9 &            19.8 &           0.171 &             103 &         55583.1 &            28.5 &          8.2e-2 &  1.09 &   24 \\
         &     $\pm$ 0.018 &       &       $\pm$ 0.3 &       $\pm$ 0.5 &     $\pm$ 0.018 &         $\pm$ 6 &       $\pm$ 1.8 &       $\pm$ 0.7 &    $\pm$ 0.6e-2 &       &      \\
    6171 &          8.8276 & 125.4 &           -41.9 &            35.9 &           0.025 &              33 &         54189.8 &            4.36 &         4.24e-2 &  0.97 &   14 \\
         &    $\pm$ 0.0013 &       &       $\pm$ 0.3 &       $\pm$ 0.4 &     $\pm$ 0.011 &        $\pm$ 24 &       $\pm$ 0.6 &      $\pm$ 0.05 &    $\pm$ 1.3e-3 &       &      \\
    6292 &         14.5988 &  70.1 &          -43.31 &           30.99 &           0.012 &              20 &         50859.9 &            6.22 &         4.50e-2 &  0.62 &   21 \\
         &    $\pm$ 0.0013 &       &      $\pm$ 0.14 &      $\pm$ 0.19 &     $\pm$ 0.007 &        $\pm$ 30 &       $\pm$ 1.3 &      $\pm$ 0.04 &    $\pm$ 0.8e-3 &       &      \\
6325$^a$ &           259.3 &  11.8 &           -42.1 &            13.3 &            0.36 &              25 &           52995 &            44.3 &          5.1e-2 &  0.98 &   12 \\
         &       $\pm$ 0.4 &       &       $\pm$ 0.4 &       $\pm$ 0.5 &      $\pm$ 0.05 &         $\pm$ 6 &         $\pm$ 6 &       $\pm$ 1.8 &    $\pm$ 0.6e-2 &       &      \\
    6586 &           239.1 &   5.1 &          -43.24 &           15.58 &           0.422 &           181.9 &         50562.2 &            46.5 &          7.0e-2 &  0.57 &   29 \\
         &       $\pm$ 0.4 &       &      $\pm$ 0.12 &      $\pm$ 0.22 &     $\pm$ 0.013 &       $\pm$ 1.5 &       $\pm$ 1.0 &       $\pm$ 0.7 &    $\pm$ 0.3e-2 &       &      \\
    8406 &            3080 &   1.3 &          -42.02 &            2.24 &            0.38 &              79 &           53210 &              88 &          2.8e-3 &  0.56 &   34 \\
         &        $\pm$ 80 &       &      $\pm$ 0.15 &      $\pm$ 0.24 &      $\pm$ 0.06 &        $\pm$ 16 &       $\pm$ 120 &        $\pm$ 10 &    $\pm$ 0.9e-3 &       &      \\
\enddata
\scriptsize
\begin{flushleft}
$^a$ We caution the reader that, though these orbital solutions appear to be robust, these binaries have poor phase coverage.
\end{flushleft}
\normalsize
\end{deluxetable}

\begin{deluxetable}{l c c c c c r c r}
\tablewidth{0pt}
\tablecaption{Physical Properties of NGC 188 Single-Lined Binaries \label{SB1masstab}}
\tablehead{\colhead{ID} & \colhead{$V$} & \colhead{$(\bv)$} & \colhead{R} & \colhead{P$_{RV}$} & \colhead{P$_{PM}$} & \colhead{M$_1$} & \colhead{M$_2$ min} & \colhead{M$_2$} \\
\colhead{} & \colhead{} & \colhead{} & \colhead{(arcmin)} & \colhead{(\%)} & \colhead{(\%)} & \colhead{(\Msolar)} & \colhead{(\Msolar)} & \colhead{(\Msolar)}}
\startdata
~880 &  15.383 & 0.684 &  16.67 & 97 & 98 &                 1.04 &                 0.23 &                 0.69 \\
3171 &  15.077 & 0.911 &  15.79 & 98 & 98 &             $<$ 1.12 &                 0.24 &             $<$ 0.80 \\
3732 &  14.935 & 0.687 &  11.89 & 98 & 96 &                 1.08 &                 0.15 &                 0.75 \\
3871 &  15.300 & 0.648 &  12.78 & 97 & 89 &                 1.05 &                 0.16 &                 0.45 \\
4080 &  15.524 & 0.674 &   7.76 & 98 & 98 &                 1.03 &                 0.39 &                 0.60 \\
4147 &  15.161 & 0.649 &  13.56 & 98 & 98 &             $<$ 1.07 &                 0.21 &             $<$ 0.79 \\
4289 &  15.303 & 0.937 &   5.97 & 98 & 98 &             $<$ 1.12 &                 0.58 &             $<$ 0.78 \\
4333 &  15.857 & 0.676 &   6.11 & 98 & 98 &             $<$ 1.00 &                 0.23 &             $<$ 0.72 \\
4369 &  15.549 & 0.660 &   6.61 & 98 & 98 &             $<$ 1.03 &                 0.25 &             $<$ 0.75 \\
4386 &  15.442 & 0.660 &   7.03 & 98 & 63 &                 1.04 &                 0.35 &                 0.55 \\
4390 &  15.662 & 0.741 &   6.52 & 98 & 97 &                 0.99 &                 0.31 &                 0.80 \\
4392 &  15.191 & 0.647 &   6.01 & 98 & 97 &                 1.04 &                 0.44 &                 0.55 \\
4524 &  12.434 & 1.165 &   5.44 & 98 & 97 &                 1.14 &                 0.34 &                 1.14 \\
4560 &  15.061 & 0.667 &   3.78 & 97 & 98 &                 1.08 &                 0.16 &                 0.33 \\
4565 &  12.416 & 1.273 &   3.72 & 98 & 96 &             $<$ 1.14 &                 0.27 &             $<$ 1.10 \\
4585 &  15.814 & 0.666 &   4.41 & 98 & 98 &             $<$ 1.00 &                 0.40 &             $<$ 0.73 \\
4595 &  15.055 & 0.660 &   2.56 & 98 & 97 &             $<$ 1.08 &                 0.40 &             $<$ 0.80 \\
4618 &  15.772 & 0.712 &   3.10 & 98 & 98 &                 1.00 &                 0.75 &                 0.75 \\
4673 &  14.880 & 0.716 &   2.06 & 95 & 98 &                 1.11 &                 0.16 &                 0.64 \\
4688 &  15.079 & 0.636 &   3.33 & 98 & 98 &             $<$ 1.07 &                 0.45 &             $<$ 0.80 \\
4710 &  14.846 & 0.667 &   2.89 &  1 & 96 &                 1.09 &                 0.19 &                 0.60 \\
4726 &  15.302 & 0.671 &   4.46 & 98 & 98 &             $<$ 1.05 &                 0.22 &             $<$ 0.78 \\
4734 &  14.950 & 0.997 &   4.74 & 96 & 98 &                 1.13 &                 0.55 &                 0.64 \\
4843 &  11.541 & 1.302 &   8.77 & 98 & 75 &                 1.14 &                 0.34 &                 1.14 \\
4865 &  14.930 & 0.783 &   7.64 & 98 & 95 &             $<$ 1.12 &                 0.66 &             $<$ 0.81 \\
4904 &  15.930 & 0.733 &   5.48 & 98 & 96 &                 0.97 &                 0.11 &                 0.60 \\
4965 &  15.282 & 0.693 &   2.36 & 96 & 85 &                 1.03 &                 0.24 &                 0.71 \\
4999 &  15.792 & 0.664 &   0.99 & 97 & 97 &             $<$ 1.00 &                 0.22 &             $<$ 0.73 \\
5040 &  15.009 & 0.667 &   0.68 & 92 & 97 &                 1.09 &                 0.11 &                 0.49 \\
5048 &  13.869 & 1.069 &   0.84 & 98 & 94 &                 1.14 &                 0.33 &                 0.74 \\
5052 &  15.916 & 0.710 &   0.33 & 98 & 98 &                 0.99 &                 0.07 &                 0.62 \\
5065 &  15.754 & 0.746 &   1.55 & 97 & 95 &                 0.98 &                 0.37 &                 0.78 \\
5095 &  14.992 & 0.939 &   2.07 & 98 & 98 &                 1.12 &                 0.67 &                 0.80 \\
5242 &  14.880 & 0.697 &   9.36 & 98 & 98 &                 1.09 &                 0.28 &                 0.77 \\
5268 &  15.317 & 0.663 &   6.32 & 98 & 91 &                 1.05 &                 0.06 &                 0.58 \\
5309 &  14.965 & 0.706 &   4.39 & 98 & 98 &                 1.10 &                 0.24 &                 0.72 \\
5332 &  15.047 & 0.672 &   2.74 & 98 & 98 &                 1.08 &                 0.13 &                 0.46 \\
5356 &  14.272 & 0.999 &   2.94 & 98 & 98 &                 1.13 &                 0.44 &                 0.88 \\
5373 &  14.380 & 1.050 &   3.59 & 96 & 95 &             $<$ 1.14 &                 0.26 &             $<$ 0.87 \\
5381 &  15.022 & 0.676 &   4.94 & 98 & 89 &                 1.09 &                 0.12 &                 0.39 \\
5438 &  13.634 & 1.112 &   3.21 & 97 & 98 &             $<$ 1.14 &                 0.53 &             $<$ 0.96 \\
5463 &  14.978 & 0.664 &   4.42 & 98 & 98 &                 1.08 &                 0.07 &                 0.40 \\
5467 &  15.738 & 0.640 &   5.51 & 98 & 98 &             $<$ 1.02 &                 0.20 &             $<$ 0.74 \\
5599 &  15.101 & 0.946 &   8.48 & 98 & 98 &             $<$ 1.13 &                 0.36 &             $<$ 0.80 \\
5601 &  15.597 & 0.682 &   9.44 & 94 & 97 &                 1.03 &                 0.06 &                 0.67 \\
5647 &  16.000 & 0.743 &   7.37 & 95 & 95 &                 0.97 &                 0.73 &                 0.73 \\
5666 &  15.216 & 0.669 &   6.02 & 96 & 94 &                 1.06 &                 0.29 &                 0.51 \\
5700 &  14.448 & 1.010 &   5.59 & 97 & 98 &                 1.13 &                 0.41 &                 0.89 \\
5738 &  15.315 & 0.675 &   6.77 & 98 & 96 &                 1.05 &                 0.07 &                 0.44 \\
5797 &  15.853 & 0.691 &   9.13 & 98 & 95 &                 1.00 &                 0.26 &                 0.60 \\
5855 &  13.356 & 1.155 &   5.62 & 98 & 98 &             $<$ 1.14 &                 0.07 &             $<$ 0.99 \\
5887 &  12.126 & 1.300 &   3.42 & 98 &  1 &             $<$ 1.14 &                 0.64 &             $<$ 1.13 \\
6171 &  16.309 & 0.743 &   9.71 & 98 & 98 &             $<$ 0.94 &                 0.43 &             $<$ 0.68 \\
6292 &  15.524 & 0.691 &  12.70 & 97 & 98 &                 1.02 &                 0.46 &                 0.65 \\
6325 &  15.678 & 0.722 &  11.25 & 97 &  2 &                 1.00 &                 0.48 &                 0.72 \\
6586 &  14.616 & 1.037 &  19.21 & 97 & 95 &             $<$ 1.13 &                 0.59 &             $<$ 0.85 \\
8406 &  14.791 & 1.002 &  10.67 & 97 & 96 &                 1.13 &                 0.17 &                 0.73 \\
\enddata
\end{deluxetable}

\clearpage

\subsection{Double-Lined Orbital Solutions} \label{SB2}

The RV measurements for the primary and secondary stars of a given SB2 binary are found using a TwO Dimensional CORelation (TODCOR) 
technique formulated by \citet{zuk94}.  TODCOR uses two template spectra to derive the two RVs of an SB2 binary simultaneously, greatly 
increasing our ability to recover reliable RVs even for those observations that appear highly blended in a one-dimensional 
cross-correlation function.  As all of our detected SB2 binaries have mass ratios $\gtrsim$0.7, we choose to use the same
solar template that we use to derive RVs for all single stars and SB1 binaries as both template spectra in TODCOR.  
Our procedure in deriving the orbital solutions is to first solve for 
the orbit of the primary in the manner discussed in Section~\ref{SB1} and then use the derived orbital elements to solve for 
the full SB2 orbit (including the RVs of the secondary star).
We provide the plotted orbital solutions in Figure 2; the plots are of the same format as for the SB1 binaries, 
except here, the primary RVs are plotted using filled circles while secondary RVs are 
plotted with open circles.  Additionally, we present the tabulated orbital elements in Table~\ref{SB2tab}, in similar 
format to Table~\ref{SB1tab}, except here, in place of the mass function, we provide the quantity $m$ $\sin^3$ $i$ and the 
mass ratio ($q$).

We also include Table~\ref{SB2masstab} that contains similar information on the SB2 binaries as we provide 
in Table~\ref{SB1masstab} for the SB1 binaries.  Here we do not quote a lower limit on the secondary mass as 
the mass ratio can be calculated directly from the orbital solution.  We use the same photometric deconvolution procedure as 
for the SB1 binaries to derive the photometric mass estimates, except, here, we keep the mass ratio fixed.  
For the red-giant binary 3118, we cannot use this technique, as the system is observed to lie redward of the giant branch.  
Therefore, we use the Padova isochrone to formulate a mass-luminosity relation of $L \propto M^{11}$, valid for this region
on the NGC 188 giant branch, to derive the appropriate correction to the observed $V$ magnitude, from which we can estimate 
the primary mass. (Specifically, we observe a mass ratio for 3118 of $q$ = 0.795, which implies a correction to the 
observed $V$ magnitude of $V_1 = V + 0.08$, and we use this $V_1$ to estimate the mass of the primary.)  Given this 
primary mass estimate and the mass ratio, we can easily derive the secondary mass.

Again, we utilize a Monte Carlo technique to estimate the uncertainties on our mass estimates in a similar manner to 
Section~\ref{SB1}. The mean uncertainty on the primary mass estimates is similar to that of the SB1 binaries.  We can then use the mass ratio, 
primary mass and their respective uncertainties to derive a mean uncertainty on the secondary-mass estimates of 
0.09 \Msolar, with a standard deviation about this mean of 0.02 \Msolar.
Additionally, we utilize our SB2 binaries to check the accuracy of this photometric deconvolution technique by first estimating masses with the 
mass ratio fixed and then estimating masses for the same binaries without fixing the mass ratio (essentially, treating the systems as 
SB1 binaries and using the technique described in Section~\ref{SB1}).  For the primary mass, we find a mean difference between these two 
techniques of 0.01 \Msolar, and for the secondary mass estimates, we find a mean difference of 0.03 \Msolar.  The standard deviations about 
these means are 0.02 \Msolar~and 0.06 \Msolar, respectively.  These values lie within our estimated uncertainties, and demonstrate the 
robustness of the mass estimates for both SB1 and SB2 binaries derived using our photometric deconvolution technique.

\begin{center}
\begin{longtable}{ccc}
\epsfig{file=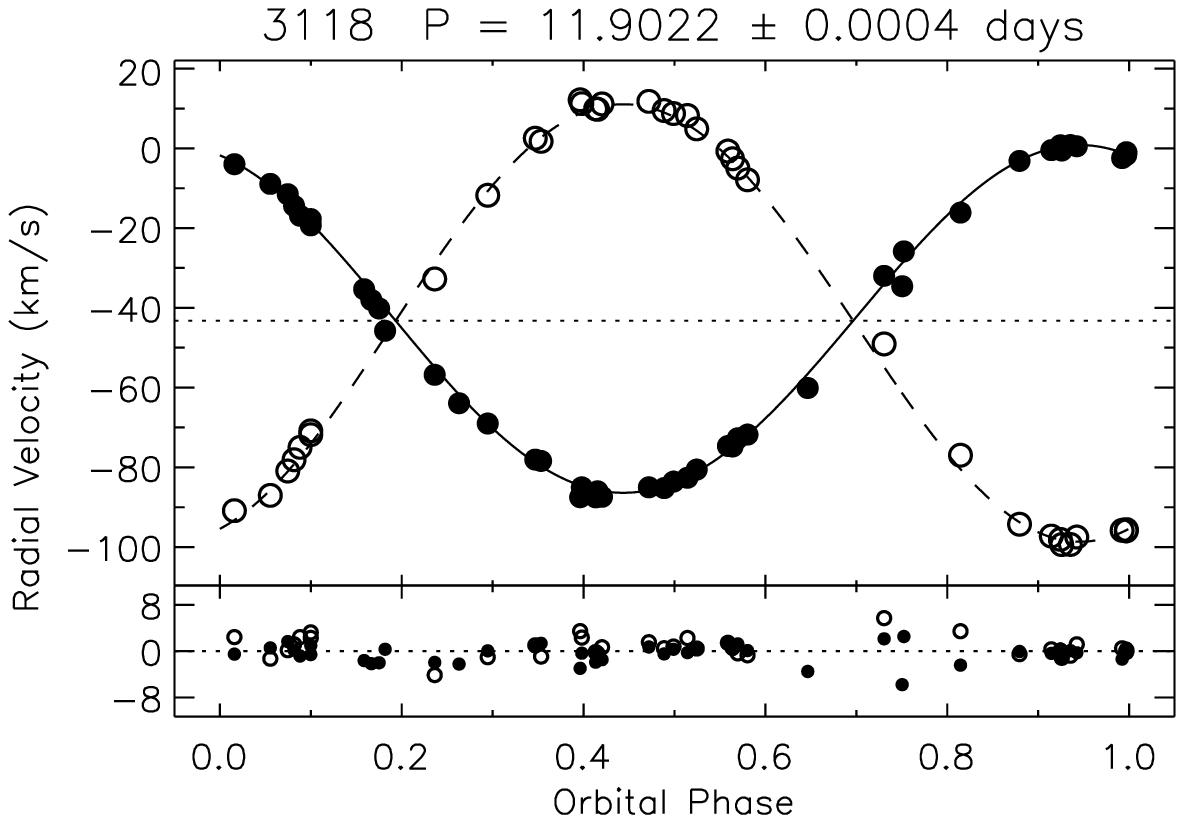,width=0.3\linewidth} & \epsfig{file=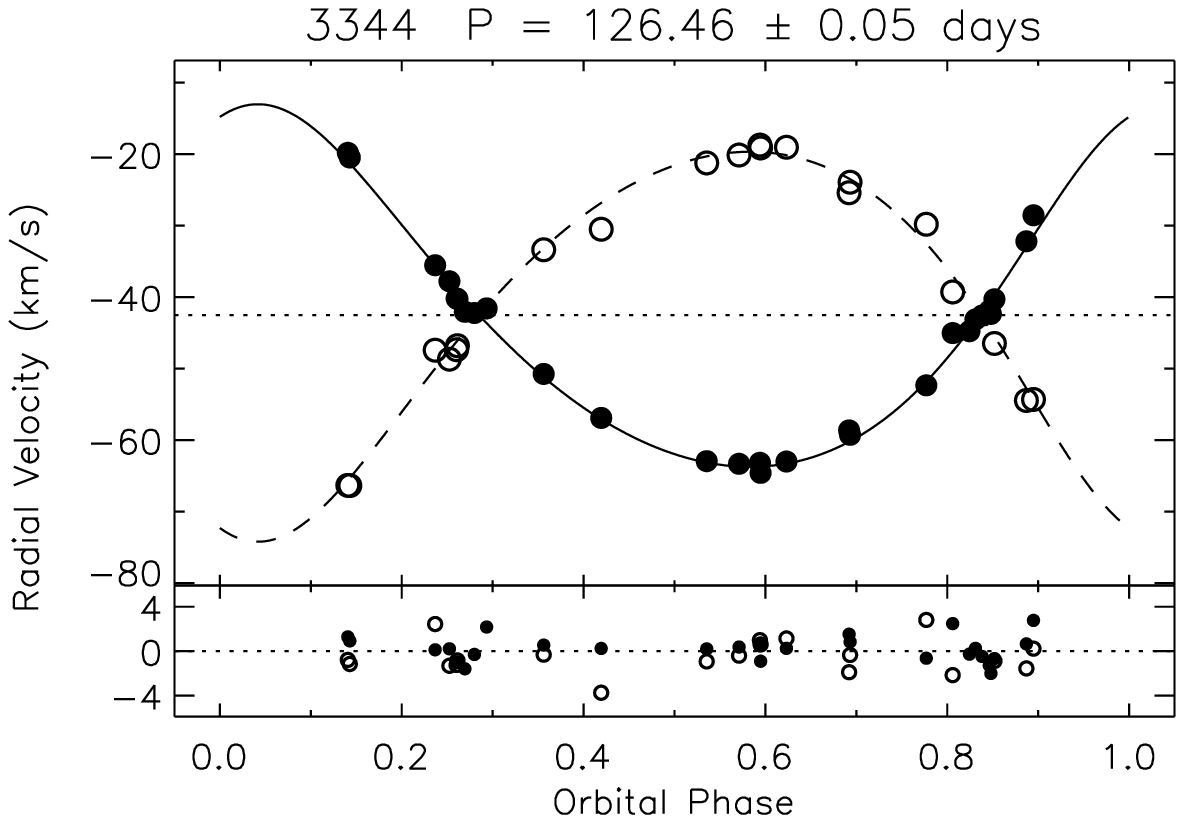,width=0.3\linewidth} & \epsfig{file=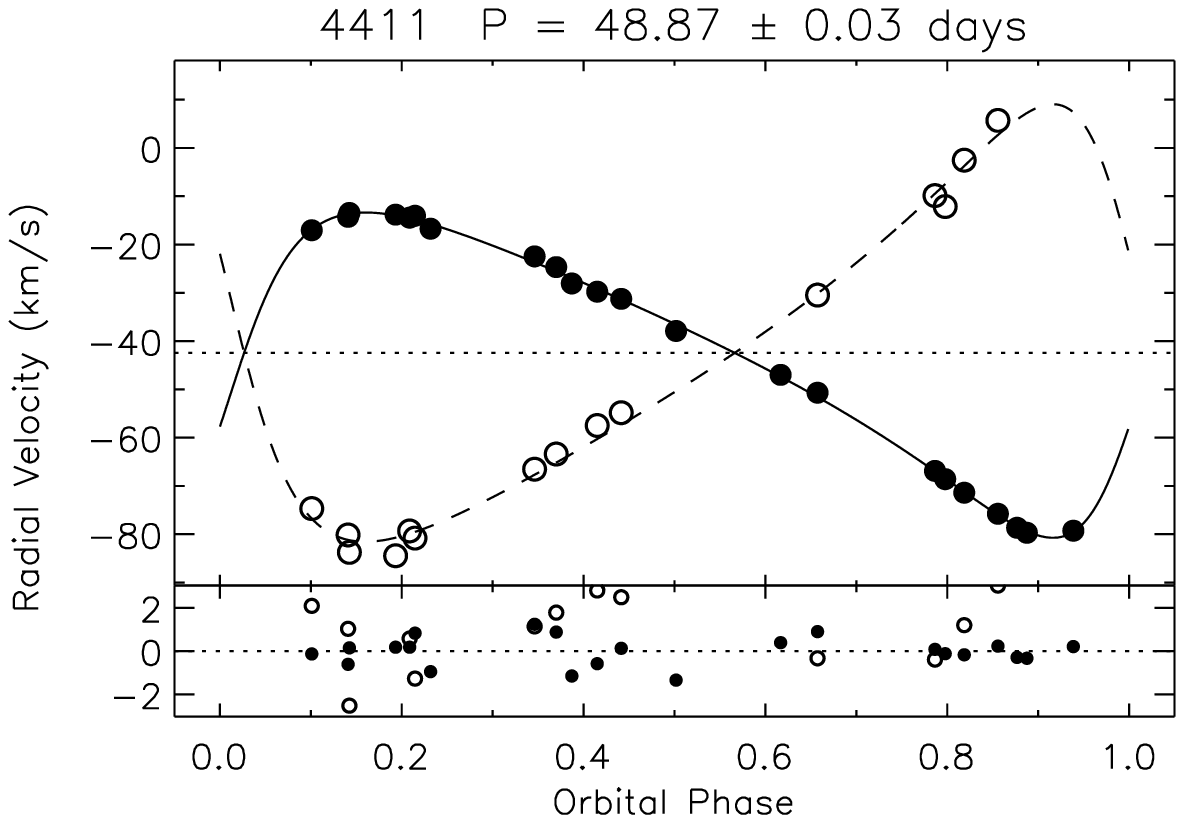,width=0.3\linewidth} \\
\epsfig{file=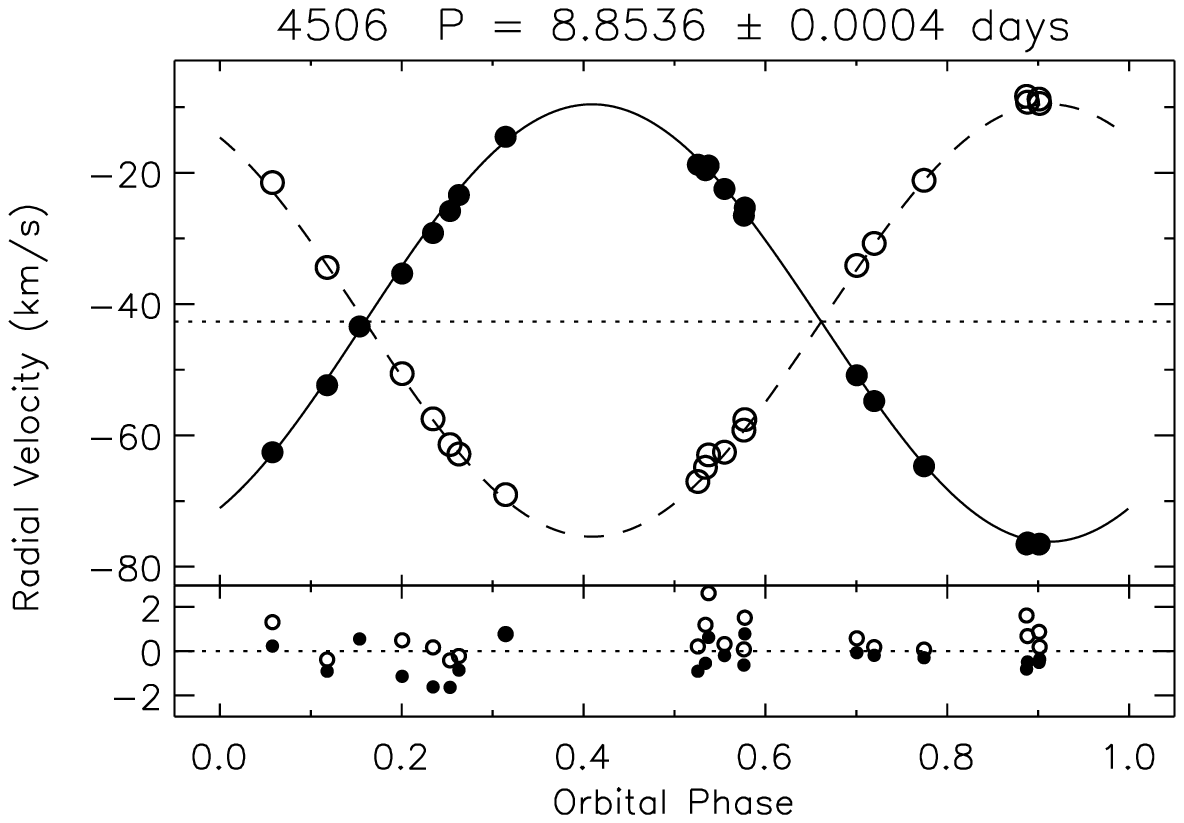,width=0.3\linewidth} & \epsfig{file=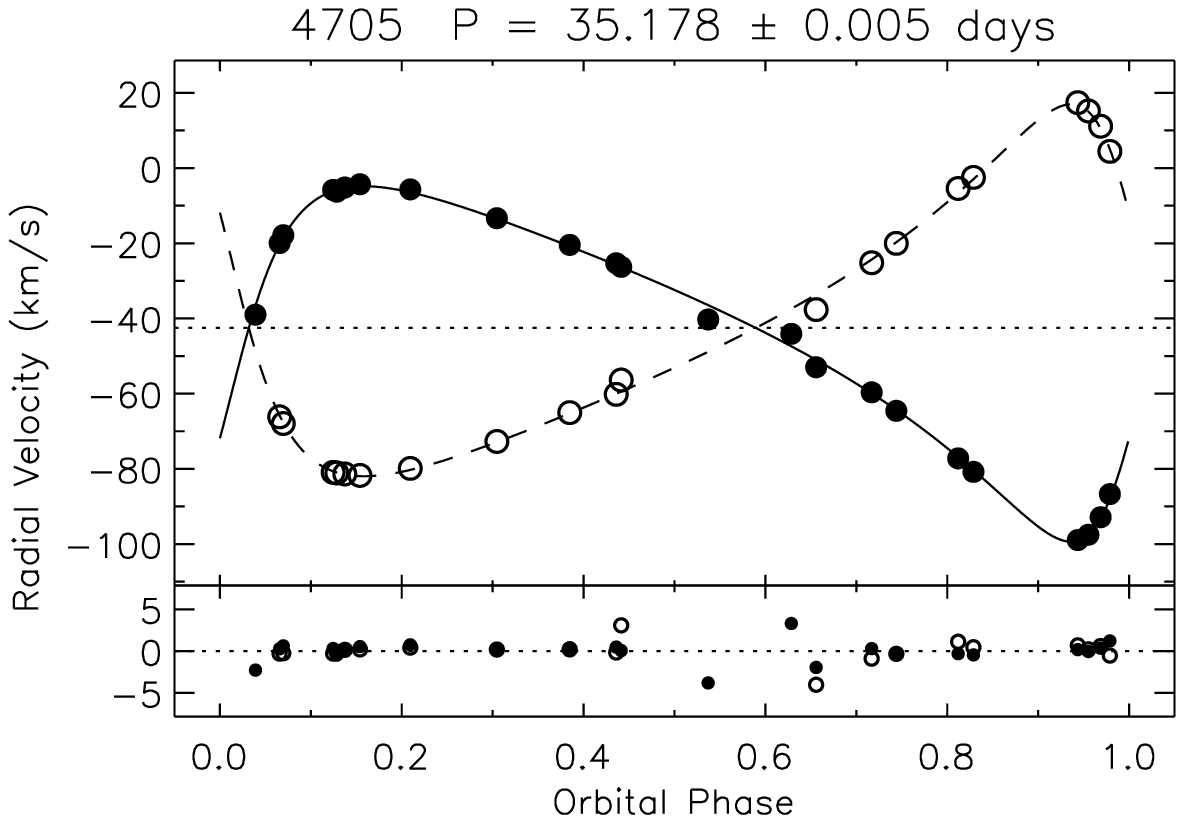,width=0.3\linewidth} & \epsfig{file=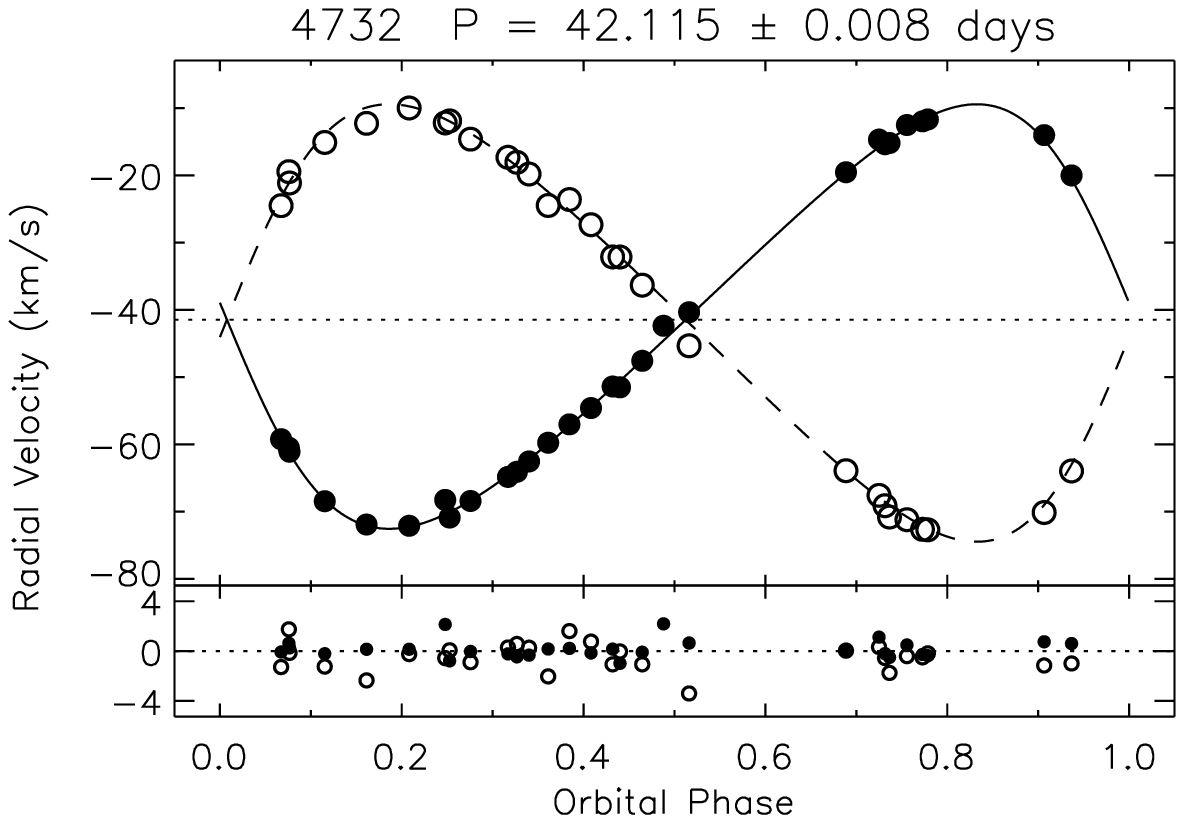,width=0.3\linewidth} \\
\epsfig{file=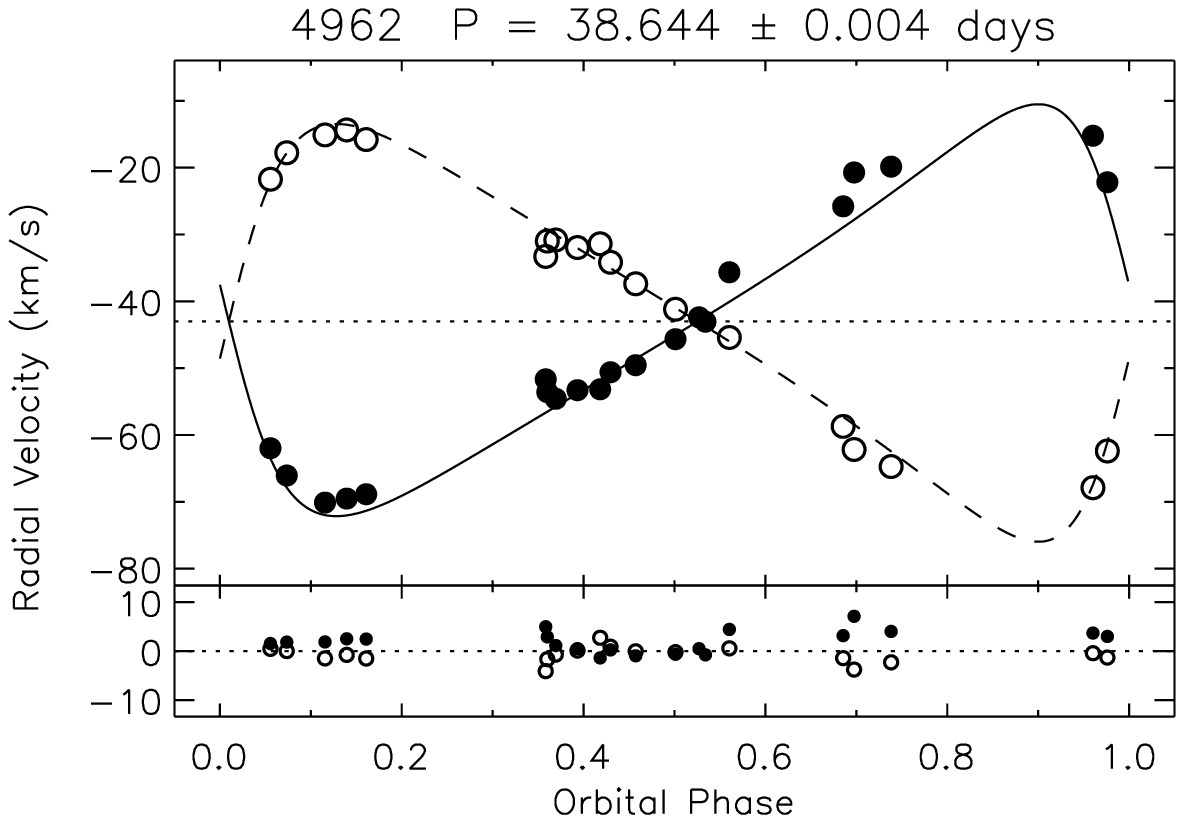,width=0.3\linewidth} & \epsfig{file=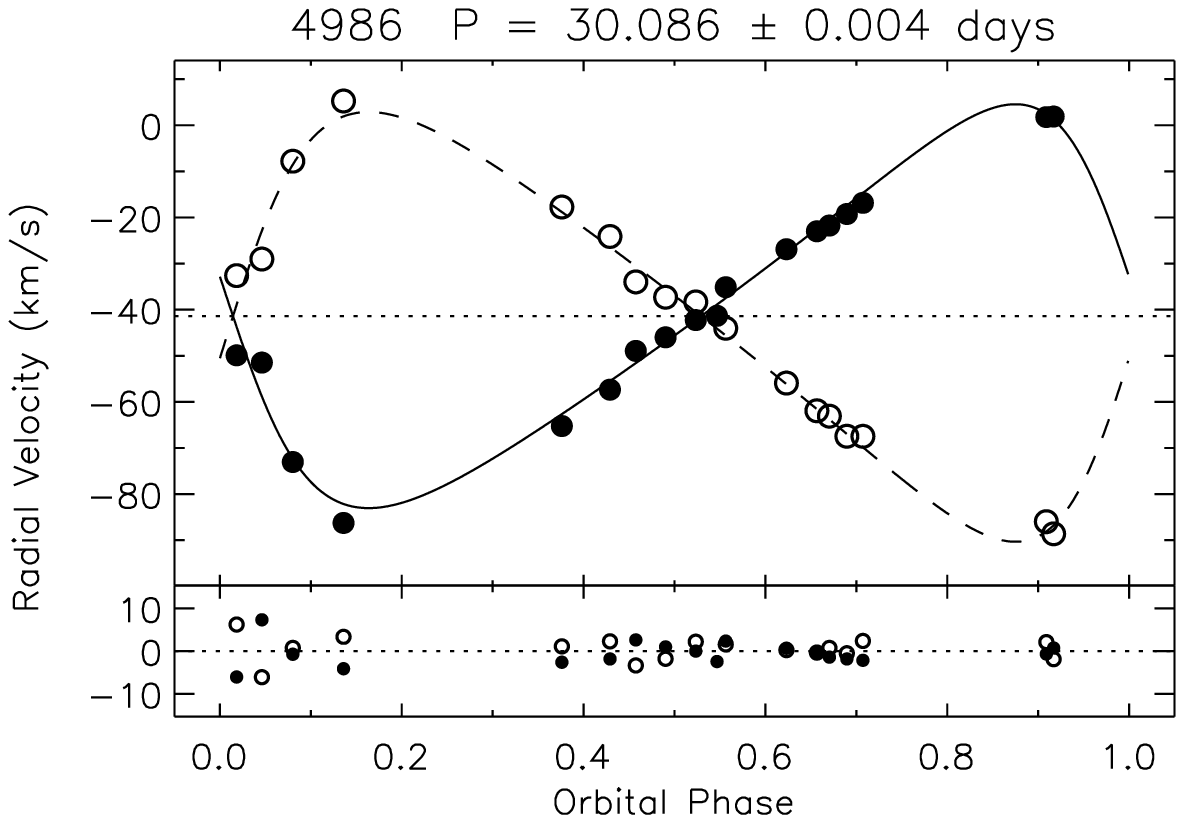,width=0.3\linewidth} & \epsfig{file=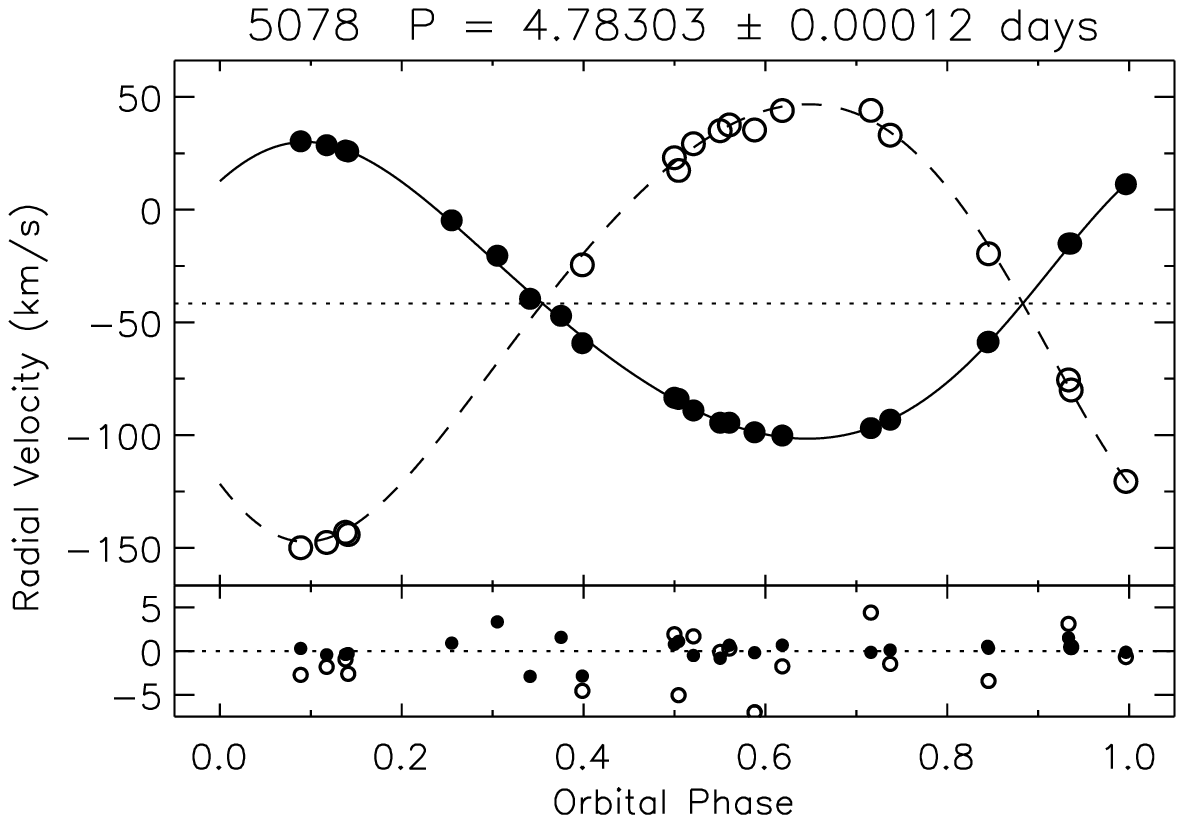,width=0.3\linewidth} \\
\epsfig{file=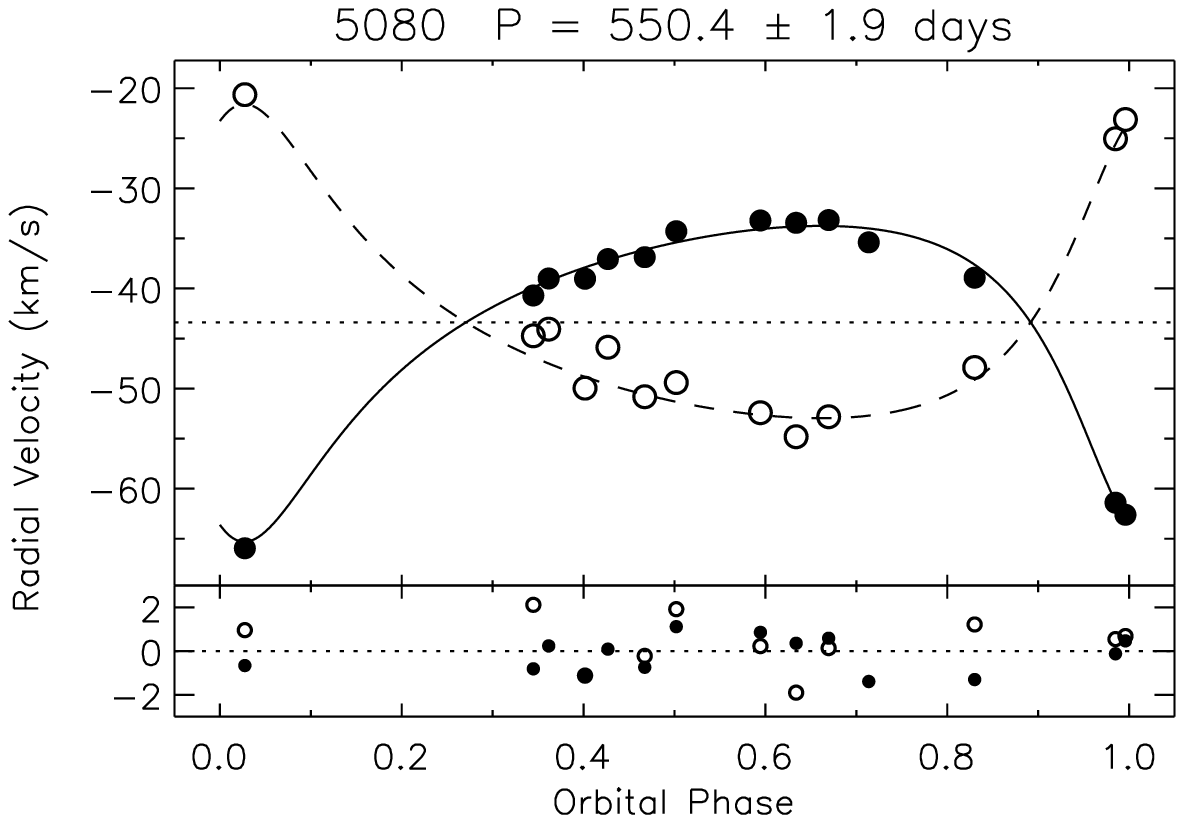,width=0.3\linewidth} & \epsfig{file=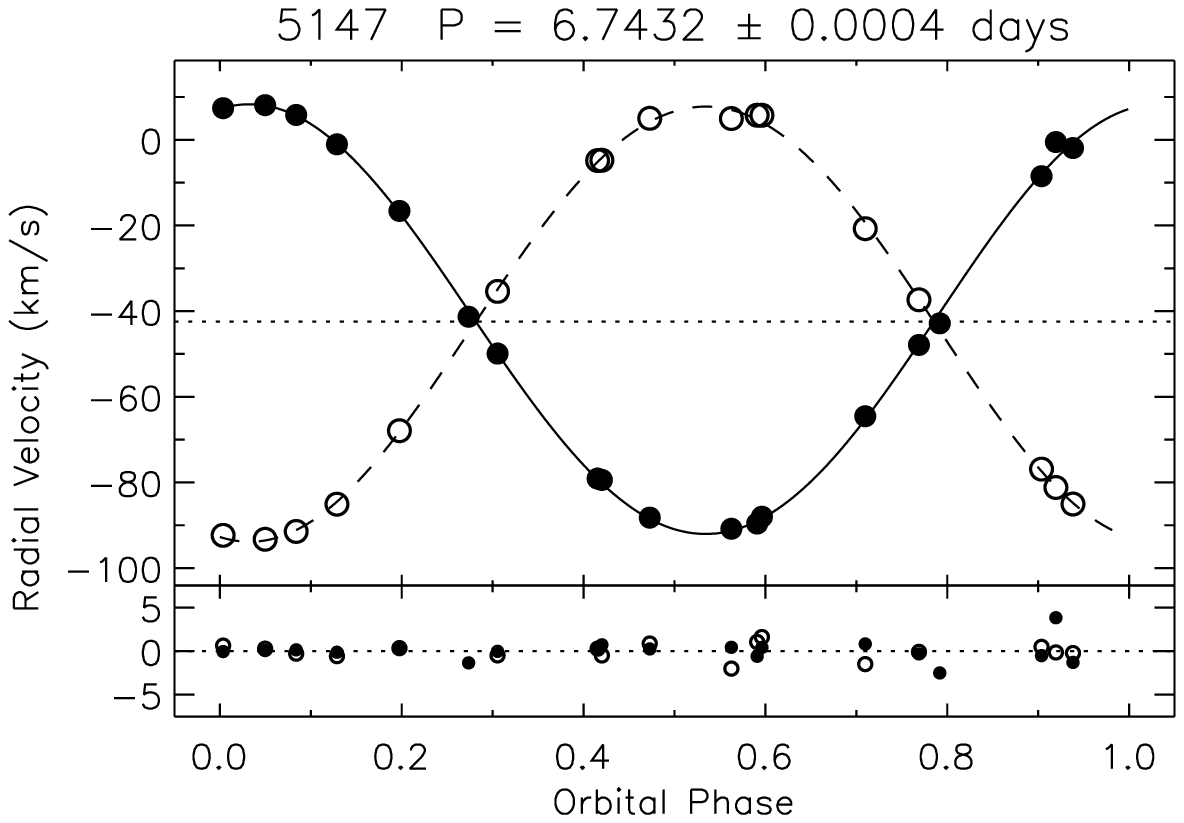,width=0.3\linewidth} & \epsfig{file=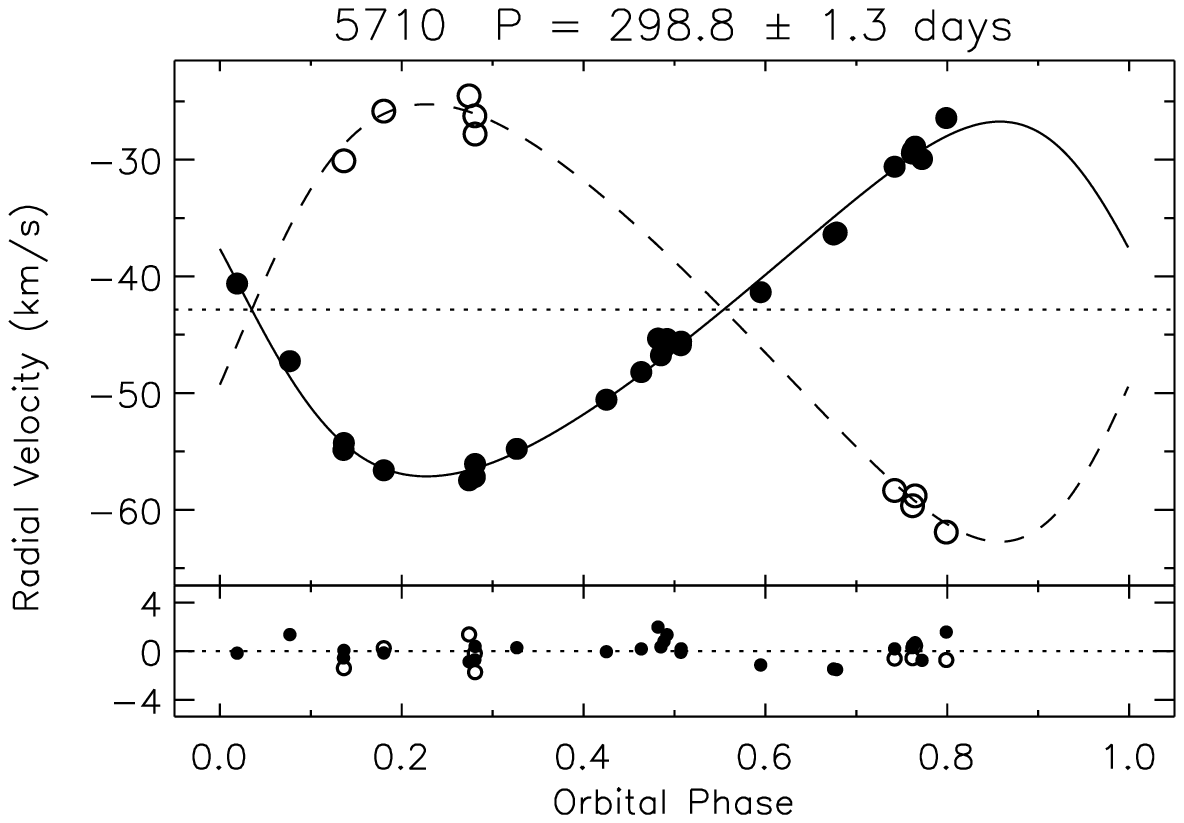,width=0.3\linewidth} \\
\epsfig{file=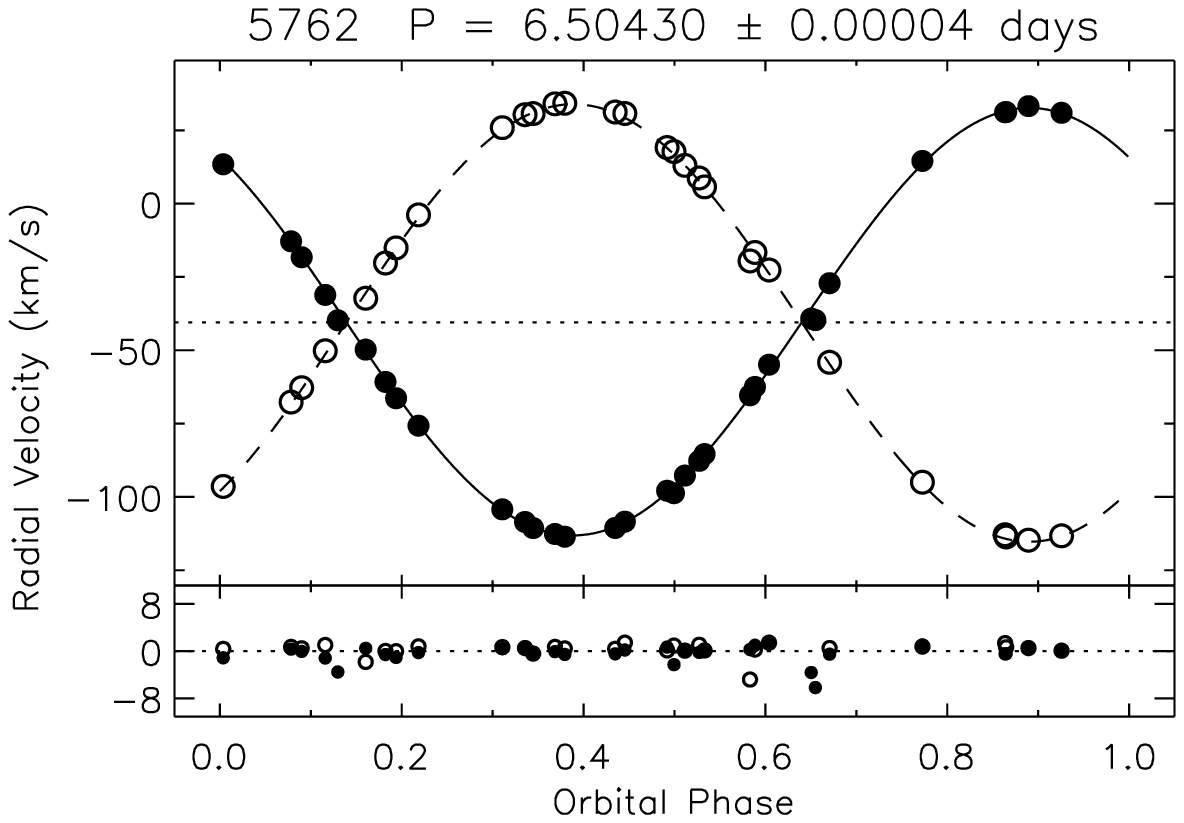,width=0.3\linewidth} & \epsfig{file=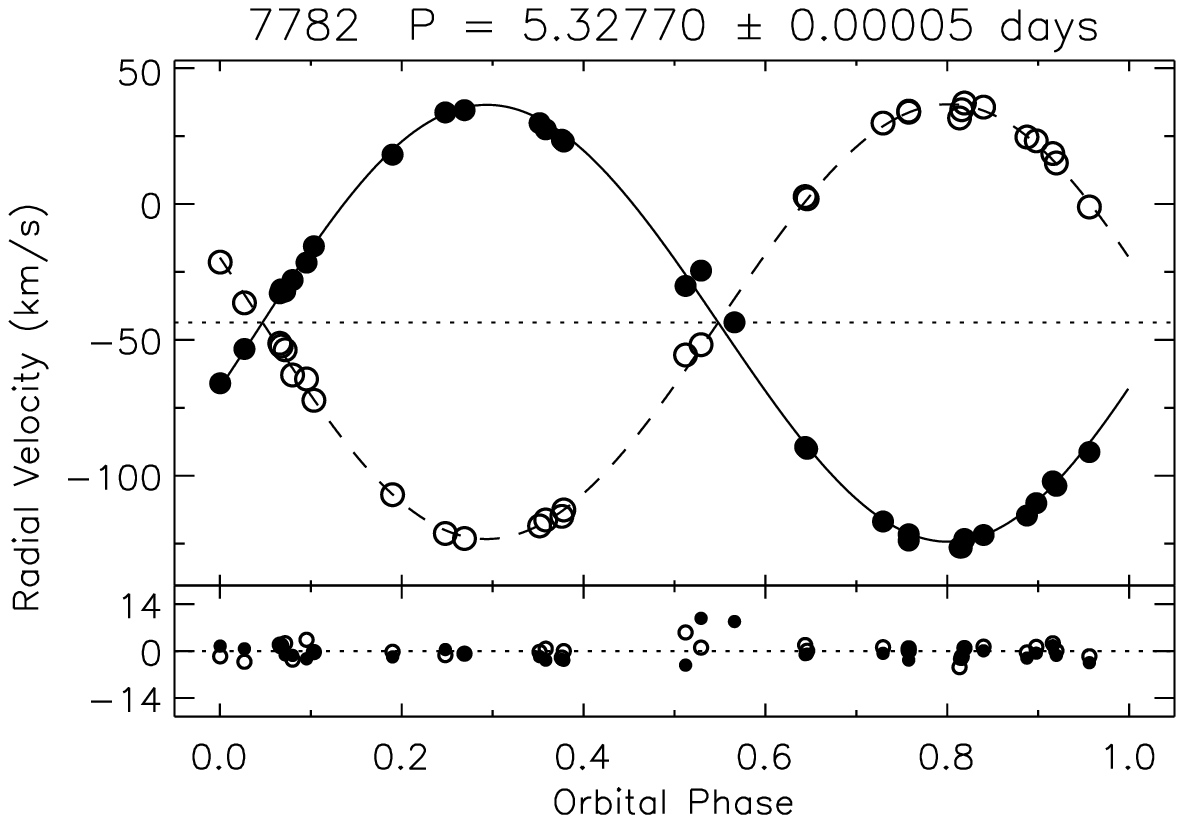,width=0.3\linewidth} & \epsfig{file=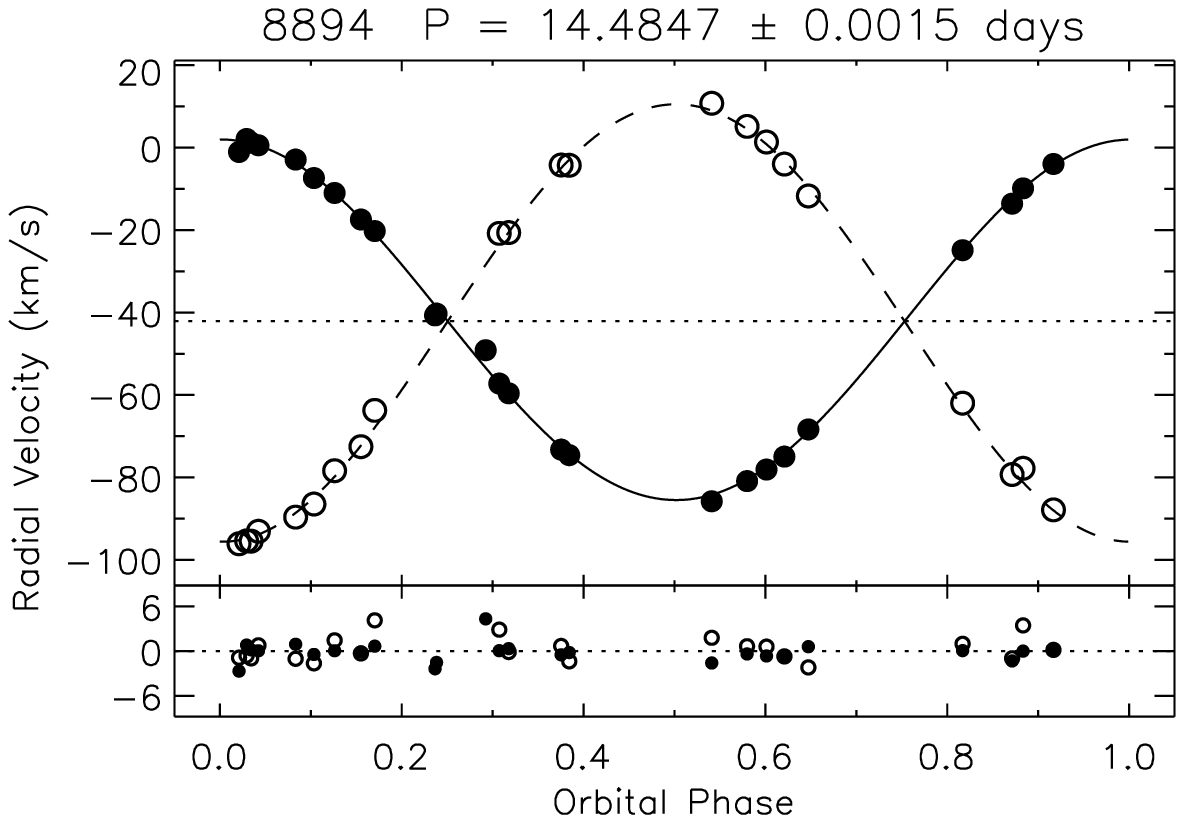,width=0.3\linewidth} \\
\end{longtable}
\end{center}
Fig. 2. --- \footnotesize NGC 188 SB2 orbit plots.  For each binary, we plot RV against orbital phase, showing the primary data points with filled circles and the secondary data points with open circles. The orbital fits to the data are plotted in the solid and dashed lines for the primary and secondary stars, respectively; the dotted line marks the $\gamma$-velocity.  Beneath each orbit plot, we show the residuals from the fit.  Above each plot, we give the binary ID and orbital period.\normalsize

\addtocounter{figure}{+1}
\addtocounter{table}{-1}

\begin{deluxetable}{l r c r r r r r r r r c c}
\tabletypesize{\scriptsize}
\tablewidth{0pt}
\rotate
\tablecaption{Orbital Parameters For NGC 188 Double-Lined Binaries\label{SB2tab}}
\tablehead{\colhead{ID} & \colhead{P} & \colhead{Orbital} & \colhead{$\gamma$} & \colhead{K} & \colhead{e} & \colhead{$\omega$} & \colhead{T$_\circ$} & \colhead{a$\sin$ i} & \colhead{m$\sin^3$ i} & \colhead{q} & \colhead{$\sigma$} & \colhead{N} \\
\colhead{} & \colhead{(days)} & \colhead{Cycles} & \colhead{(\kms)} & \colhead{(\kms)} & \colhead{} & \colhead{(deg)} & \colhead{(HJD-2400000 d)} & \colhead{(10$^6$ km)} & \colhead{(\Msolar)} & \colhead{} & \colhead{(\kms)} & \colhead{}}
\startdata
    3118 &         11.9022 & 133.8 &          -43.23 &            43.6 &           0.011 &              20 &         50900.3 &            7.14 &           0.656 &           0.795 &  1.68 &   44 \\
         &    $\pm$ 0.0004 &       &      $\pm$ 0.20 &       $\pm$ 0.3 &     $\pm$ 0.006 &        $\pm$ 30 &       $\pm$ 1.0 &      $\pm$ 0.06 &     $\pm$ 0.012 &     $\pm$ 0.009 &       &      \\
         &                 &       &                 &            54.9 &                 &                 &                 &            8.98 &           0.521 &                 &  1.98 &   36 \\
         &                 &       &                 &       $\pm$ 0.4 &                 &                 &                 &      $\pm$ 0.07 &     $\pm$ 0.009 &                 &       &      \\
    3344 &          126.46 &  24.2 &          -42.51 &            25.4 &           0.175 &             338 &         50889.0 &            43.4 &            0.95 &            0.93 &  1.26 &   28 \\
         &      $\pm$ 0.05 &       &      $\pm$ 0.23 &       $\pm$ 0.6 &     $\pm$ 0.014 &         $\pm$ 4 &       $\pm$ 1.6 &       $\pm$ 1.1 &      $\pm$ 0.06 &      $\pm$ 0.03 &       &      \\
         &                 &       &                 &            27.3 &                 &                 &                 &            46.7 &            0.88 &                 &  1.70 &   20 \\
         &                 &       &                 &       $\pm$ 0.7 &                 &                 &                 &       $\pm$ 1.3 &      $\pm$ 0.05 &                 &       &      \\
    4411 &           48.87 &  11.8 &           -42.4 &            33.7 &           0.433 &           251.6 &        50888.82 &           20.40 &            1.05 &           0.743 &  0.72 &   22 \\
         &      $\pm$ 0.03 &       &       $\pm$ 0.3 &       $\pm$ 0.3 &     $\pm$ 0.008 &       $\pm$ 1.2 &      $\pm$ 0.21 &      $\pm$ 0.16 &      $\pm$ 0.05 &     $\pm$ 0.018 &       &      \\
         &                 &       &                 &            45.3 &                 &                 &                 &            27.4 &            0.78 &                 &  2.55 &   15 \\
         &                 &       &                 &       $\pm$ 1.0 &                 &                 &                 &       $\pm$ 0.6 &      $\pm$ 0.03 &                 &       &      \\
    4506 &          8.8536 & 115.7 &          -42.69 &            33.3 &           0.007 &             210 &         50943.7 &            4.06 &           0.133 &           1.012 &  0.87 &   21 \\
         &    $\pm$ 0.0004 &       &      $\pm$ 0.15 &       $\pm$ 0.3 &     $\pm$ 0.008 &        $\pm$ 60 &       $\pm$ 1.4 &      $\pm$ 0.04 &     $\pm$ 0.004 &     $\pm$ 0.015 &       &      \\
         &                 &       &                 &            32.9 &                 &                 &                 &            4.01 &           0.134 &                 &  1.04 &   20 \\
         &                 &       &                 &       $\pm$ 0.4 &                 &                 &                 &      $\pm$ 0.05 &     $\pm$ 0.003 &                 &       &      \\
    4705 &          35.178 &  25.7 &          -42.53 &            47.3 &           0.487 &           245.4 &        50752.07 &           19.98 &            1.13 &           0.956 &  1.41 &   23 \\
         &     $\pm$ 0.005 &       &      $\pm$ 0.21 &       $\pm$ 0.4 &     $\pm$ 0.005 &       $\pm$ 1.0 &      $\pm$ 0.07 &      $\pm$ 0.22 &      $\pm$ 0.03 &     $\pm$ 0.013 &       &      \\
         &                 &       &                 &            49.5 &                 &                 &                 &           20.90 &            1.08 &                 &  1.33 &   20 \\
         &                 &       &                 &       $\pm$ 0.4 &                 &                 &                 &      $\pm$ 0.21 &      $\pm$ 0.03 &                 &       &      \\
    4732 &          42.115 &  17.6 &          -41.46 &           31.56 &           0.232 &            86.3 &        50945.87 &           17.78 &           0.537 &           0.970 &  0.78 &   29 \\
         &     $\pm$ 0.008 &       &      $\pm$ 0.13 &      $\pm$ 0.22 &     $\pm$ 0.006 &       $\pm$ 1.8 &      $\pm$ 0.20 &      $\pm$ 0.13 &     $\pm$ 0.013 &     $\pm$ 0.013 &       &      \\
         &                 &       &                 &            32.5 &                 &                 &                 &           18.32 &           0.521 &                 &  1.29 &   28 \\
         &                 &       &                 &       $\pm$ 0.3 &                 &                 &                 &      $\pm$ 0.20 &     $\pm$ 0.009 &                 &       &      \\
    4962 &          38.644 &  80.4 &           -43.0 &            30.8 &           0.445 &              83 &         51636.6 &            14.7 &            0.35 &            0.99 &  3.22 &   21 \\
         &     $\pm$ 0.004 &       &       $\pm$ 0.4 &       $\pm$ 1.4 &     $\pm$ 0.017 &         $\pm$ 4 &       $\pm$ 0.3 &       $\pm$ 0.7 &      $\pm$ 0.03 &      $\pm$ 0.06 &       &      \\
         &                 &       &                 &            31.3 &                 &                 &                 &            14.9 &            0.34 &                 &  1.91 &   19 \\
         &                 &       &                 &       $\pm$ 1.0 &                 &                 &                 &       $\pm$ 0.5 &      $\pm$ 0.04 &                 &       &      \\
    4986 &          30.086 & 103.3 &           -41.4 &            43.8 &           0.342 &              82 &        53568.95 &            17.0 &            0.99 &            0.94 &  3.22 &   18 \\
         &     $\pm$ 0.004 &       &       $\pm$ 0.5 &       $\pm$ 1.5 &     $\pm$ 0.020 &         $\pm$ 3 &      $\pm$ 0.25 &       $\pm$ 0.7 &      $\pm$ 0.09 &      $\pm$ 0.05 &       &      \\
         &                 &       &                 &            46.6 &                 &                 &                 &            18.1 &            0.93 &                 &  3.10 &   17 \\
         &                 &       &                 &       $\pm$ 1.4 &                 &                 &                 &       $\pm$ 0.7 &      $\pm$ 0.09 &                 &       &      \\
    5078 &         4.78303 & 191.3 &           -41.6 &            65.8 &           0.121 &             317 &        50704.13 &            4.30 &            1.25 &           0.678 &  1.41 &   23 \\
         &   $\pm$ 0.00012 &       &       $\pm$ 0.3 &       $\pm$ 0.4 &     $\pm$ 0.006 &         $\pm$ 3 &      $\pm$ 0.04 &      $\pm$ 0.03 &      $\pm$ 0.03 &     $\pm$ 0.009 &       &      \\
         &                 &       &                 &            97.0 &                 &                 &                 &            6.34 &           0.846 &                 &  3.34 &   18 \\
         &                 &       &                 &       $\pm$ 1.0 &                 &                 &                 &      $\pm$ 0.07 &     $\pm$ 0.016 &                 &       &      \\
    5080 &           550.4 &   6.2 &           -43.4 &            15.8 &            0.44 &             153 &           52273 &             107 &            0.65 &            1.01 &  0.95 &   14 \\
         &       $\pm$ 1.9 &       &       $\pm$ 0.3 &       $\pm$ 0.5 &      $\pm$ 0.03 &         $\pm$ 4 &         $\pm$ 7 &         $\pm$ 4 &      $\pm$ 0.09 &      $\pm$ 0.07 &       &      \\
         &                 &       &                 &            15.7 &                 &                 &                 &             107 &            0.65 &                 &  2.08 &   13 \\
         &                 &       &                 &       $\pm$ 0.9 &                 &                 &                 &         $\pm$ 7 &      $\pm$ 0.07 &                 &       &      \\
    5147 &          6.7432 &  66.5 &          -42.45 &            50.2 &           0.012 &             350 &         54356.3 &            4.65 &           0.362 &           0.988 &  1.33 &   19 \\
         &    $\pm$ 0.0004 &       &      $\pm$ 0.19 &       $\pm$ 0.4 &     $\pm$ 0.006 &        $\pm$ 30 &       $\pm$ 0.5 &      $\pm$ 0.04 &     $\pm$ 0.006 &     $\pm$ 0.011 &       &      \\
         &                 &       &                 &            50.8 &                 &                 &                 &            4.71 &           0.357 &                 &  0.96 &   17 \\
         &                 &       &                 &       $\pm$ 0.3 &                 &                 &                 &      $\pm$ 0.03 &     $\pm$ 0.007 &                 &       &      \\
    5710 &           298.8 &   4.1 &          -42.86 &            15.2 &           0.215 &              74 &           50775 &            61.0 &            0.62 &            0.81 &  0.98 &   26 \\
         &       $\pm$ 1.3 &       &      $\pm$ 0.20 &       $\pm$ 0.4 &     $\pm$ 0.023 &         $\pm$ 5 &         $\pm$ 4 &       $\pm$ 1.7 &      $\pm$ 0.04 &      $\pm$ 0.03 &       &      \\
         &                 &       &                 &            18.7 &                 &                 &                 &            75.2 &            0.51 &                 &  1.08 &    9 \\
         &                 &       &                 &       $\pm$ 0.5 &                 &                 &                 &       $\pm$ 2.2 &      $\pm$ 0.03 &                 &       &      \\
    5762 &         6.50430 & 421.7 &          -40.48 &            72.9 &           0.004 &              40 &         51061.2 &            6.52 &           1.093 &           0.977 &  1.69 &   32 \\
         &   $\pm$ 0.00004 &       &      $\pm$ 0.19 &       $\pm$ 0.4 &     $\pm$ 0.004 &        $\pm$ 50 &       $\pm$ 1.0 &      $\pm$ 0.04 &     $\pm$ 0.013 &     $\pm$ 0.008 &       &      \\
         &                 &       &                 &            74.6 &                 &                 &                 &            6.67 &           1.067 &                 &  1.26 &   29 \\
         &                 &       &                 &       $\pm$ 0.3 &                 &                 &                 &      $\pm$ 0.03 &     $\pm$ 0.015 &                 &       &      \\
    7782 &         5.32770 & 426.5 &           -43.6 &            80.4 &           0.013 &             250 &         52206.8 &            5.89 &           1.136 &           1.005 &  3.12 &   32 \\
         &   $\pm$ 0.00005 &       &       $\pm$ 0.3 &       $\pm$ 0.8 &     $\pm$ 0.006 &        $\pm$ 30 &       $\pm$ 0.4 &      $\pm$ 0.06 &     $\pm$ 0.020 &     $\pm$ 0.013 &       &      \\
         &                 &       &                 &            80.0 &                 &                 &                 &            5.86 &            1.14 &                 &  2.12 &   31 \\
         &                 &       &                 &       $\pm$ 0.5 &                 &                 &                 &      $\pm$ 0.04 &      $\pm$ 0.03 &                 &       &      \\
    8894 &         14.4847 &  43.5 &          -42.11 &            43.7 &           0.008 &             360 &         51017.9 &            8.71 &           0.746 &           0.824 &  1.41 &   25 \\
         &    $\pm$ 0.0015 &       &      $\pm$ 0.24 &       $\pm$ 0.4 &     $\pm$ 0.010 &        $\pm$ 40 &       $\pm$ 1.8 &      $\pm$ 0.09 &     $\pm$ 0.019 &     $\pm$ 0.012 &       &      \\
         &                 &       &                 &            53.1 &                 &                 &                 &           10.57 &           0.615 &                 &  1.78 &   22 \\
         &                 &       &                 &       $\pm$ 0.5 &                 &                 &                 &      $\pm$ 0.12 &     $\pm$ 0.015 &                 &       &      \\
\enddata
\end{deluxetable}

\begin{deluxetable}{l c c c c c c c}
\tablewidth{0pt}
\tablecaption{Physical Properties of NGC 188 Double-Lined Binaries\label{SB2masstab}}
\tablehead{\colhead{ID} & \colhead{$V$} & \colhead{$(\bv)$} & \colhead{R} & \colhead{P$_{RV}$} & \colhead{P$_{PM}$} & \colhead{M$_1$} & \colhead{M$_2$} \\
\colhead{} & \colhead{} & \colhead{} & \colhead{(arcmin)} & \colhead{(\%)} & \colhead{(\%)} & \colhead{(\Msolar)} & \colhead{(\Msolar)}}
\startdata
3118 &  14.652 & 1.123 & 18.73 & 95 & 34 &                 1.14 &                 0.90 \\
3344 &  15.290 & 0.704 & 14.58 & 98 & 98 &                 1.01 &                 0.94 \\
4411 &  15.693 & 0.734 &  6.97 & 98 & 98 &                 0.99 &                 0.73 \\
4506 &  14.865 & 0.650 &  7.11 & 98 & 97 &                 1.04 &                 1.05 \\
4705 &  13.933 & 0.938 &  3.16 & 98 & 98 &                 1.14 &                 1.09 \\
4732 &  15.467 & 0.720 &  4.07 & 96 & 98 &                 0.97 &                 0.94 \\
4962 &  15.286 & 0.692 &  2.21 & 98 & 98 &                 0.99 &                 0.98 \\
4986 &  15.274 & 0.671 &  1.42 & 96 & 97 &                 1.03 &                 0.97 \\
5080 &  14.624 & 0.668 &  1.55 & 96 & 98 &                 1.02 &                 1.02 \\
5147 &  15.343 & 0.701 &  4.58 & 98 & 98 &                 0.98 &                 0.96 \\
5710 &  14.823 & 0.726 &  7.50 & 98 & 98 &                 1.11 &                 0.90 \\
5762 &  14.759 & 0.673 &  8.00 & 66 & 97 &                 1.03 &                 1.00 \\
8894 &  15.290 & 0.684 & 15.95 & 98 & 91 &                 1.05 &                 0.87 \\
\enddata
\end{deluxetable}

\clearpage

\section{Binaries of Note} \label{anom}

In the following section, we discuss the properties of various intriguing binaries that we have discovered in NGC 188.  
We first discuss three binaries that contain potential encounter products.  We then include our photometric variables 
and X-ray sources, and present evidence that 5015 is in fact a quadruple system composed of two SB1 binary cluster members.

\subsection{Binaries Containing Potential Encounter Products}

\paragraph{5078:}
5078 has a period of 4.78303 $\pm$ 0.00012 days, well below the circularization period of 14.5 days in NGC 188 \citep{mei05}.
However this binary has a significantly higher than circular eccentricity, at 0.121 $\pm$ 0.006.  5078 is a particularly intriguing 
binary as it is a BS with an SB2 orbital solution.  This relatively high eccentricity 
may be a sign of a recent dynamical interaction or an additional companion \citep{maz90}.  Triple systems are 
not uncommon within binary populations, with observational evidence ranging from 5-50\% \citep{may87,duq91,pou04,tok06}.
Furthermore, \citet{tok06} showed that for solar-type binaries, the frequency of additional companions increases towards 
shorter inner-binary periods, finding a frequency of tertiary companions for binaries with periods $\sim$5 days of $\sim$65\%.

\paragraph{5080 :}
5080 is a SB2 binary found right above the main-sequence turnoff at $V$ = 14.624 and $(\bv)$ = 0.668.  The system is located 
at 0.7 core radii from the cluster center, and has a \PRV~= 96\% and a \PPM~= 98\%.   From our orbital solution, we find a mass 
ratio of 1.01 $\pm$ 0.07, and we estimate that both stars have masses of $\sim$1.02 \Msolar.  
However, from inspection of the cross-correlation functions, it is clear that the two stars have
different luminosities.  We checked for a potential template mismatch using a set of solar-metallicity synthetic
spectral templates ranging from a 0.5 \Msolar~main-sequence star to a 1.14 \Msolar~star at the tip of the giant branch.
For all spectra of 5080 in which we detect the secondary, a combination of two solar templates returns the highest 
two-dimensional correlation peak height and therefore the best fit to the data.  
Hence we proceed to use our standard solar spectrum as the template for both the primary and secondary stars in order to 
derive the luminosity ratio.  The majority of the correlation functions are highly blended.  Consequently we ran TODCOR on the four 
observations that show the largest RV separations and derive a luminosity ratio ($L_2/L_1$) of 0.32, with a 
standard deviation of 0.04.  Thus the secondary star appears to be under-luminous for its mass.  We note that, if we take the 
lowest value for the mass ratio allowed by the error, of 0.94, then we could be observing a binary containing a primary star that has 
evolved just past the turnoff with a main-sequence secondary star.  If we take the mass of the primary star to be 1.02 \Msolar, as derived in 
Section~\ref{SB2}, then the secondary star could have a mass as low as 0.96 \Msolar.  Using these values with the Padova isochrone, we derive a 
luminosity ratio of 0.65, which is certainly much larger than what we observe.

\paragraph{7782 :}
7782 is a BS SB2 binary located at 9.7 core radii with a \PRV~= 95\% and a \PPM~= 11\%.  7782
is the second bluest of our detected BSs in NGC 188 with a $(\bv)$ = 0.494.  Interestingly, we find the 
system to have a mass ratio of 1.005 $\pm$ 0.013, meaning that both stars in the system are likely more massive than the main-sequence
turnoff mass. Utilizing TODCOR, we select the 11 observations with well separated peaks to find a luminosity ratio of 0.739 with 
a standard deviation of 0.026. We suggest that 7782 may be a BS - BS binary system.

\subsection{Photometric Variables and X-ray Sources}

\paragraph{4289 : }
4289 is a SB1 binary found at the base of the giant branch at a radius of 2.5 core radii.  The binary is a 
secure cluster member with both \PRV~and \PPM~= 98 \%.  We derive an orbital solution with a period of 11.4877 $\pm$ 0.0009
days and an eccentricity consistent with circular of 0.012 $\pm$ 0.010.  We estimate that the primary star is likely a red giant with a mass 
of $<$ 1.12 \Msolar, and the secondary star is on the main sequence with a mass of $<$ 0.78 \Msolar.  
This binary was observed to be one of the brightest X-ray sources, GX28, in the \citet{gon05} survey. 
They point out that one would not expect a giant star in NGC 188 to show rapid rotation or surface activity unless the star is a member 
of a tight binary system in which rapid rotation has been maintained by synchronization.
We do not see any evidence for line broadening due to rotation in our spectra, which corresponds to an upper limit of $\sim$10 
\kms~(derived from similar analysis to that of \citet{rho01}).  With a period of $\sim$11.5 days and assuming an appropriate radius for 
the primary star of $\sim$2.3 \Rsolar, we would expect a maximum rotational velocity of $\sim$10 \kms~resulting from tidal 
synchronization.  According to \citet{gon05a}, even this relatively slow rotation may be sufficient to increase the surface coverage 
of magnetic-loop structures in giants like 4289 enough to produce the observed X-ray emission.

\paragraph{4705 : }
This SB2 binary is found at 1.3 core radii, and lies near the giant branch with $V$ = 13.933 and $(\bv)$ = 0.938.  The 
binary is a high-probability cluster member with both \PRV~and \PPM~= 98 \%.  We derive a kinematic orbital solution with a 
period of 35.178 $\pm$ 0.005 days and an eccentricity of 0.487 $\pm$ 0.005.  This star was observed as a photometric variable,
V11, by \citet{kal90}, who noted a dimming of almost 0.4 magnitudes over the course of the night of December 13, 1986. 
Kaluzny et al.~conjecture that this variability and the location of 4705 on the CMD can be explained if 4705 is an eclipsing binary 
with a relatively unevolved red-giant primary and an upper-main-sequence secondary star.  The observed photometric dimming occurred 
at a phase of $\sim$0.02 in our derived orbit (when the RVs of both the primary and secondary stars were very near the $\gamma$-velocity of 
the system).
We used the program NIGHTFALL\footnote{NIGHTFALL is copyright (c) 1998-2002 Rainer Wichmann, (c) 2001-2002 Markus Kuster, (c) 2001- 
2002 Patrick Risse and can be downloaded from http://www.hs.uni-hamburg.de/DE/Ins/Per/Wichmann/Nightfall.html.}
 to determine the phase at which one would expect to observe an eclipse in this 
system, and find that we would indeed expect an eclipse to occur at a phase of $\sim$0.02.  Thus 4705 may be an eclipsing 
binary system in NGC 188.  Furthermore, we estimate the primary mass to be 1.14 \Msolar~and 
find a mass ratio of 0.956 $\pm$ 0.013.  This would allow for an upper main-sequence secondary star as predicted.  Additionally, 4705
was found to be an X-ray variable, GX18, by \citet{gon05}, who observed low-amplitude brightness variations on the time 
scale of weeks.  They suggest that these variations are due to slow rotation, as rotating giants can produce high 
X-ray luminosities, possibly related to the existence of magnetic fields induced by turbulent motion in their deepening 
convective zones.  It has also been suggested by \citet{zha02} and \citet{gon05} that 4705 may be an RS CVn system.  

\paragraph{5379 : }
This SB1 BS binary lies at 1.6 core radii from the cluster center and is a secure cluster member with both \PPM~and \PRV~= 98 \%.
This binary is a BS, with a $V$ magnitude of 15.373 and a $(\bv)$ color of 0.542. We derive a period of 120.21 
$\pm$ 0.04 days with an eccentricity of 0.24 $\pm$ 0.03.  Additionally, \citet{kaf03} found this binary to be a photometric 
variable (WV3) with a period of 0.18148 days.  We cannot derive a kinematic orbital solution with this short period.  
We do observe signs of above average rotation in the 5379 spectra, and we have used the procedure of \citet{rho01} to derive a 
$v \sin i$ of 15.4 $\pm$ 0.5 \kms.  If this photometric variability is due to chromospheric activity or star spots at 
this short period, we would expect a rotational velocity for the star of $>$250 \kms, which can be ruled out for all inclination 
angles greater than $\sim$3.5\degr.  \citet{kaf03} suggested that 5379 may be a member of the short-period end of the NGC 188 
W UMa population.  This now seems less likely given our lack of observed rapid rotation.  We note that the photometric period, 
amplitude of the oscillations, and the observed $v \sin i$ lie within the observed range of $\delta$ Sct variable stars 
\citep{rod00}.  However 5379 does not lie near the instability strip. 

\paragraph{5762 : }
5762 is a SB2 binary found at the main-sequence turnoff at 3.4 core radii from the cluster center.
The binary has a \PPM~= 97\% and \PRV~= 66\%.  We derive a circular orbit with a period of 6.50430 $\pm$ 0.00004 days,
a mass ratio near unity of 0.977 $\pm$ 0.008, and a minimum separation between the primary and secondary of 18.95 $\pm$ 0.08 \Rsolar.  
Zhang et al.~(2002, 2004) identified this system as an eclipsing binary (V12).
The observed photometric eclipse in \citet{zha02} occurred at a phase of 0.88 in our orbital 
solution, when both stars in the system were moving near the $\gamma$-velocity.  This provides further evidence for the eclipsing 
nature of the system.  \citet{mei09} discuss this eclipsing binary in detail.  We simply point out that 
even if we are viewing this system at a low inclination angle, the true separation between the two stars will likely be very 
favorable to mass transfer as both stars evolve up the giant branch.  As such, 5762 may be a pre-mass-transfer system which 
could represent a BS precursor.

\subsection{A Possible Quadruple System : 5015}

\begin{figure}[!ht]
\label{5015aborbs}
\plottwo{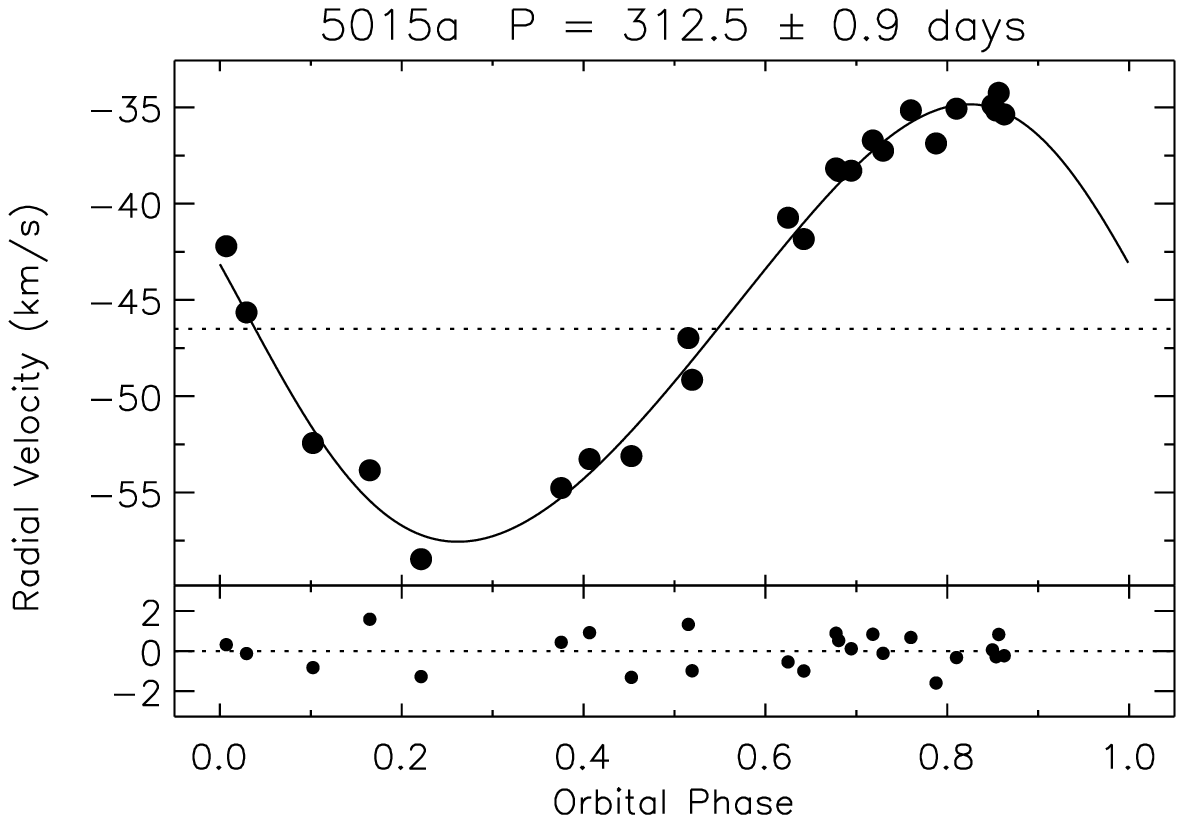}{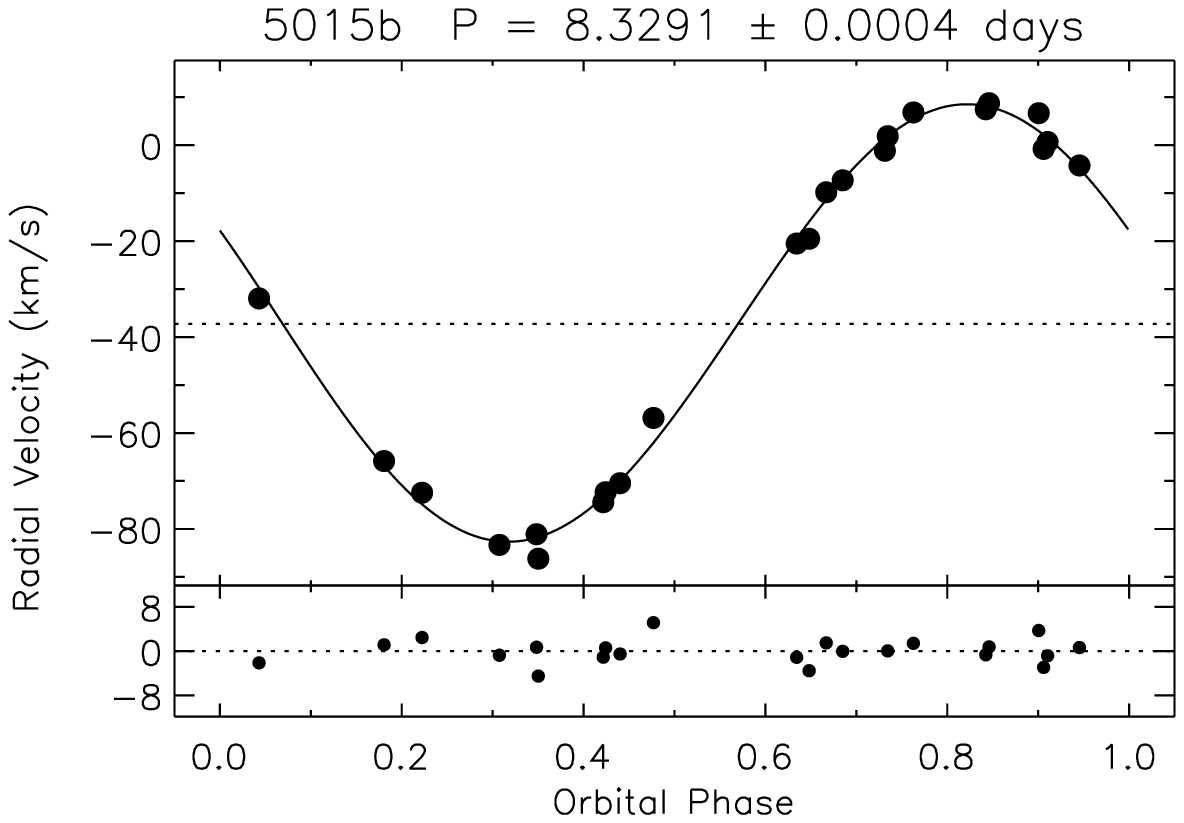}
\caption{\footnotesize SB1 orbital solutions for the two binaries 5015a (left) and 5015b (right) that likely reside in a quadruple system.
In the top panels we plot the observed data with dots and the orbital fits in the solid lines; the dotted lines mark the 
$\gamma$-velocities.  Below the orbital plots, we show the RV residuals, and above the plots we provide the IDs and periods.}
\end{figure}

5015 is a 90\% PM member, and upon preliminary inspection of the observed spectra and the resulting cross-correlation functions,
we presumed that 5015 was a typical SB2 binary.  
There are two clear peaks in most of the 1D correlation functions, and both 
RVs are easily recovered using TODCOR for all but one observation.  We followed the usual procedure of fitting an orbital 
solution to the primary, then using the derived orbital parameters to fit the full orbital solution, including the secondary
velocities.  However, we were unable to derive an SB2 orbit using the parameters from the fit to the primary.  We then
proceeded to fit a separate orbital solution to the secondary RVs, and found that the two solutions had entirely different parameters.  
We show the individual orbits in Figure~\ref{5015aborbs} and give the respective orbital parameters in Table~\ref{5015ab}. 
Individually, each of the derived $\gamma$-velocities results in a \PRV~= 0\%.  Interestingly, though, if 
we take the average of the two $\gamma$-velocities, we get -41.9 $\pm$ 0.3 \kms, which is very close to the cluster mean RV of 
\mrv~(Paper 1).
Thus we have two options:  either the two observed binaries are a chance superposition of two field binaries, or we are observing
a quadruple system that is a likely member of NGC 188.

\begin{deluxetable}{l r@{\hspace{0.5em}}c@{\hspace{0.5em}}l r@{\hspace{0.5em}}c@{\hspace{0.5em}}l}
\tabletypesize{\small}
\tablewidth{0pt}
\tablecaption{Orbital Parameters for 5015a and 5015b\label{5015ab}}
\tablehead{\colhead{} & \multicolumn{3}{c}{5015a} & \multicolumn{3}{c}{5015b}}
\startdata
P (days)               & 312.5 & $\pm$ & 0.9    & 8.3291 & $\pm$ & 0.0004 \\
$\gamma$ (\kms)        & -46.50 & $\pm$ & 0.24    & -37.2 & $\pm$ & 0.6 \\
K (\kms)               & 11.4 & $\pm$ & 0.4     & 45.6 & $\pm$ & 0.7 \\
e                      & 0.10 & $\pm$ & 0.03    & 0.008 & $\pm$ & 0.016 \\
$\omega$ (deg)         & 74 & $\pm$ & 21          & 70 & $\pm$ & 150 \\
T$_\circ$ (HJD-2400000 d) & 51599 & $\pm$ & 19       & 51875 & $\pm$ & 4 \\
a$\sin$ i (10$^6$ km)   & 48.6 & $\pm$ & 1.5       & 5.22 & $\pm$ & 0.08 \\
f(m) (\Msolar)         & 4.7e-2 & $\pm$ & 0.4e-2 & 8.2e-2 & $\pm$ & 0.4e-2 \\
$\sigma$ (\kms)        & \multicolumn{3}{c}{1.0} & \multicolumn{3}{c}{2.55} \\
N                      & \multicolumn{3}{c}{23}   & \multicolumn{3}{c}{22} \\
\enddata
\end{deluxetable}

If we assume that the two binaries are not cluster members, then we can ask what is the likelihood that we are observing a superposition of 
two binaries in the field.  To answer this question, we utilized the theoretical Besan\c{c}on model of the Milky Way \citep{rob03} 
to derive the expected number of field stars within one square degree, covering our observed magnitude range, towards the direction 
of NGC 188.  We then assume that the locations of these field stars are described by a Poisson distribution and proceed to 
calculate the conditional probability that we would observe two field stars within a three arcsecond diameter fiber, given that we observe 
at least one, and find a 0.04\% probability.  Furthermore,  since 5015 contains two binaries within a three arcsecond diameter region, 
we then multiply this value twice by the field binary fraction of 51\%, as observed by \citet{duq91}.  Finally, we must account for the 
RVs of the two binaries.  To do so, we again use the Besan\c{c}on model to calculate the percentage of field stars with RVs within 
five \kms~from the mean RV for NGC 188 (i.e., only including field stars with -47 \kms~$\leq$~RV~$\leq$~-37 \kms), and find these stars 
to populate 20\% of the field towards NGC 188.  Including these constraints, the probability of observing two field binaries 
in the direction of NGC 188 within a three arcsecond diameter fiber that have RVs within five \kms~from the mean RV for NGC 188 is 
decidedly small, at 0.002\%.  To date, we have observed a total of 1116 stars in the direction of NGC 188.  Though this is a 
relatively large number of stars, it is certainly not enough for us to expect to observe such a chance superposition of two field 
binaries.  Therefore, this scenario seems unlikely.

Conversely, we can assume that these two binaries are members of a quadruple system in which the two binaries orbit each other 
about the system's center of mass.  Observations of field solar-type binary populations find the frequency of triples and higher-order systems 
to be 5-50\% \citep[e.g.][]{may87,duq91,tok97}.  Additionally, there is observational evidence for the presence of multiple-star systems 
in a few well studied open clusters (e.g., M67, \citet{mat90}; Praesepe, \citet{mer94}; Pleiades, \citet{bou97}; Hyades, \citet{pat98}).  
Recent $N$-body simulations by \citet{hur05} suggest that in an old open cluster, we might expect up to $\sim$7\% of the sources to 
reside in dynamically-formed triple or higher-order systems.  Thus we should not be surprised to find a few such star systems in NGC 188.

Using TODCOR, we derive a luminosity ratio of 0.36 $\pm$ 0.02.  From the Padova isochrone, we find a luminosity 
ratio of $L \propto M^{4.5}$, valid for this region of the NGC 188 main-sequence, which results in a mass ratio of 0.80 $\pm$ 0.04.
Therefore, the true center-of-mass RV of the quadruple system would be 
-42.4 $\pm$ 0.3 \kms, which would result in a \PRV~= 98\%.  This along with the \citet{pla03} \PPM~= 90\% provides strong evidence 
for cluster membership.

\section{Summary}

In this paper, we present \orb~binary orbits resulting from our ongoing RV survey of the old open cluster 
NGC 188.  This is the second paper in a series aimed at characterizing the solar-type single- and 
binary-star populations within the cluster.  These data will enable us to investigate the formation mechanisms and evolution 
of anomalous stars, like BSs, as they are influenced by the binary population, through comparison 
with detailed theoretical models of the cluster.  

We provide our complete current RV database for NGC 188 in Table~\ref{RVtable}, including the measured RVs for 
all stars observed in the direction of NGC 188 over the course of our RV survey of the cluster.  We use 
these data to derive the \SBone~SB1 (Section~\ref{SB1}) and \SBtwo~SB2 (Section~\ref{SB2}) orbital solutions for the 
NGC 188 cluster member binaries presented in this paper, and provide the results 
both graphically and as tabulated orbital elements.  For the main-sequence, sub-giant and giant binaries we use a 
photometric deconvolution technique to estimate the masses of the primary and secondary stars relative to a 7 Gyr 
solar-metallicity isochrone, and we provide the SB1 results in Table~\ref{SB1masstab} and the SB2 results in 
Table~\ref{SB2masstab}.  For SB1 systems, we also provide a lower limit on the secondary mass, derived using 
the orbital mass function.

In Section~\ref{anom} we identify a few binaries of note, including a likely quadruple system, 5015. 
Notably, 4705 and 5762 are both SB2 systems that may also be eclipsing binaries (5762 is studied in detail by 
\citet{mei09}). 
We also observe the BS 7782 as an SB2 system with a mass-ratio near unity, which suggests that the system may contain two BS stars.  
We use TODCOR to investigate the luminosity ratio for the equal mass SB2 binary 5080 and find that the 
secondary star appears to be under-luminous for its mass.  Finally we discuss the additional photometric variables and X-ray sources that 
are in binaries in NGC 188.  The binaries of note discussed in Section~\ref{anom} are ripe for further study.

The WIYN Open Cluster Study will continue its survey of NGC 188 in order to provide orbital solutions for 
all binaries in the cluster out to periods of 1000 days as well as a fraction of longer period binaries.  
In future papers, we will analyze the 
binary distribution in period, eccentricity and secondary mass, and constrain the cluster binary fraction.
These data will form critical constraints on future detailed $N$-body models of NGC 188 as well as other open clusters, 
allowing us to study the complex interplay of stellar evolution and dynamics amongst the single- and binary-cluster
members as they interact in the open cluster environment.

\acknowledgments
The authors would like to express their gratitude to the staff of the WIYN Observatory without whom we would
not have been able to acquire these thousands of superb stellar spectra. We also thank the many undergraduate 
and graduate students who have helped to obtain these spectra over the years at WIYN for this project.
We would like to acknowledge R. F. Griffin and J. E. Gunn for contributing their NGC 188 RVs to our project, who, 
in turn, wish to express their thanks to the Palomar Observatory for the use of the 5m telescope.  
Thanks to Murray Fletcher for his expertise in developing the DAO RVS instrument, and to Jim Hesser who acquired
a portion of the DAO NGC 188 data.   Finally, we wish to thank to anonymous referee for the helpful suggestions
in improving this paper.  This work was funded by the National Science Foundation grant AST-0406615 and
the Wisconsin Space Grant Consortium.

Facilities: \facility{WIYN 3.5m}, \facility{DAO 1.2m}, \facility{Palomar 5m}

\appendix
\section{APPENDIX}
\subsection{Field Binaries}
In our survey to find binary cluster members, we have serendipitously derived orbital solutions for \orbnm~field 
binaries, all with either \PRV~or \PPM~= 0\%.  We note that some of these binaries appear to be kinematic 
members of NGC 188 from either PM or RV evidence, but none are cluster members in all three dimensions.  In the interest of 
studies of the field binary population, we present these orbital plots (Figures 4 and 5) and parameters 
(Tables~\ref{fieldSB1tab} and~\ref{fieldSB2tab}) here.

\begin{center}
\begin{longtable}{ccc}
\epsfig{file=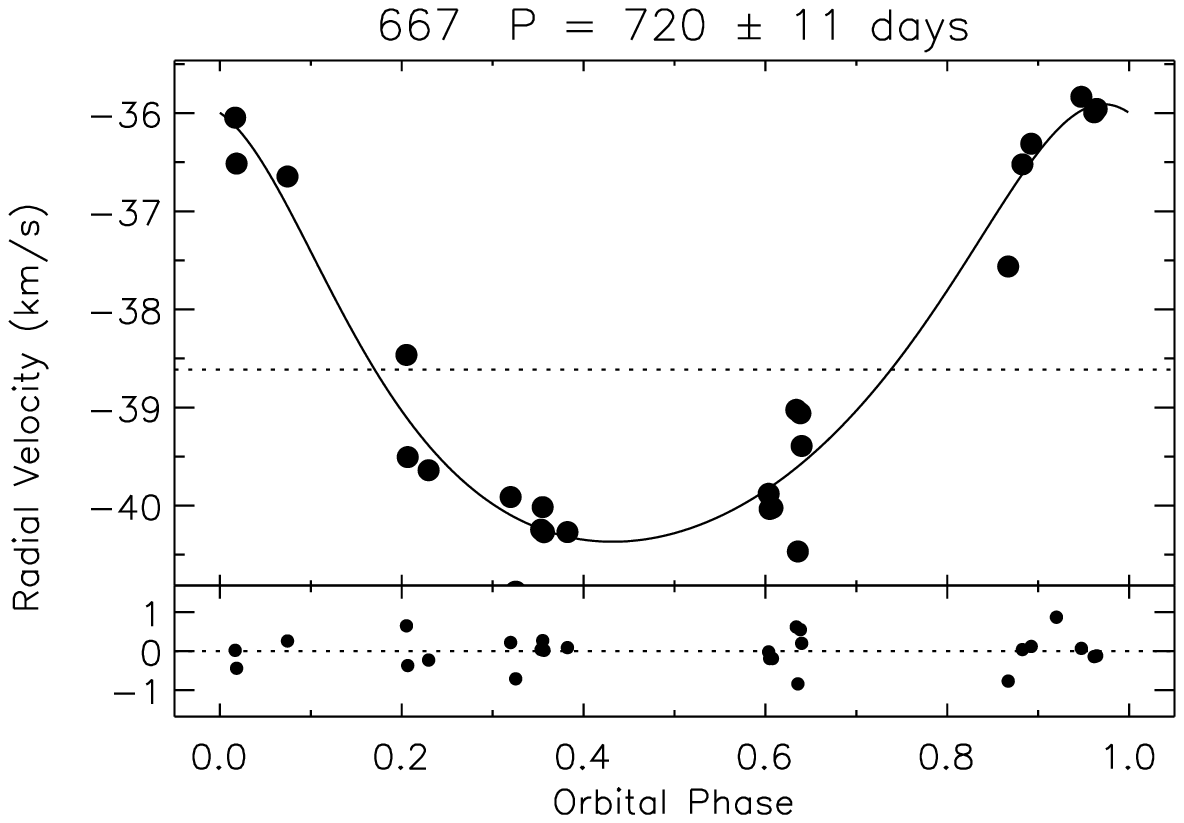,width=0.3\linewidth} & \epsfig{file=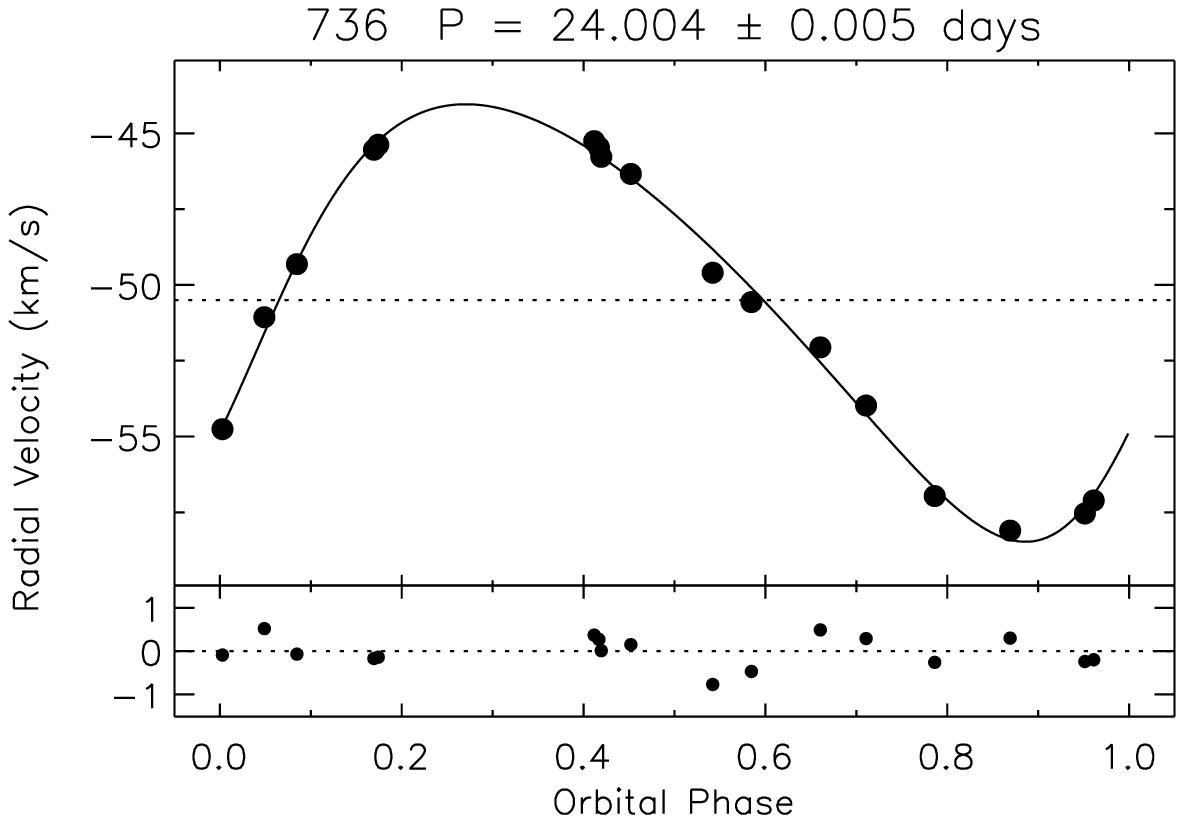,width=0.3\linewidth} & \epsfig{file=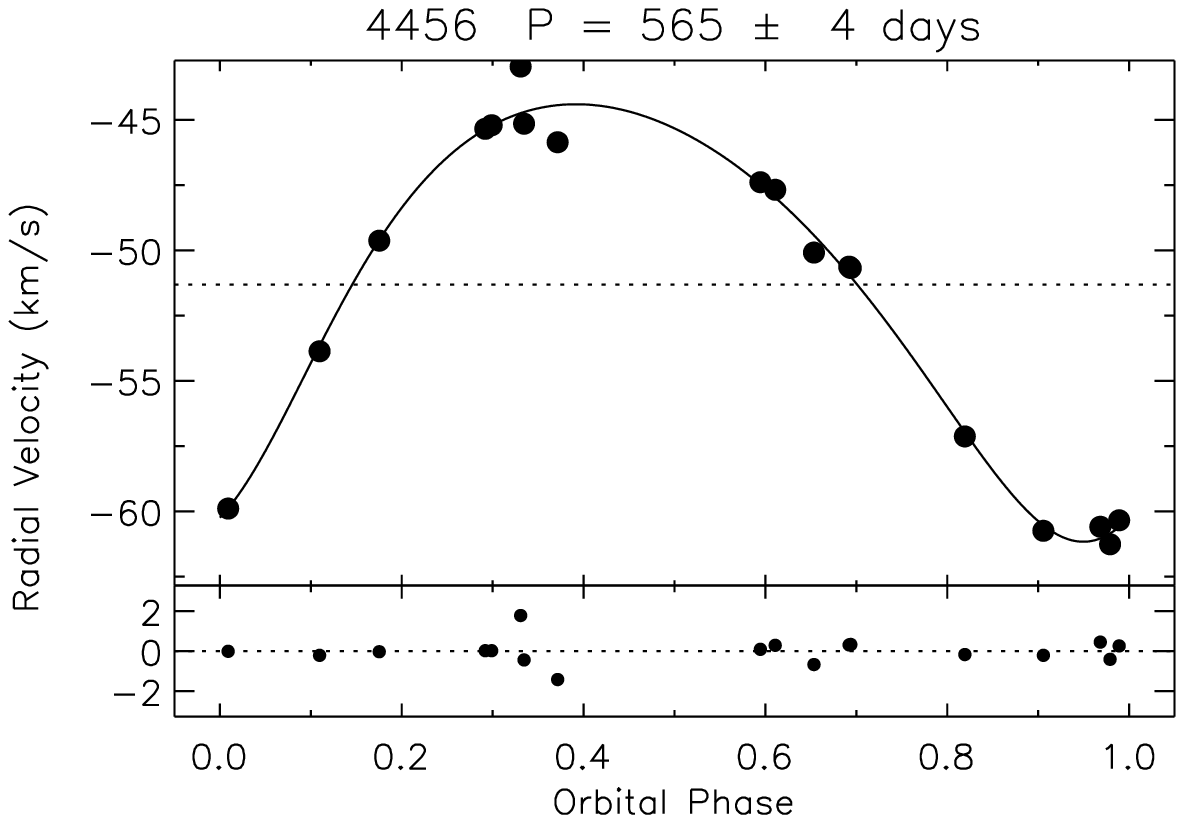,width=0.3\linewidth} \\
\epsfig{file=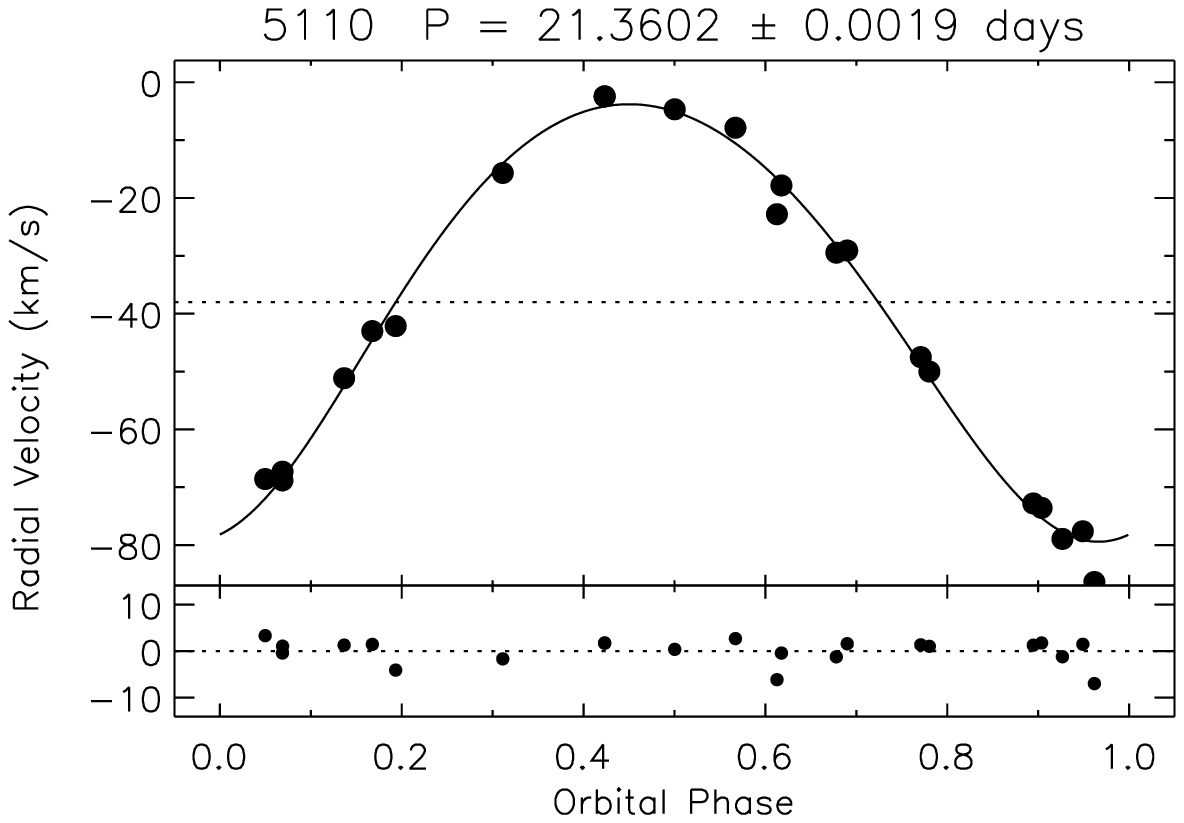,width=0.3\linewidth} & \epsfig{file=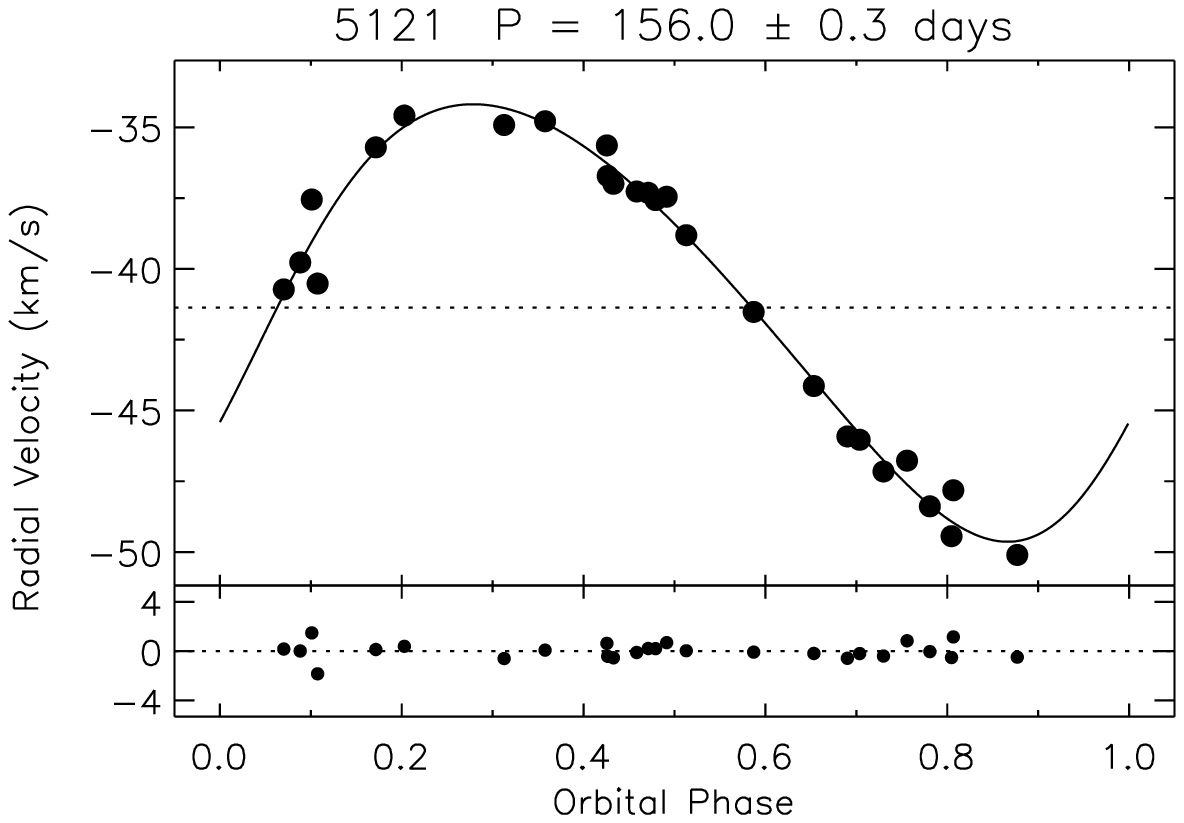,width=0.3\linewidth} & \epsfig{file=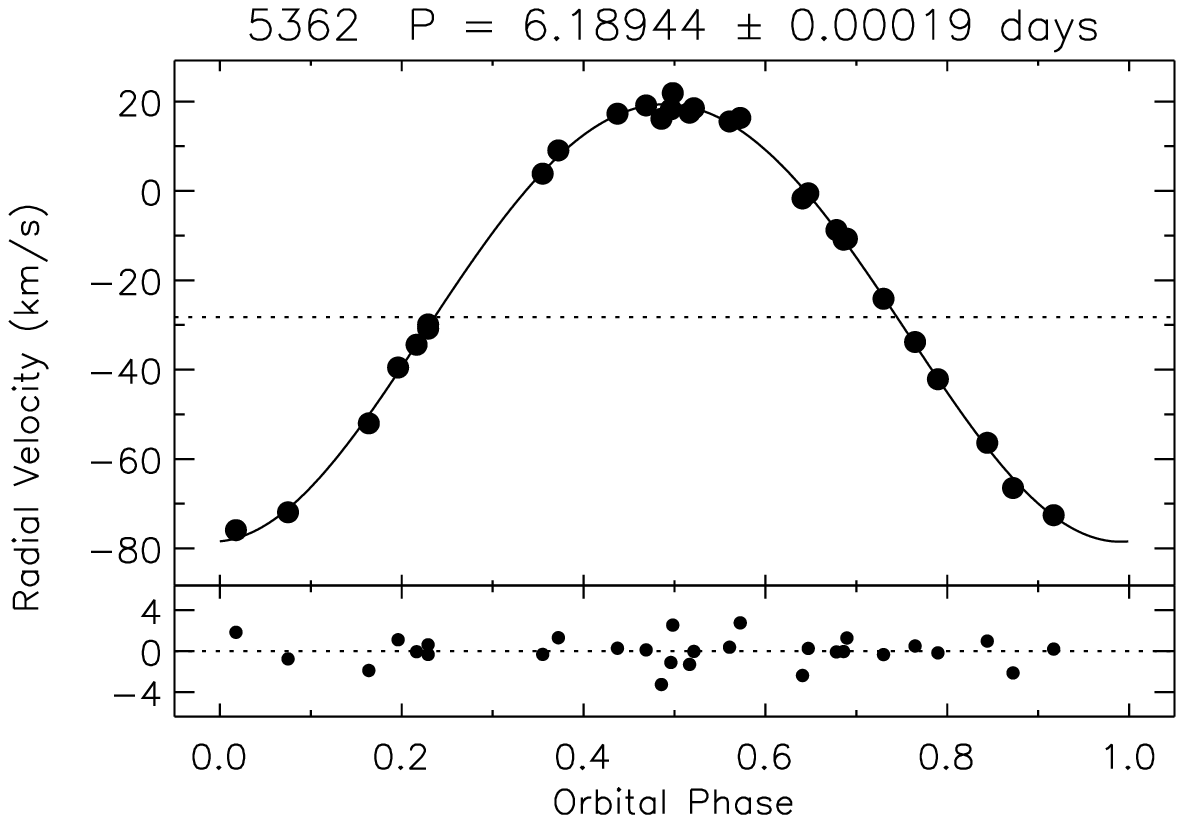,width=0.3\linewidth} \\
\epsfig{file=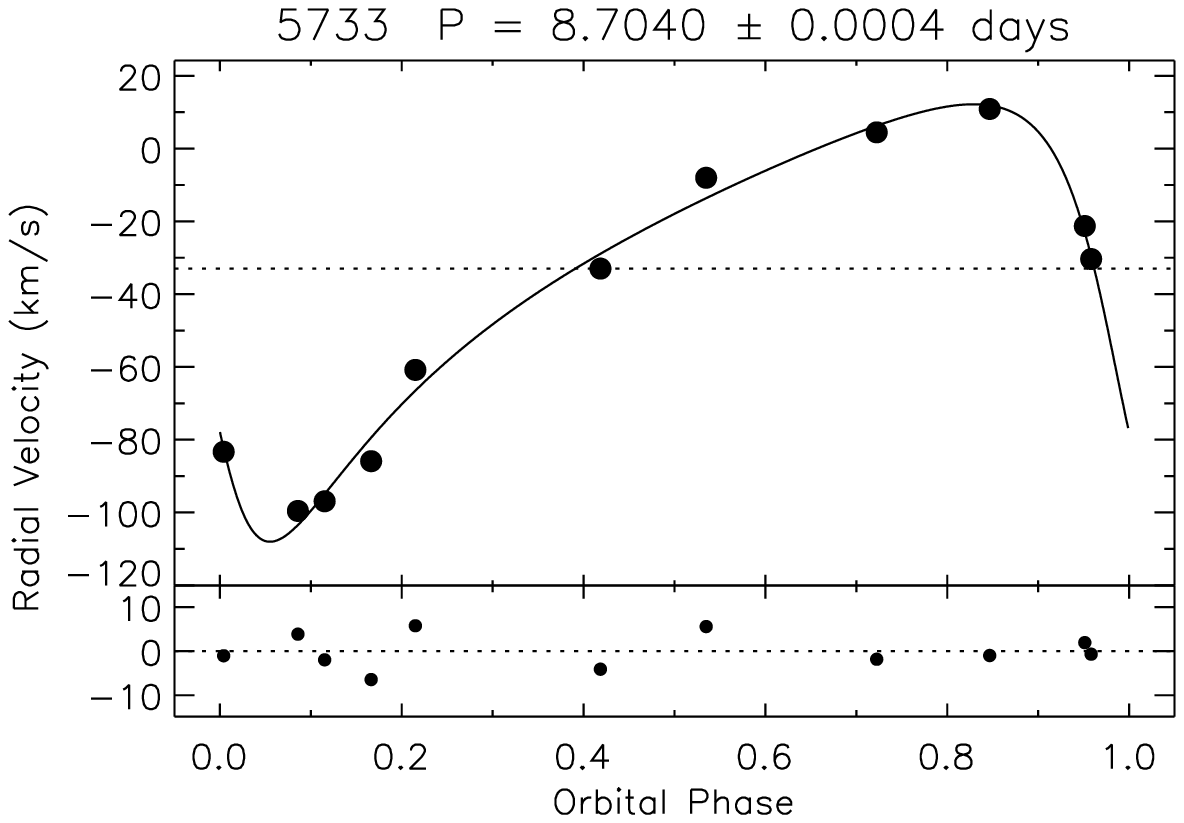,width=0.3\linewidth} & \epsfig{file=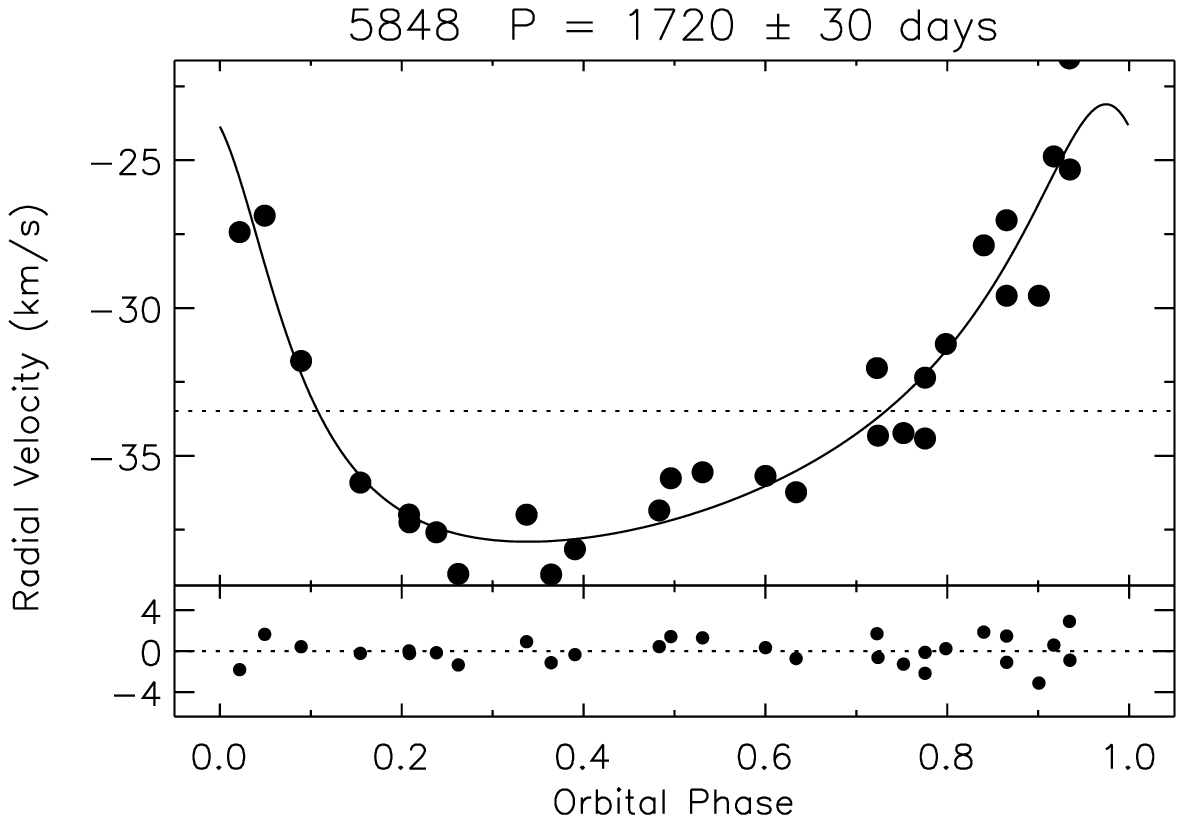,width=0.3\linewidth} & \epsfig{file=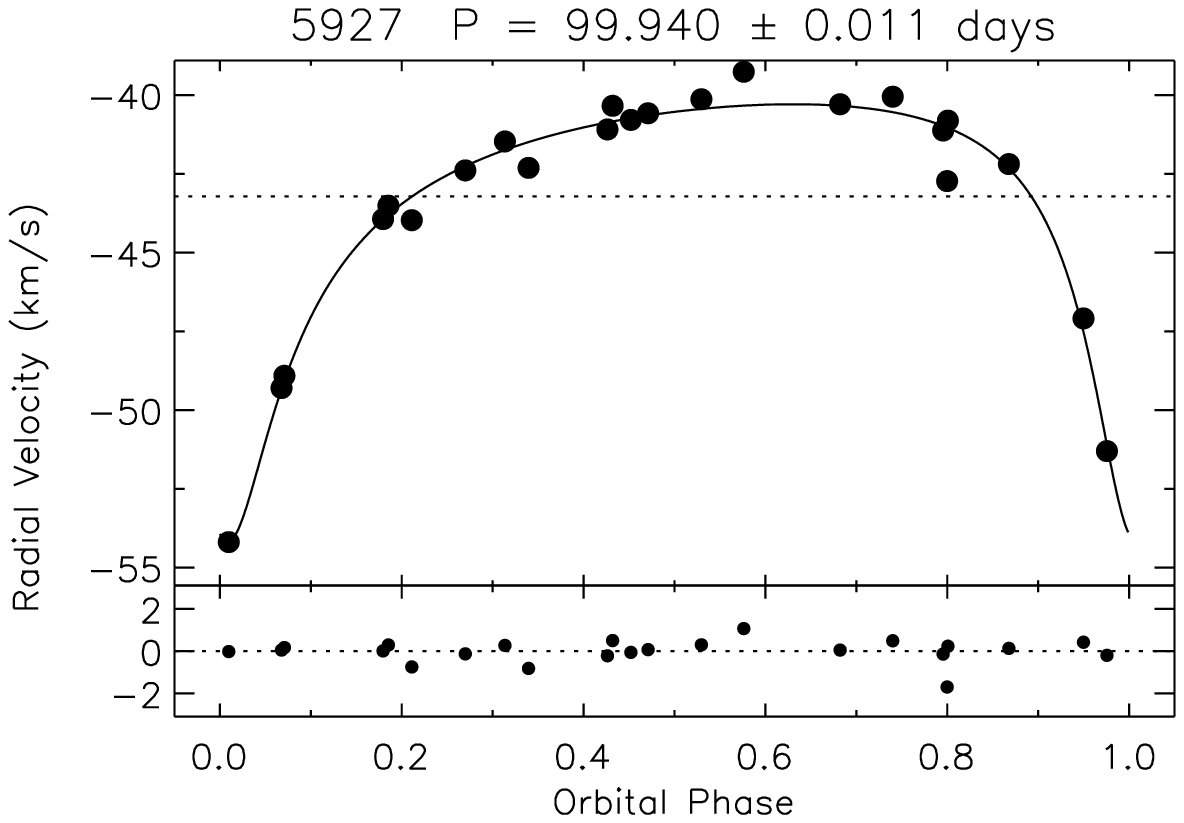,width=0.3\linewidth} \\
\epsfig{file=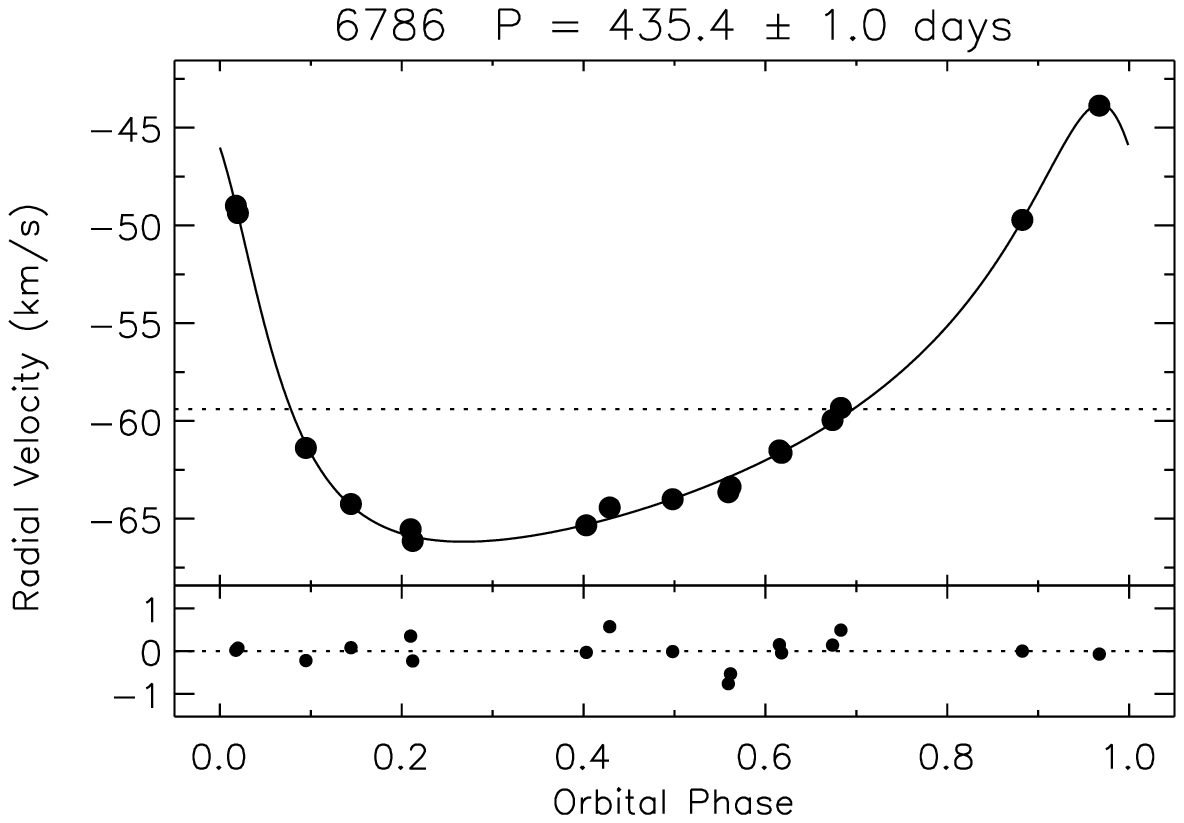,width=0.3\linewidth} & \epsfig{file=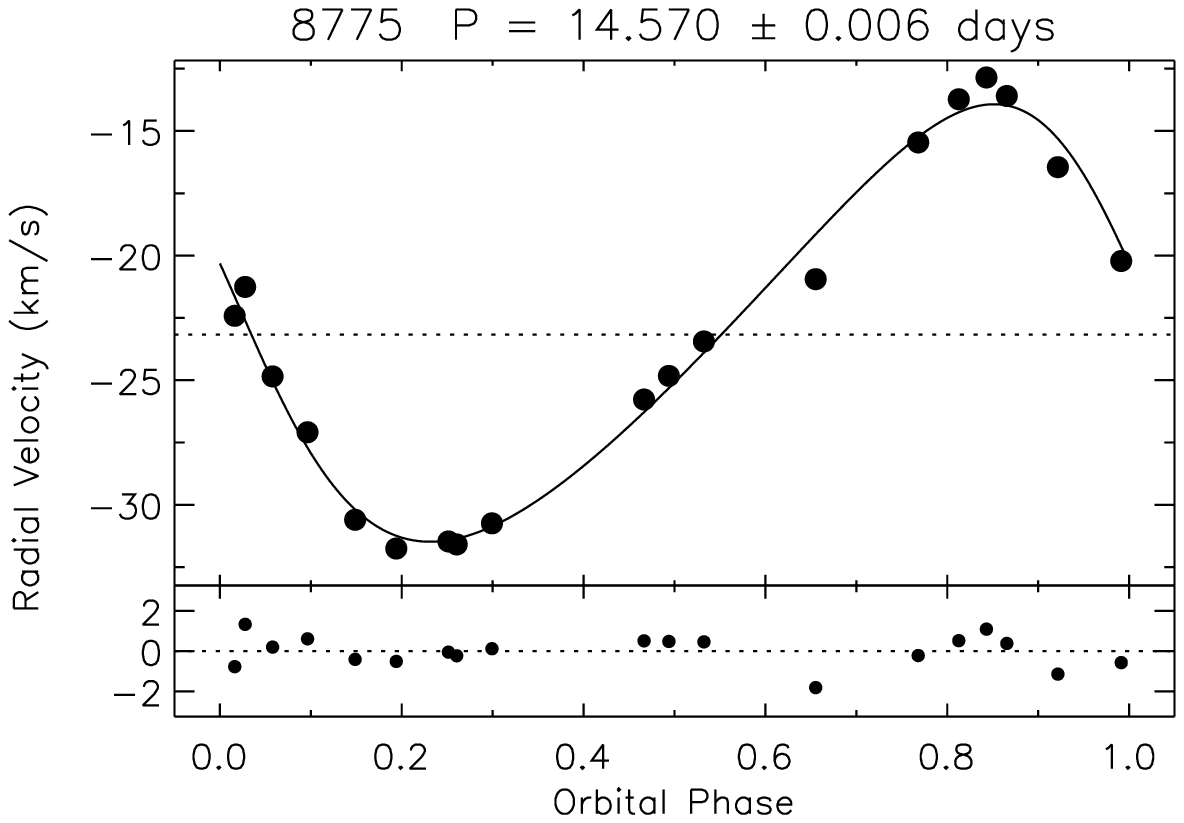,width=0.3\linewidth} & \\
\end{longtable}
\end{center}
Fig. 4. --- \footnotesize Field SB1 orbit plots.  For each binary, we plot RV against orbital phase, showing the data points with black dots and the orbital fit to the data with the solid line; the dotted line marks the $\gamma$-velocity.  Beneath each orbit plot, we show the residuals from the fit.  Above each plot, we give the binary ID and orbital period.\normalsize

\begin{center}
\begin{longtable}{ccc}
\epsfig{file=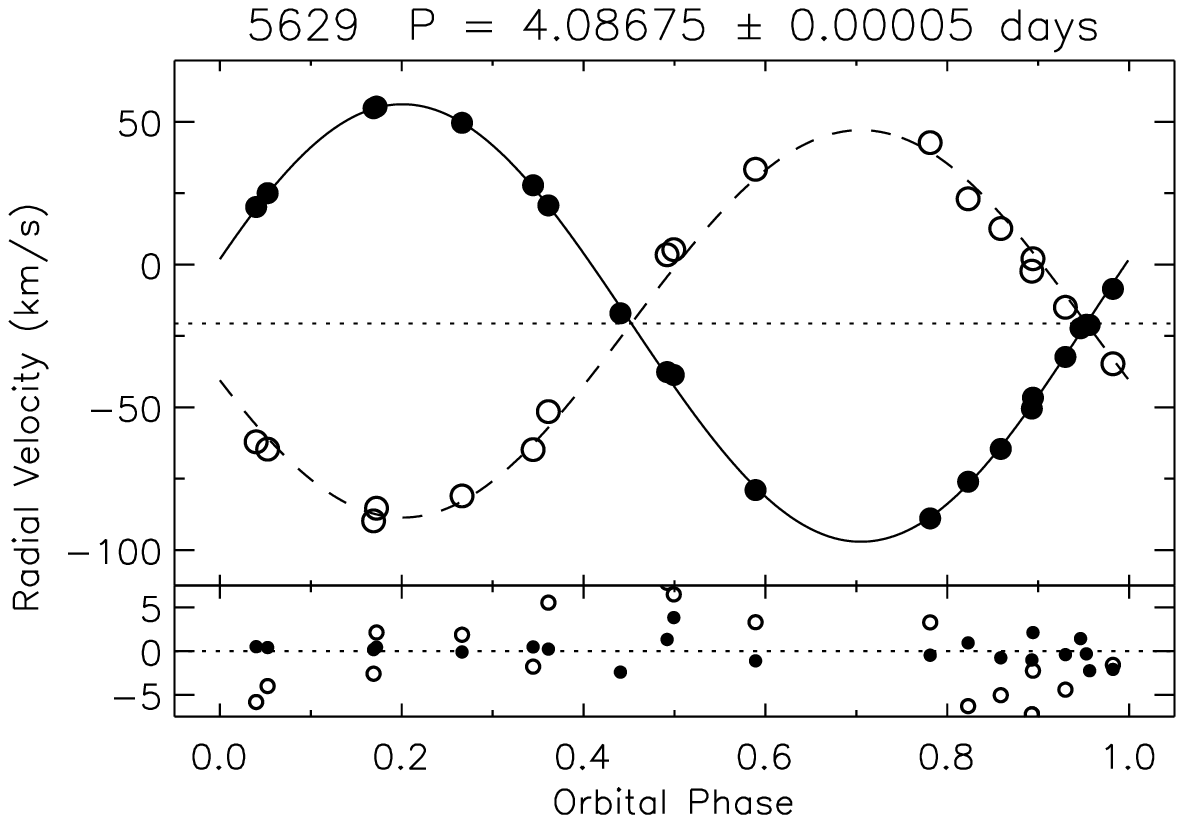,width=0.3\linewidth} & \epsfig{file=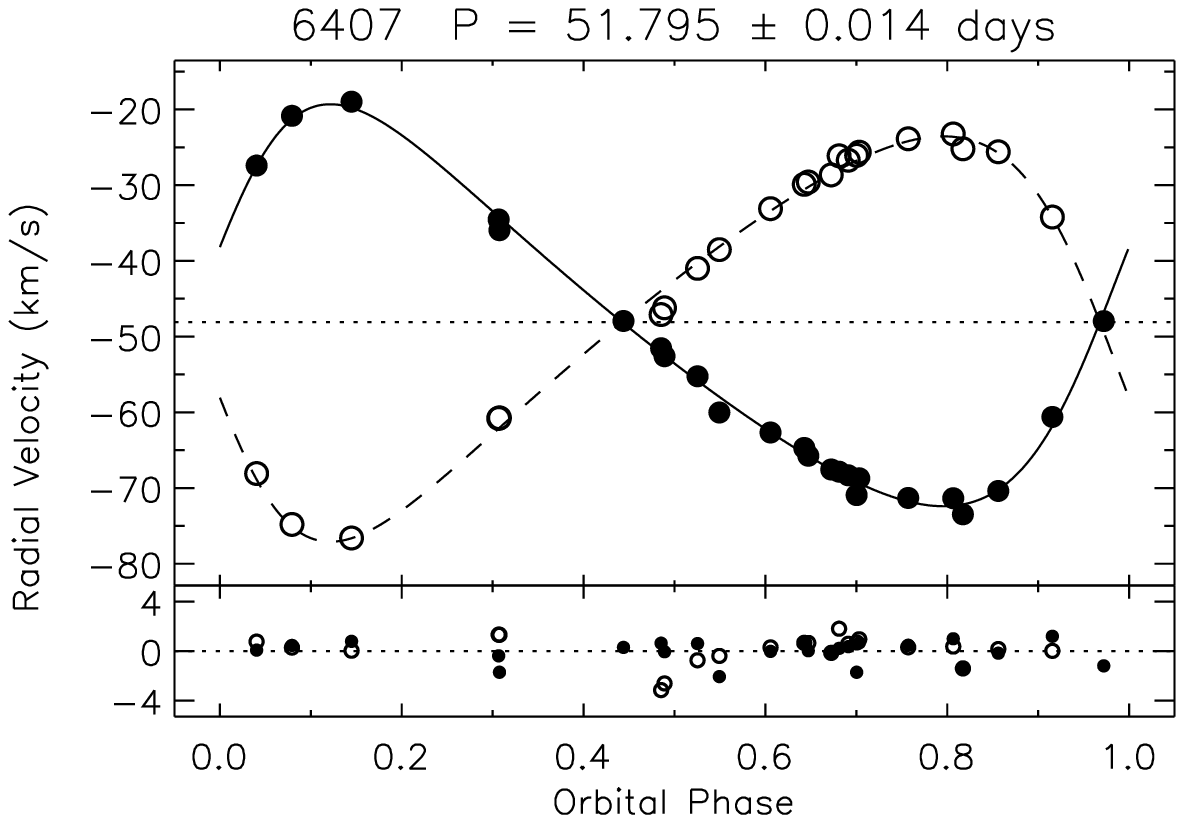,width=0.3\linewidth} & \hspace{0.3\linewidth} \\
\end{longtable}
\end{center}
Fig. 5. --- \footnotesize Field SB2 orbit plots.  For each binary, we plot RV against orbital phase, showing the primary data points with filled circles and the secondary data points with open circles. The orbital fits to the data are plotted in the solid and dashed lines for the primary and secondary stars, respectively; the dotted line marks the $\gamma$-velocity.  Beneath each orbit plot, we show the residuals from the fit.  Above each plot, we give the binary ID and orbital period.\normalsize

\addtocounter{figure}{+2}
\addtocounter{table}{-2}

\begin{deluxetable}{l r c r r r r r r r c c}
\tabletypesize{\scriptsize}
\tablewidth{0pt}
\tablecaption{Orbital Parameters For Field Single-Lined Binaries\label{fieldSB1tab}}
\tablehead{\colhead{ID} & \colhead{P} & \colhead{Orbital} & \colhead{$\gamma$} & \colhead{K} & \colhead{e} & \colhead{$\omega$} & \colhead{T$_\circ$} & \colhead{a$\sin$ i} & \colhead{f(m)} & \colhead{$\sigma$} & \colhead{N} \\
\colhead{} & \colhead{(days)} & \colhead{Cycles} & \colhead{(\kms)} & \colhead{(\kms)} & \colhead{} & \colhead{(deg)} & \colhead{(HJD-2400000 d)} & \colhead{(10$^6$ km)} & \colhead{(\Msolar)} & \colhead{(\kms)} & \colhead{}}
\startdata
    ~667 &             720 &   2.3 &          -38.61 &            2.29 &            0.22 &              16 &           52210 &            22.1 &          8.3e-4 &  0.47 &   26 \\
         &        $\pm$ 11 &       &      $\pm$ 0.10 &      $\pm$ 0.13 &      $\pm$ 0.08 &        $\pm$ 14 &        $\pm$ 30 &       $\pm$ 1.3 &    $\pm$ 1.4e-4 &       &      \\
    ~736 &          24.004 &  50.8 &          -50.50 &            7.21 &           0.210 &             240 &         52039.8 &            2.33 &          8.7e-4 &  0.42 &   17 \\
         &     $\pm$ 0.005 &       &      $\pm$ 0.11 &      $\pm$ 0.16 &     $\pm$ 0.019 &         $\pm$ 6 &       $\pm$ 0.4 &      $\pm$ 0.05 &    $\pm$ 0.6e-4 &       &      \\
    4456 &             565 &   2.9 &          -51.32 &             8.4 &            0.20 &             207 &           52117 &            63.8 &          3.2e-2 &  0.75 &   18 \\
         &         $\pm$ 4 &       &      $\pm$ 0.18 &       $\pm$ 0.3 &      $\pm$ 0.03 &        $\pm$ 10 &        $\pm$ 15 &       $\pm$ 2.1 &    $\pm$ 0.3e-2 &       &      \\
    5110 &         21.3602 & 163.2 &           -38.0 &            37.8 &            0.10 &             195 &         51197.2 &            11.1 &         1.18e-1 &  3.07 &   22 \\
         &    $\pm$ 0.0019 &       &       $\pm$ 0.7 &       $\pm$ 0.9 &      $\pm$ 0.03 &        $\pm$ 16 &       $\pm$ 0.9 &       $\pm$ 0.3 &    $\pm$ 0.9e-2 &       &      \\
    5121 &           156.0 &   7.3 &          -41.37 &             7.7 &            0.16 &             243 &           51217 &            16.4 &          7.2e-3 &  0.73 &   26 \\
         &       $\pm$ 0.3 &       &      $\pm$ 0.17 &       $\pm$ 0.3 &      $\pm$ 0.04 &        $\pm$ 12 &         $\pm$ 5 &       $\pm$ 0.7 &    $\pm$ 0.9e-3 &       &      \\
    5362 &         6.18944 & 178.9 &           -28.2 &            49.0 &           0.027 &             183 &         51035.3 &            4.17 &         7.52e-2 &  1.51 &   29 \\
         &   $\pm$ 0.00019 &       &       $\pm$ 0.3 &       $\pm$ 0.5 &     $\pm$ 0.009 &        $\pm$ 20 &       $\pm$ 0.4 &      $\pm$ 0.04 &    $\pm$ 2.2e-3 &       &      \\
    5733 &          8.7040 & 356.8 &           -33.0 &              60 &            0.50 &             120 &        51606.69 &             6.2 &          1.3e-1 &  5.49 &   11 \\
         &    $\pm$ 0.0004 &       &       $\pm$ 2.2 &         $\pm$ 5 &      $\pm$ 0.05 &         $\pm$ 7 &      $\pm$ 0.10 &       $\pm$ 0.6 &    $\pm$ 0.3e-1 &       &      \\
    5848 &            1720 &   2.0 &           -33.5 &             7.4 &            0.45 &              26 &           51090 &             156 &          5.2e-2 &  1.48 &   29 \\
         &        $\pm$ 30 &       &       $\pm$ 0.3 &       $\pm$ 0.7 &      $\pm$ 0.07 &         $\pm$ 7 &        $\pm$ 24 &        $\pm$ 15 &    $\pm$ 1.4e-2 &       &      \\
    5927 &          99.940 & 113.5 &          -43.21 &             6.9 &           0.600 &             165 &         48271.1 &             7.6 &         1.78e-3 &  0.62 &   23 \\
         &     $\pm$ 0.011 &       &      $\pm$ 0.13 &       $\pm$ 0.3 &     $\pm$ 0.023 &         $\pm$ 3 &       $\pm$ 0.6 &       $\pm$ 0.4 &    $\pm$ 2.5e-4 &       &      \\
    6786 &           435.4 &   3.7 &          -59.39 &           11.19 &           0.491 &            36.7 &         51935.2 &            58.4 &         4.18e-2 &  0.39 &   17 \\
         &       $\pm$ 1.0 &       &      $\pm$ 0.10 &      $\pm$ 0.19 &     $\pm$ 0.013 &       $\pm$ 1.8 &       $\pm$ 1.4 &       $\pm$ 1.1 &    $\pm$ 2.2e-3 &       &      \\
    8775 &          14.570 &  50.8 &          -23.18 &             8.8 &            0.20 &              74 &         50890.4 &            1.72 &          9.6e-4 &  0.90 &   19 \\
         &     $\pm$ 0.006 &       &      $\pm$ 0.22 &       $\pm$ 0.3 &      $\pm$ 0.04 &         $\pm$ 9 &       $\pm$ 0.3 &      $\pm$ 0.06 &    $\pm$ 1.0e-4 &       &      \\
\enddata
\end{deluxetable}

\begin{deluxetable}{l r c r r r r r r r r c c}
\tabletypesize{\scriptsize}
\tablewidth{0pt}
\rotate
\tablecaption{Orbital Parameters For Field Double-Lined Binaries\label{fieldSB2tab}}
\tablehead{\colhead{ID} & \colhead{P} & \colhead{Orbital} & \colhead{$\gamma$} & \colhead{K} & \colhead{e} & \colhead{$\omega$} & \colhead{T$_\circ$} & \colhead{a$\sin$ i} & \colhead{m$\sin^3$ i} & \colhead{q} & \colhead{$\sigma$} & \colhead{N} \\
\colhead{} & \colhead{(days)} & \colhead{Cycles} & \colhead{(\kms)} & \colhead{(\kms)} & \colhead{} & \colhead{(deg)} & \colhead{(HJD-2400000 d)} & \colhead{(10$^6$ km)} & \colhead{(\Msolar)} & \colhead{} & \colhead{(\kms)} & \colhead{}}
\startdata
    5629 &         4.08675 & 237.0 &           -20.6 &            76.6 &           0.007 &             290 &         51336.4 &            4.31 &            0.60 &            1.13 &  1.59 &   21 \\
         &   $\pm$ 0.00005 &       &       $\pm$ 0.4 &       $\pm$ 0.7 &     $\pm$ 0.007 &        $\pm$ 60 &       $\pm$ 0.7 &      $\pm$ 0.04 &      $\pm$ 0.04 &      $\pm$ 0.04 &       &      \\
         &                 &       &                 &            67.8 &                 &                 &                 &            3.81 &            0.68 &                 &  5.14 &   17 \\
         &                 &       &                 &       $\pm$ 2.0 &                 &                 &                 &      $\pm$ 0.12 &      $\pm$ 0.03 &                 &       &      \\
    6407 &          51.795 &  19.8 &          -48.08 &            26.5 &           0.289 &           286.8 &         50937.5 &           18.10 &           0.358 &           0.992 &  0.97 &   24 \\
         &     $\pm$ 0.014 &       &      $\pm$ 0.18 &       $\pm$ 0.3 &     $\pm$ 0.009 &       $\pm$ 2.1 &       $\pm$ 0.3 &      $\pm$ 0.24 &     $\pm$ 0.013 &     $\pm$ 0.021 &       &      \\
         &                 &       &                 &            26.8 &                 &                 &                 &            18.3 &           0.355 &                 &  1.27 &   22 \\
         &                 &       &                 &       $\pm$ 0.4 &                 &                 &                 &       $\pm$ 0.3 &     $\pm$ 0.011 &                 &       &      \\
\enddata
\end{deluxetable}

\end{document}